\documentclass[iop,useAMS]{emulateapj}

\usepackage{graphicx}
\usepackage[percent]{overpic}
\usepackage{multirow}
\usepackage{color}
\usepackage{rotating}

\newcommand{\msun}{${\rm M_{\sun}}$}

\def\ltsima{$\; \buildrel < \over \sim \;$}
\def\simlt{\lower.5ex\hbox{\ltsima}}
\def\gtsima{$\; \buildrel > \over \sim \;$}
\def\simgt{\lower.5ex\hbox{\gtsima}}
\def\kpc{{\rm\,kpc}}

\def\msun{{\rm\,M_\odot}}
\def\lsun{{\rm\,L_\odot}}
\newcommand{\fmmm}[1]{\mbox{$#1$}}
\newcommand{\scnd}{\mbox{\fmmm{''}\hskip-0.3em .}}

\newcommand{\mcnp}{\mbox{\fmmm{'}}}

\def\deg{^\circ}
\def\degg{\hbox{$\null^\circ$\hskip-3pt .}}

\def\Gyr{{\rm\,Gyr}}

\def\ltsima{$\; \buildrel < \over \sim \;$}
\def\gtsima{$\; \buildrel > \over \sim \;$}

\shorttitle{The halo of M31}
\shortauthors{Ibata et al.}

\begin{document}

\title{The large-scale structure of the halo of the Andromeda Galaxy Part I:\\
global stellar density, morphology and metallicity properties\altaffilmark{1}}

\author{Rodrigo A. Ibata\altaffilmark{2}}
\author{Geraint F. Lewis\altaffilmark{3}}
\author{Alan W. McConnachie\altaffilmark{4}}
\author{Nicolas F. Martin\altaffilmark{2,5}}
\author{Michael J. Irwin\altaffilmark{6}}
\author{Annette M. N. Ferguson\altaffilmark{7}}
\author{Arif Babul\altaffilmark{8}}
\author{Edouard J. Bernard\altaffilmark{7}}
\author{Scott C. Chapman\altaffilmark{9}}
\author{Michelle Collins\altaffilmark{5}}
\author{Mark Fardal\altaffilmark{10}}
\author{A.D. Mackey\altaffilmark{11}}
\author{Julio Navarro\altaffilmark{8}}
\author{Jorge Pe\~narrubia\altaffilmark{7}}
\author{R. Michael Rich\altaffilmark{12}}
\author{Nial Tanvir\altaffilmark{13}}
\author{Lawrence Widrow\altaffilmark{14}}

\altaffiltext{1}{Based on observations obtained with MegaPrime/MegaCam, a joint project of CFHT and CEA/DAPNIA, at the Canada-France-Hawaii Telescope (CFHT) which is operated by the National Research Council (NRC) of Canada, the Institute National des Sciences de l'Univers of the Centre National de la Recherche Scientifique of France, and the University of Hawaii.}

\altaffiltext{2}{Observatoire astronomique de Strasbourg, Universit\'e de Strasbourg, CNRS, UMR 7550, 11 rue de lÕUniversit\'e, F-67000 Strasbourg, France; rodrigo.ibata@astro.unistra.fr}
\altaffiltext{3}{Institute of Astronomy, School of Physics A28, University of Sydney, NSW 2006, Australia}
\altaffiltext{4}{NRC Herzberg Institute of Astrophysics, 5071 West Saanich Road, Victoria, BC, V9E 2E7, Canada}
\altaffiltext{5}{Max-Planck-Institut f\"ur Astronomie, K\"onigstuhl 17, D-69117 Heidelberg, Germany}
\altaffiltext{6}{Institute of Astronomy, University of Cambridge, Madingley Road, Cambridge CB3 0HA, UK}
\altaffiltext{7}{Institute for Astronomy, University of Edinburgh, Blackford Hill, Edinburgh EH9 3HJ, UK}
\altaffiltext{8}{Department of Physics and Astronomy, University of Victoria, 3800 Finnerty Road, Victoria, British Columbia, Canada V8P 5C2}
\altaffiltext{9}{Department of Physics and Atmospheric Science, Dalhousie University, 6310 Coburg Rd., Halifax, NS  B3H 4R2 Canada}
\altaffiltext{10}{University of Massachusetts, Department of Astronomy, LGRT 619-E, 710 N. Pleasant Street, Amherst, Massachusetts 01003-9305, USA.}
\altaffiltext{11}{RSAA, The Australian National University, Mount Stromlo Observatory, Cotter Road, Weston Creek, ACT 2611, Australia}
\altaffiltext{12}{Department of Physics and Astronomy, University of California, Los Angeles, PAB, 430 Portola Plaza, Los Angeles, California 90095-1547, USA}
\altaffiltext{13}{Department of Physics and Astronomy, University of Leicester, University Road, Leicester, LE1 7RH, UK}
\altaffiltext{14}{Department of Physics, Engineering Physics, and Astronomy Queen's University, Kingston, Ontario, Canada K7L 3N6}

\begin{abstract}
We present an analysis of the large-scale structure of the halo of the Andromeda galaxy, based on the Pan-Andromeda Archeological Survey (PAndAS), currently the most complete map of resolved stellar populations in any galactic halo. Despite the presence of copious substructure, the global halo populations follow closely power law profiles that become steeper with increasing metallicity. We divide the sample into stream-like populations and a smooth halo component (defined as the population that cannot be resolved into spatially distinct substructure with PAndAS). Fitting a three-dimensional halo model reveals that the most metal-poor populations (${\rm [Fe/H] < -1.7}$) are distributed approximately spherically (slightly prolate with ellipticity $c/a=1.09\pm0.03$), with only a relatively small fraction residing in discernible stream-like structures ($f_{stream} = 42\%$). The sphericity of the ancient smooth component strongly hints that the dark matter halo is also approximately spherical. More metal-rich populations contain higher fractions of stars in streams, with $f_{stream}$ becoming as high as 86\% for ${\rm [Fe/H] > -0.6}$. The space density of the smooth metal-poor component has a global power-law slope of $\gamma=-3.08 \pm 0.07$, and a non-parametric fit shows that the slope remains nearly constant from $30\kpc$ to $\sim 300\kpc$. The total stellar mass in the halo at distances beyond $2\deg$ is $\sim 1.1\times 10^{10}\msun$, while that of the smooth component is $\sim 3\times 10^{9}\msun$. Extrapolating into the inner galaxy, the total stellar mass of the smooth halo is plausibly $\sim 8\times 10^{9}\msun$. We detect a substantial metallicity gradient, which declines from $\langle[Fe/H]\rangle=-0.7$ at $R=30\kpc$ to  $\langle[Fe/H]\rangle=-1.5$ at $R=150\kpc$ for the full sample, with the smooth halo being $\sim0.2$~dex more metal poor than the full sample at each radius. While qualitatively in-line with expectations from cosmological simulations, these observations are of great importance as they provide a prototype template that such simulations must now be able to reproduce in quantitative detail.
\end{abstract}

\keywords{galaxies: halos Ð galaxies: individual (M31) Ð galaxies: structure}

\section{Introduction}
\label{sec:Introduction}

The stellar halos of giant galaxies are repositories of much of the stellar material that was accreted during the formation of their host galaxy. It is in these regions alone that we can resolve and identify (in extra-Galactic systems) the remnants of the numerous low-mass accretion events that contributed to the build-up of present-day galaxies. Halo studies are further motivated by the fact that modern simulations in the standard $\Lambda$CDM cosmology make predictions for the spatial distribution, age and metallicity of stars in the halo \citep[e.g.,][]{2006MNRAS.365..747A,Bullock:2005is,Johnston:2008jp,Font:2008jh,2010MNRAS.406..744C,2010ApJ...721..738Z}, so that detailed observations provide a means to test this cosmological paradigm.

It was for these reasons that we began, a full decade ago, an in-depth panoramic study of the Andromeda galaxy on the Canada-France-Hawaii Telescope (CFHT), an effort that morphed into the Pan-Andromeda Archeological Survey (PAndAS), which was set up as a ``Large Program'' on the CFHT in 2008. As the closest giant galaxy, the choice of Andromeda (M31) for the target is obvious; at present (before the advent of Gaia or homogenous digital full-sky optical surveys) one can also argue that for matters of homogenous spatial coverage the M31 halo is a far better and easier specimen for study than the halo of our own Milky Way. 

The aim of the present paper is to update and complement an earlier contribution on this subject \citep[][hereafter paper I]{Ibata:2007jr}. In that article we gave a full exposition of the CFHT survey from data obtained up to 2006, with almost all of the studied fields located in the southern quadrant of M31. That contribution also provides a detailed review of the literature prior to 2007 on galactic halos, focussing particularly on M31, to which we refer the interested reader. 

Since 2007 significant progress has been made in the photometric and kinematic mapping of the haloes of nearby galaxies. Perhaps the greatest surprises have come from the analyses of the Milky Way halo, which has turned out to be much more complex than previously imagined. \citet{Carollo:2007fw} showed that the Sloan Digital Sky Survey (SDSS) halo sample can be deconstructed into two sub-components, the inner and outer halos, that have different morphological and chemical properties, with the outer halo being more metal poor and possessing slightly retrograde mean rotation, but see also \citet{2011MNRAS.415.3807S}. In addition, the inner halo appears to contain itself a highly-flattened sub-component \citep{Morrison:2009bb}, with axis ratio $c/a \sim 0.2$. Nearby main sequence halo dwarfs in the SDSS extending out to $\sim 20\kpc$ (i.e. encompassing both the ``inner'' and ``outer'' halos) are found to be distributed as a smooth oblate structure (axis ratio $c/a \sim 0.5$--$0.8$) with space density that follows a power law of index $-2.5$ to $-3$, once the known halo substructures are removed \citep{Juric:2008ie}. It is very interesting that the incidence of the halo substructures appears to be similar to that found in cosmologically-motivated stellar halo formation simulations \citep{Bell:2008bc}. Note however, that recent analyses of distant blue horizontal-branch (BHB) stars \citep{2011ApJ...738...79X,2011MNRAS.416.2903D}, which represent very old stellar populations, and also of main sequence turn-off stars \citep{2011ApJ...733L...7H}, suggest that the incidence of substructure may be slightly lower than expected from simulations.

Great strides have also been made in the mapping of the halos of external galaxies. Through the use of novel detection techniques, a great wealth of substructure has been identified via integrated surface brightness measurements \citep{2008ApJ...689..184M,2010AJ....140..962M,Paudel:2013ww} despite the extremely faint nature of the features (extending down to $\sim 29 \, {\rm mag \, arcsec^{-2}}$). While these studies are still in progress, integrated surface brightness surveys appear to be a promising avenue to identify and quantify fossil remnants of intermediate mass accretions that have occurred in the last few Gyr out to a distance of several tens of Mpc.

Halo surveys based on resolved stellar populations are necessarily limited to nearer galaxies, although the information that can be recovered is significantly more detailed. In this endeavor the Hubble Space Telescope has been a powerful tool in recent years (see, e.g., \citealt{2009ApJS..183...67D}, and references therein). A particularly interesting case is NGC~891, a Milky Way analogue, for which deep HST imaging showed evidence for clear signs of spatial substructure in its halo, with similar statistical significance as was identified in the Milky Way by \citet{Bell:2008bc}, but in addition strong spatial variations in chemical abundance were also found \citep{2009MNRAS.395..126I}. Modern wide-field cameras on 8~m telescopes are now beginning to open up this field of study to ground-based observatories \citep{2009AJ....138.1469B,Mouhcine:2010cz,Jablonka:2010gp,2011ApJ...736...24B,Crnojevic2013} for galaxies within $\sim 10$~Mpc.

Nevertheless, M31 remains an important and unique stepping stone in our efforts to explore those more distant stellar halos, due to our current ability to probe the stellar populations in M31 both to extreme photometric depths \citep{2009MNRAS.396.1842R}, including down to the main sequence turnoff in a handful of HST fields \citep{Brown:2006iq,Brown:2007gq,Bernard:2012kz}, and over an extremely wide area, such as in the present study. A previous contribution \citep{2013MNRAS.428.1248C}, based also on PAndAS data, detected the halo of M33 (a satellite of M31) out to very large radii, revealing hints of asymmetric morphology. Recent spectroscopic studies have shown evidence of a metal-poor halo in M31 \citep{Chapman:2006ia,Kalirai:2006ix,2008ApJ...689..958K}, which had been largely missed in earlier work due to the presence of substantial contamination from the so-called southern Giant Stellar Stream \citep{Ibata:2001vs}, which pollutes almost all the inner halo region. The large spectroscopic study  of \citet[][hereafter G12]{2012ApJ...760...76G}, which probed the halo at the location of 38 fields with Keck/DEIMOS, found that the projected stellar density falls off with a power-law index of $-2.2 \pm 0.2$, and that globally the halo is slightly prolate. However, the effectively pencil-beam nature of the (relatively) small $5\mcnp \times 16\mcnp$ DEIMOS fields means that it is unclear to what extent these results are affected by the substructure found in the M31 halo \citep{2009Natur.461...66M,Tanaka:2011hm}. 

The layout of the paper is as follows. Section \ref{sec:Survey} presents an overview of the data and reductions, \S3 shows the spatial distribution of the halo stars as a function of metallicity, while \S4 presents a structural analysis of the halo. We discuss the implication of these findings in \S5, and draw our conclusions in \S6.

\begin{figure*}
\begin{center}
  \begin{overpic}[viewport= 90 30 700 580, clip, width=\hsize]{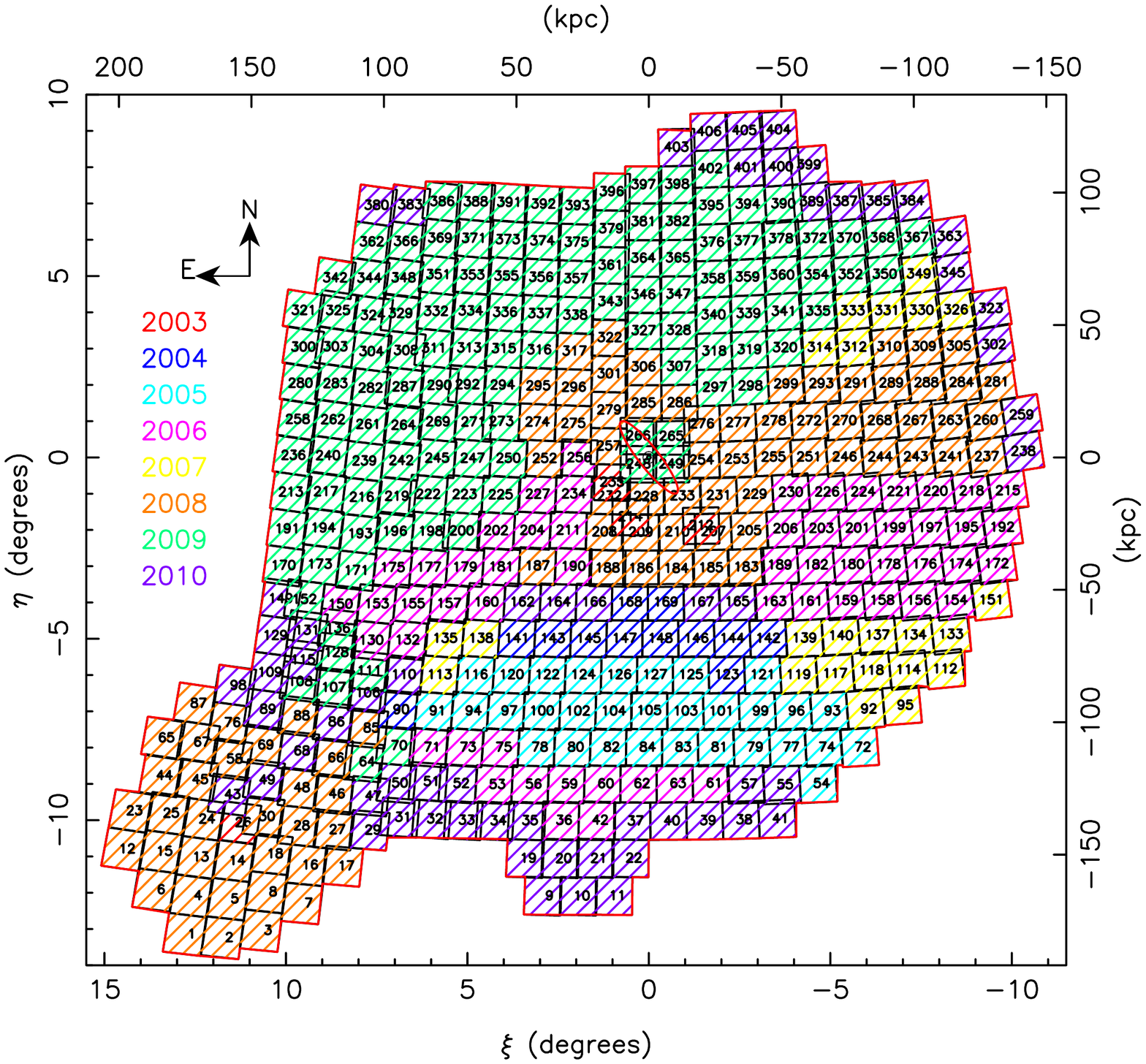}
        \put(12,73){\includegraphics[viewport= 22 114 589 680, clip, width=1.3cm]{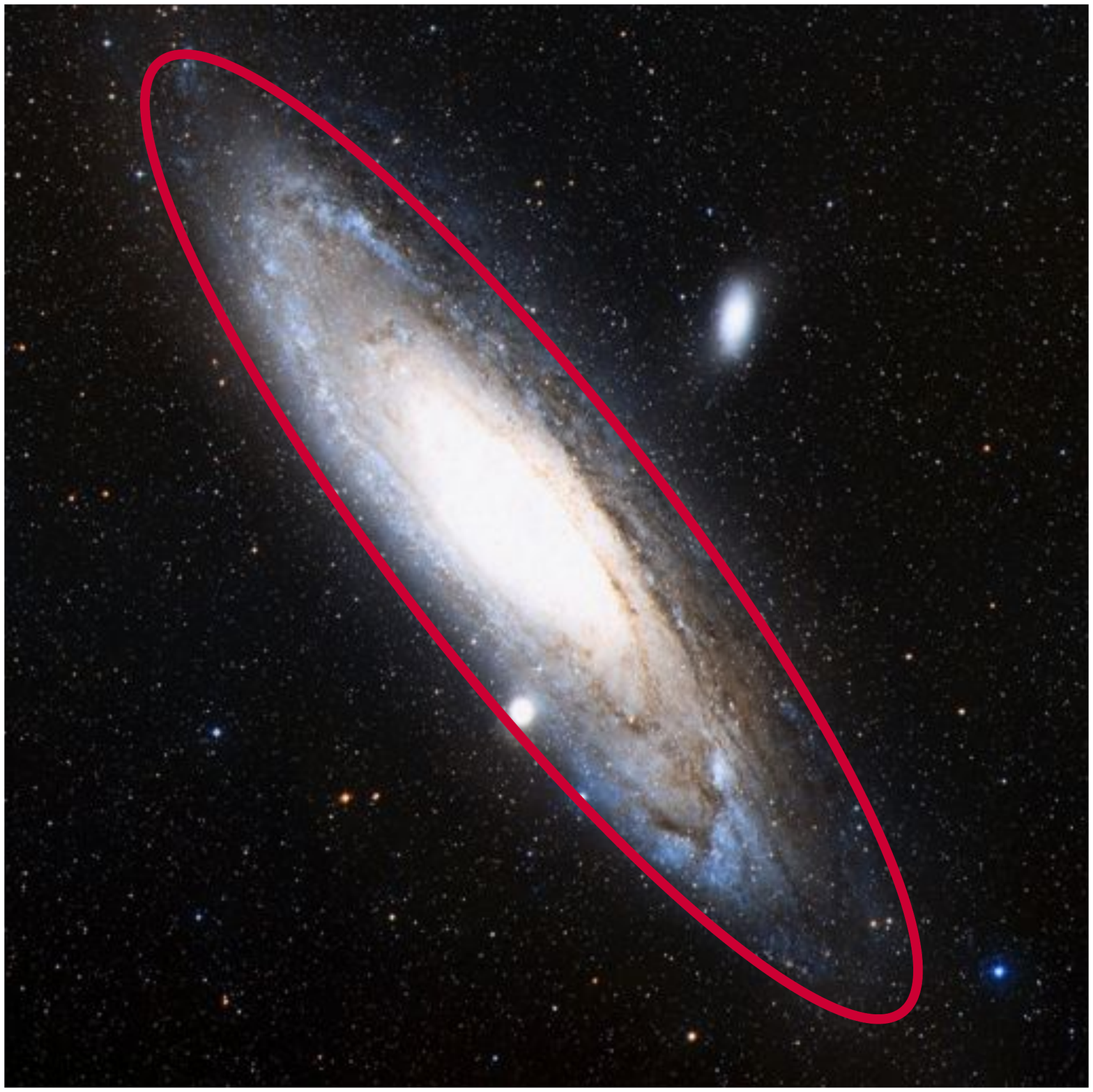}}  
  \end{overpic}
\end{center}
\caption{The chosen CFHT tiling pattern of the region around M31 (on which the coordinates are centered) is displayed, with the colors marking the year of observation of the i-band images from 2003 to 2010, as indicated in the legend. The g-band data have an almost identical tiling pattern and temporal (observing year) distribution. Each tile represents a CFHT/MegaPrime field. The inner red ellipse represents a disk of inclination $77\deg$ and radius $1\degg25$ ($17\kpc$), the approximate edge of the ``classical'' regular stellar disk. This same ellipse is reproduced in the image insert on the top left, a view of M31 constructed from Palomar photographic plates.}
\label{fig:field_numbers}
\end{figure*}

\section{The Survey}
\label{sec:Survey}

The imaging data that provided the foundations for the PAndAS survey were obtained with the MegaCam wide-field camera at the CFHT from 2003 to 2010. All fields were observed in both the g and i bands. Although the bulk of the fields were taken as part of the dedicated CFHT Large Program, a substantial number of fields were taken from earlier programs (PIs Ibata and McConnachie). A full account of the data and the data processing will be provided in a future contribution (McConnachie et al., in preparation), but we provide here a summary necessary for the present analysis. 

The CFHT MegaCam imager is a mosaic of 36 (usable) individual $2048\times4612$ CCDs (340 Megapixels) arranged in a $9\times4$ grid. With a pixel scale of $0.187\, {\rm arcsec/pixel}$, its field of view is $0\degg96 \times 0\degg94$. An overview of the chosen tiling pattern is shown in Fig.~\ref{fig:field_numbers}, where the color codes the year of observation. As can be seen from the diagram, the survey covers an extremely large field around the M31 galaxy almost fully encompassing a $150\kpc$ radius circle in projection, along with an extension out to M33 (located at $\xi=11\degg3$, $\eta=-10\degg1$ on this map).

\begin{figure}
\begin{center}
\includegraphics[angle=0, viewport= 90 30 710 580, clip, width=\hsize]{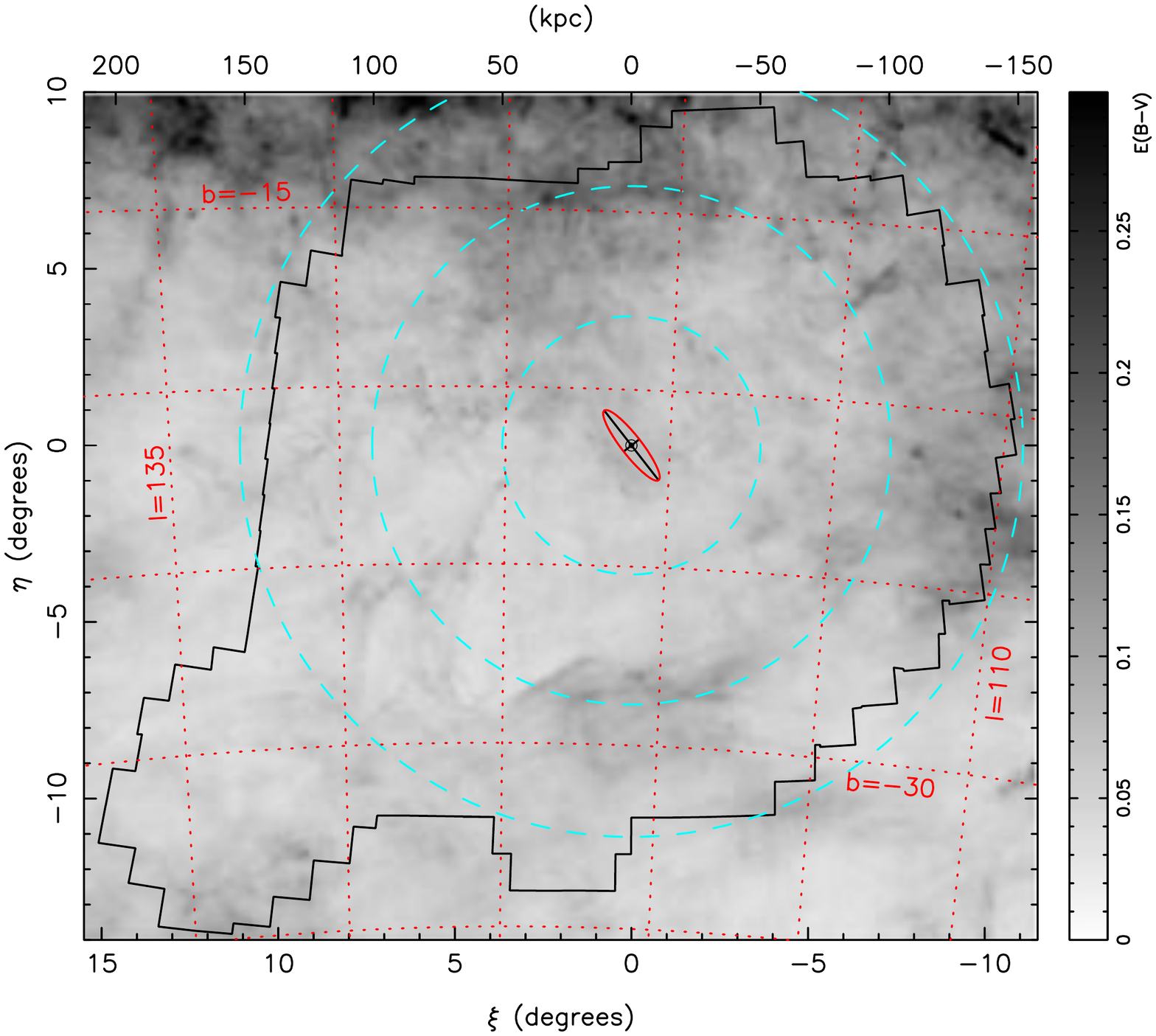}
\end{center}
\caption{The ${\rm E(B-V)}$ reddening over the survey region according to the \citep{Schlegel:1998fw} maps. Extinction increases northward toward the Galactic plane, and is typically ${\rm E(B-V) \sim 0.07}$ in the vicinity of M31. The irregular polygon marks the footprint of the PAndAS survey. The inner ellipse has the same meaning as in Fig.~\ref{fig:field_numbers}, while the dashed circles centered on M31 correspond to projected radii of $50\kpc$, $100\kpc$ and $150\kpc$.}
\label{fig:extinction}
\end{figure}

The Andromeda galaxy is located at relatively low Galactic latitude ($\ell=121\degg2$, $b=-21\degg6$), and consequently its environment suffers from non-negligible foreground extinction. Figure~\ref{fig:extinction} shows the distribution of the interstellar reddening ${\rm E(B-V)}$ over the PAndAS survey region, as estimated by \citet{Schlegel:1998fw}. It can be appreciated from this map that we refrained from extending the survey further to the North so as to avoid more highly extincted regions of sky. Within the PAndAS footprint (but avoiding the central $2\deg$ around M31), the minimum and maximum values of the extinction are 0.034 and 0.24, with an average value of 0.077, and an r.m.s. scatter of 0.028.
The reddening was converted into extinction in the MegaCam $g$ and $i$ bands, using the following relations:
\begin{eqnarray}
g_0  &=& g-E(B-V) \times 3.793 \nonumber \\
i_0  &=& i-E(B-V) \times 2.086 \, .
\end{eqnarray}

\begin{figure}
\begin{center}
\vbox{
\includegraphics[angle=0, viewport= 140 50 770 580, clip, width=9cm]{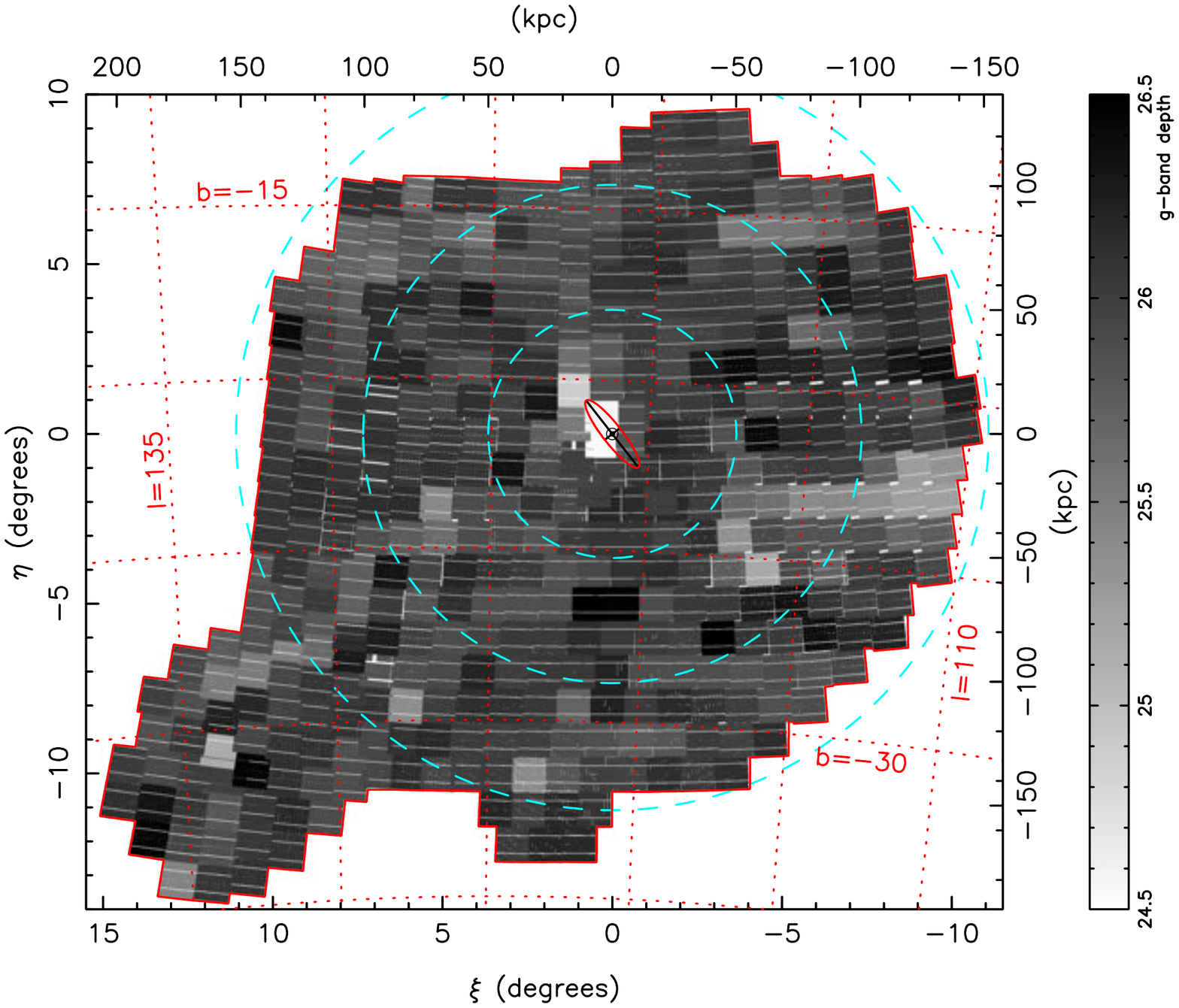}
\includegraphics[angle=0, viewport= 140 50 770 580, clip, width=9cm]{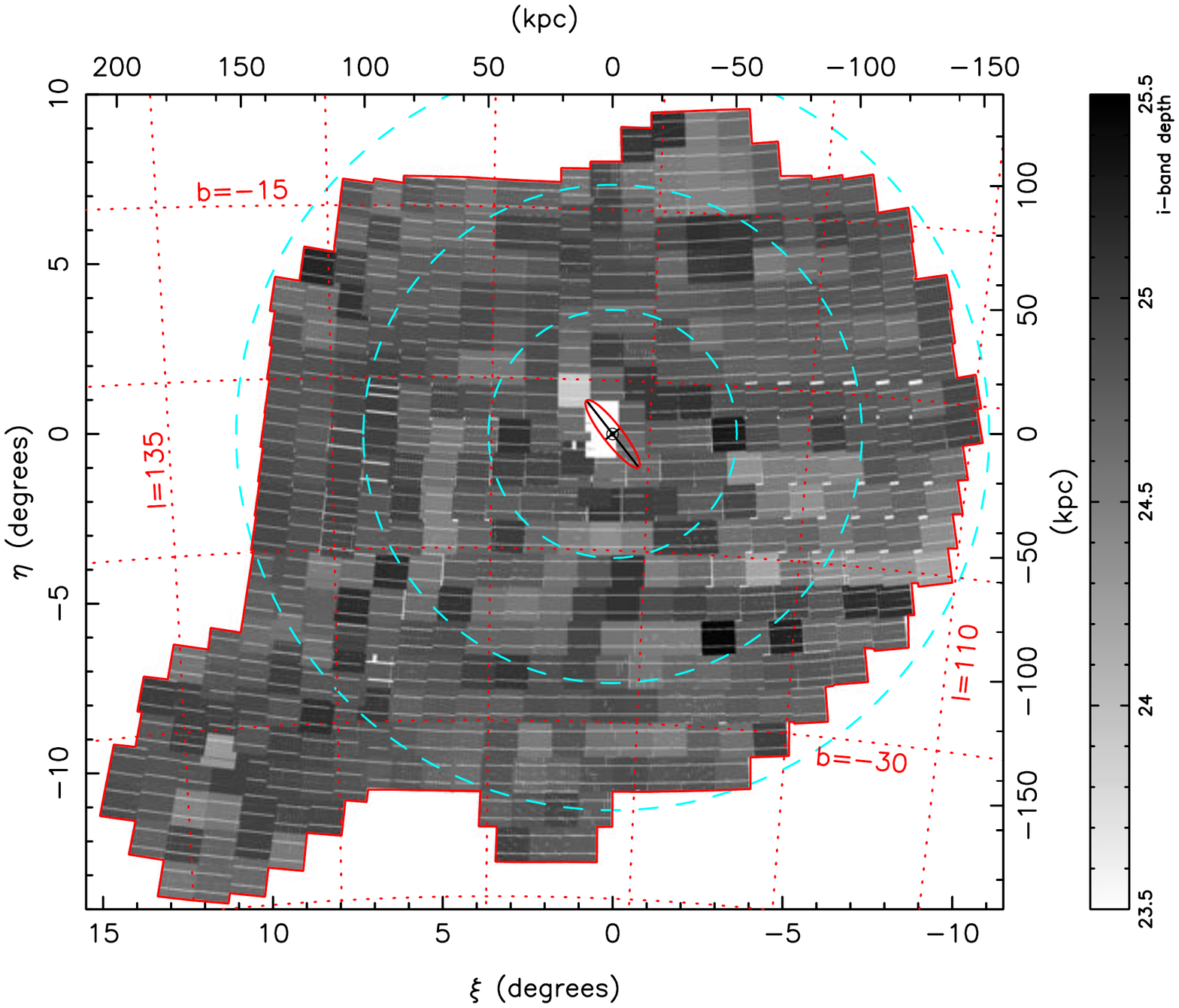}
}
\end{center}
\caption{Map of the photometric depths (5$\sigma$ limit) for point sources in the g- (top) and i-band (bottom). Moderate inhomogeneities are apparent over the survey region. The circles, lines and ellipse have the same meaning as in Fig.~\ref{fig:extinction}. (The light-colored region in the M31 disk is not a hole in the survey, but marks fields where the limiting magnitude is shallower).}
\label{fig:depth}
\end{figure}

All observations were undertaken in queue mode by CFHT staff. Generally, the observing conditions were extremely good, and only a minor fraction of frames had poor image quality. During the final observing season in 2010, most of these poorer fields were re-observed in good seeing conditions. The final set of fields used in the analysis below have good image quality, with a g-band mean of $0\scnd67$ (rms scatter $0\scnd10$) and i-band mean $0\scnd60$ (rms scatter $0\scnd10$). However, some good and bad outliers are present in the sample (in g, the best and worst seeing was $0\scnd41$ and $0\scnd93$, respectively, while in the i-band the best and worst seeing frames has $0\scnd35$ and $0\scnd92$, respectively). As a consequence of these variations, the photometric depth is not uniform over the survey. This can be appreciated in Fig.~\ref{fig:depth}, where we display the limiting magnitude in the g- (left panel) and i-band (right panel). The median (5-$\sigma$) depth is $26.0$ in the g-band and $24.8$ in the i-band. With only a few exceptions, the chosen exposure times for the observation of these fields were $3\times450{\rm s}$ in both the g and i bands.

The queue mode observations ensure reasonably good photometric calibration of the data, as the camera is typically operated for a ``run'' lasting several weeks, during which time a substantial number of photometric standards are observed at varying airmass. The transparency during the night is also monitored. The data products from the CFHT pre-processing pipeline ``Elixir'' \citep{2004PASP..116..449M} include de-biassing, flat-fielding and an estimate of the photometric zero-point for each observation.

The pre-processed CFHT images were passed through the Cambridge Astronomical Survey Unit (CASU) pipeline \citep{Irwin:2001eq}, as discussed in Paper~I. The software combines the individual exposures, and then proceeds CCD by CCD, detecting sources and measuring their photometry, image profile and shape. Based upon the information contained in the curve of growth (for each CCD), the algorithm classifies the objects into noise detections, galaxies, and probable stars. As in Paper~I, we select objects with classifications of either -1 or -2 in both g and i, which includes point-sources up to $2\sigma$ from the stellar locus.

In addition, sources in the stacked images were also measured using the photometric package DAOPHOT and ALLSTAR \cite{Stetson:1987fx} so as to provide point spread function (PSF) fitted photometry and additional morphological measurements to improve star/galaxy discrimination. While the DAOPHOT photometry and fit parameters were found to be very useful in crowded regions, in the outer halo the resulting catalogue was less homogenous (due to significant PSF variations between fields) than that derived from the CASU pipeline. In the analysis below we therefore use only the CASU photometry.

Unfortunately, the Elixir data products received over the course of the project suffered from the several photometric problems that are discussed at length in \citet{Regnault:2009bk}. Those authors propose corrections to account for non-uniformities of the detector response as well as for spatial variations in the effective passbands of the MegaCam filters. While we initially implemented the corrections of \citet{Regnault:2009bk}, which were developed for the Supernova Legacy Survey (SNLS) observations, it became apparent from comparison to the Sloan Digital Sky Survey (SDSS) that the corrections were not applicable to our fields. 

Part of the problem was that the newer Elixir recipes developed for the SNLS were only applied to a subset of our images, but even those images had obvious residual patterns. We therefore decided to proceed to calibrate our data empirically using the SDSS as a reference, constructing two-dimensional correction functions over the MegaCam focal plane. Nevertheless, in order to minimize the amplitude of the correction, we first applied to all the images the most recent\footnote{As of December 2011.} flat-field correction images derived by the CFHT for each observing run.

\begin{figure}
\begin{center}
\includegraphics[width=\hsize]{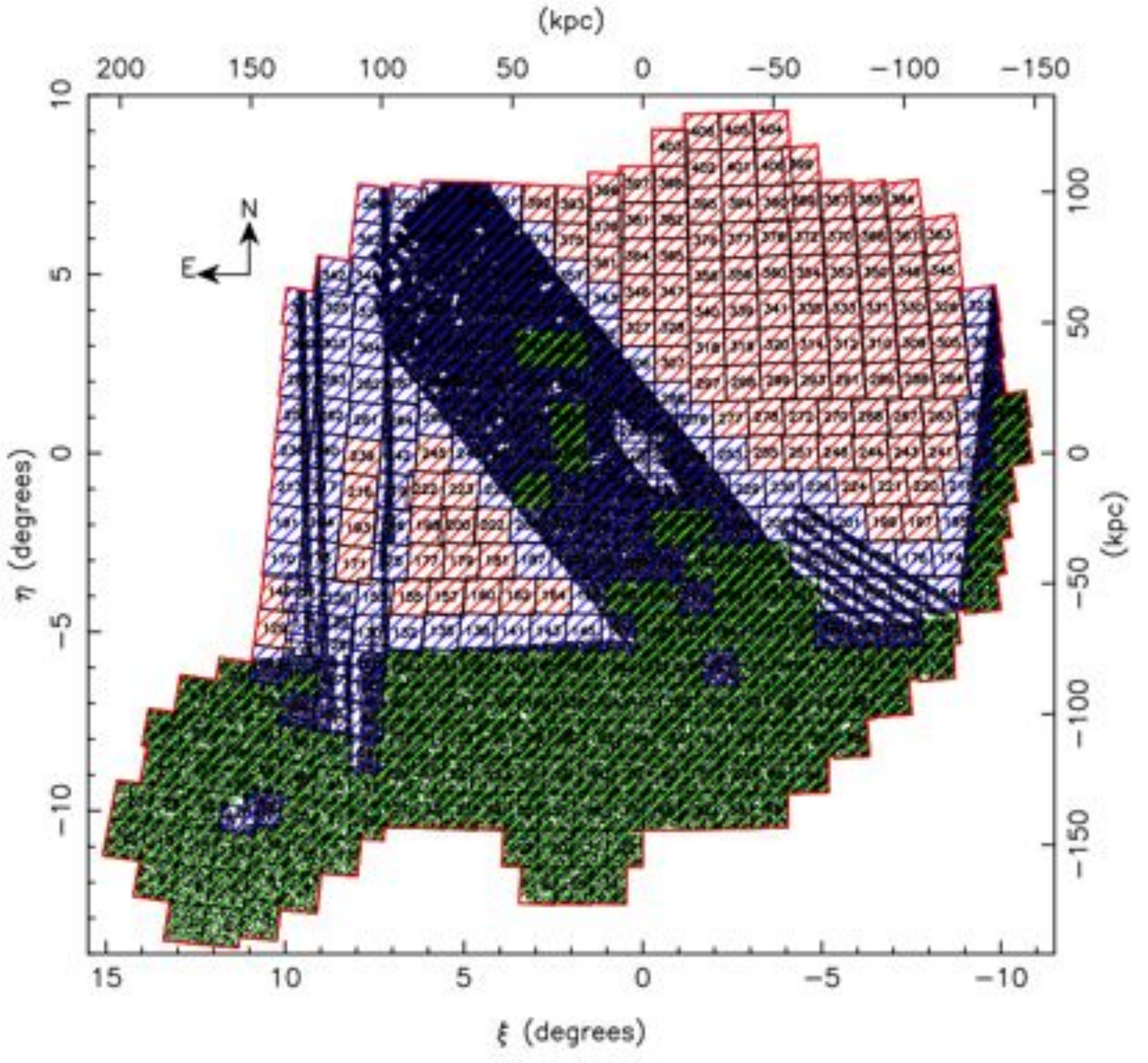}
\end{center}
\caption{The shallower SDSS survey (whose sources are marked with black dots) overlaps a significant fraction of the PAndAS field. This provides a very convenient means to refine the calibration of the CFHT photometric zero-points. The CFHT fields hatched in green mark those fields where the SDSS calibration appears to be of good quality. Red fields have no SDSS counterparts, while blue fields contain either too few SDSS stars to be useful, or the SDSS calibration was found to be unreliable.}
\label{fig:comp_SDSS}
\end{figure}

In Fig.~\ref{fig:comp_SDSS}, we display the overlap of PAndAS with the SDSS Data Release 8 (DR8). Although the DR8 data are substantially shallower than the MegaCam photometry presented in this contribution, they have the advantage of being (mostly) well-calibrated. Furthermore, the DR8 observing pattern in ``stripes'' is very different to the orientation of the MegaCam camera (aligned with ecliptic coordinates), which makes it easy to identify regions in the SDSS with suspicious calibration. These suspicious areas have the size of the SDSS CCDs and follow an SDSS stripe rather than equatorial cardinal directions. Figure~\ref{fig:comp_SDSS} shows the areas without SDSS comparison stars (red), where the SDSS calibration is good (green), and where the SDSS calibration appears incorrect or unreliable (blue).

\begin{figure}
\begin{center}
\includegraphics[width=\hsize]{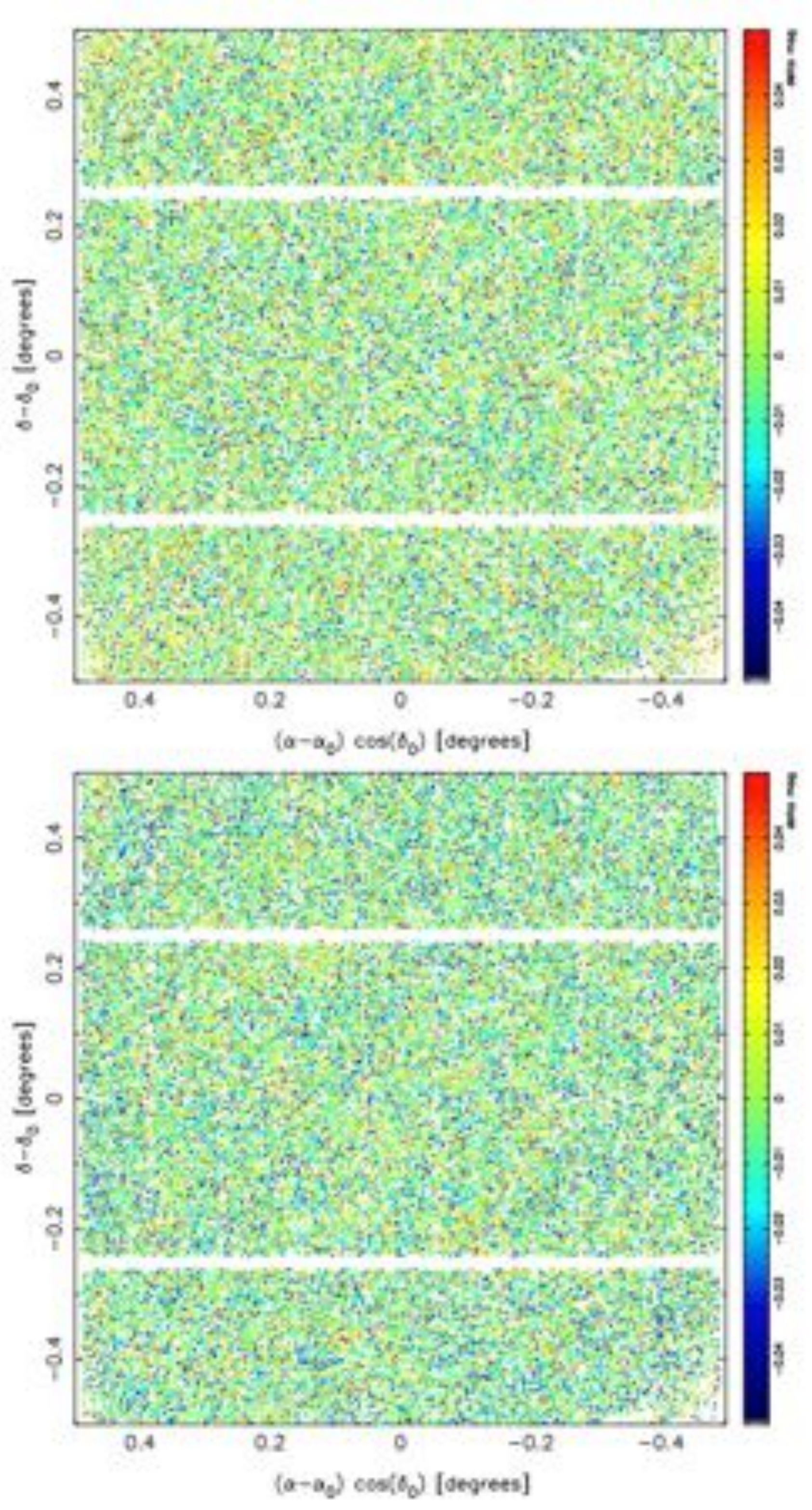}
\end{center}
\caption{The spatial distribution over the CFHT field of view of the photometric differences in the g- (top) and i-band (bottom) with respect to well-measured stars in the SDSS after the CFHT flattening operation has been applied. (The samples consist of stars with $18<g<20$ or $18<i<19.5$). The residual maps are clearly very flat, and have an r.m.s. scatter lower than 0.02~magnitudes (a significant fraction of this scatter is due to the photometric errors in the SDSS).}
\label{fig:SDSS_difference}
\end{figure}

For each observing season we constructed the empirical photometric correction function over the MegaCam focal plane using overlapping SDSS stars with magnitudes in the range $18 < g < 20$ or $18 < i < 19.5$. For those seasons without SDSS reference data, we adopted the correction closest in time. Note that SDSS g,i passbands are not identical to those of MegaCam, so the color transformations given in \citet{Regnault:2009bk} were used to convert between the two systems. It is also noteworthy that the original MegaCam i-band filter was damaged in June 2007, and its replacement has a slightly different transmission. From PAndAS regions observed with both i-band filters, we deduced the following simple transformation:
\begin{displaymath}
i_{new} = 
\left\{
\begin{array}{rl}
-0.010 + 0.031 \times (g - i_{old}) + i_{old}  & \\
                                                                     \textrm{for} \, \, \, \, (g - i_{old}) & < 1.9 \, , \\
                                                                     & \\
-0.081 + 0.069 \times (g - i_{old}) + i_{old}  & \\
                                                                     \textrm{for} \, \, \, \,  (g - i_{old}) & > 1.9 \, .
\end{array} \right.
\end{displaymath}
In the analysis below g,i refer to AB magnitudes in the CFHT system, with old i-band data transformed into the new filter using this equation. 

Figure~\ref{fig:SDSS_difference} shows the resulting distribution of residuals between the PAndAS and SDSS data over the optical field of MegaCam, after the empirical correction functions were applied (and the field-to-field zero-points). The distributions in both g and i are flat (with r.m.s. scatter below $0.02$) and substantially better than the initial ``elixir'' calibrations that had up to $\sim 0.1$~magnitude peak-to-peak residuals following a large-scale pattern. 

With each MegaCam field thus flattened we then initially proceeded to refine the individual zero-points of the CFHT fields. This was achieved by solving a system of equations for the photometric differences between the overlapping fields. We included the SDSS data in the solution, taking the contiguous region of good SDSS data (marked green in Fig.~\ref{fig:comp_SDSS}) as the zero-point reference. Over the regions marked in red in Fig.~\ref{fig:comp_SDSS} the zero-point could only be constrained via the field-to-field overlaps; in this situation it is quite easy to imagine that gradients in the photometric calibration could give rise to a large-scale zero-point error that could render our photometry useless in regions far from anchoring SDSS data (e.g. at the Northern boundary of our data). For this reason we requested the Pan-STARRS consortium to allow us to check the calibration of our data using their photometry \citep{2012ApJ...756..158S} of bright stars (${\rm 18.5 < g < 20}$, ${\rm 18 < i < 19}$) in this region, which they generously granted. It transpired that, within the uncertainties, our calibration agreed very well with Pan-STARRS, with only 5 fields showing substantial errors (and which we corrected to be in-line with Pan-STARRS, as will be discussed in McConnachie et al. 2013). The large-scale calibration of the photometric zero-point appears to be within $\sim 0.01$~magnitudes, judging from the small change in the zero-point between our original photometric calibration and the Pan-STARRS calibration, although we are aware that this test is not entirely independent since Pan-STARRS also bootstrapped their calibration from the SDSS.

\section{Global Color-Magnitude and spatial distribution}
\label{sec:CMD_spatial}

\begin{figure}
\begin{center}
\includegraphics[angle=0, viewport= 20 20 560 480, clip, width=\hsize]{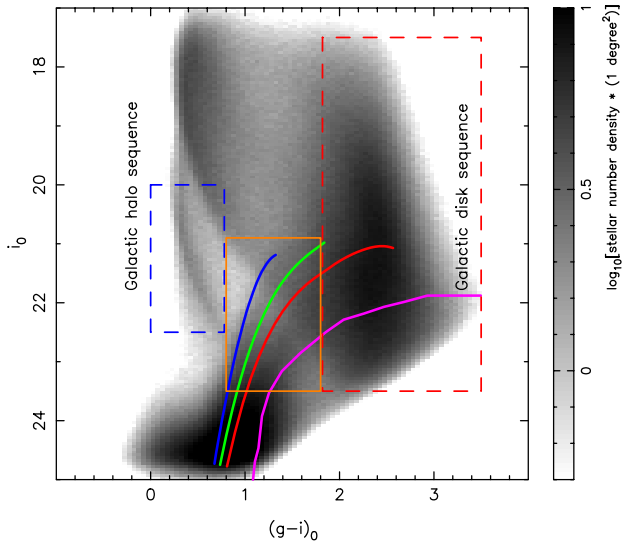}
\end{center}
\caption{The combined CMD of the PAndAS survey at distances beyond $2\deg$ of M31, showing the main RGB feature of interest as well as the contamination from foreground and background sources that must be overcome. The fiducial RGBs \citep{Ibata:2001vs}  correspond to, from left to right, NGC~6397, NGC~1851, 47~Tuc, NGC~6553, which have metallicity of ${\rm [Fe/H] = -1.91}$, $-1.29$, $-0.71$, and $-0.2$, respectively. The sequences have been shifted to a distance modulus of ${\rm (m-M)_0=24.47}$. The dashed-line rectangles highlight CMD regions where the contamination is especially prominent: these are the foreground Galactic halo (blue) and Galactic disk (red). The orange box shows the adopted color-magnitude selection region, inside which we estimate the photometric metallicities of the stars in the survey (as discussed in \S3).}
\label{fig:Hess_overall}
\end{figure}

In Fig.~\ref{fig:Hess_overall} we display the color-magnitude distribution of point sources over the entire survey area, and the expected loci of plausible stellar populations in the M31 halo are marked by the fiducial RGB tracks. A large number of these sources are Galactic contaminants, principally from the Galactic thin and thick disks, and halo. Their color-magnitude distribution is complex and varies spatially, but for illustrative purposes in Fig.~\ref{fig:Hess_overall} we mark the color magnitude diagram (CMD) regions where they are most visible. In addition to the foreground Galactic stars, the survey also contains unresolved background galaxies, which contaminate especially at faint magnitudes. In an accompanying contribution \citep[][hereafter M13]{2013arXiv1307.7626M}, we have developed a detailed empirical model to account for both of these contaminating populations. The  model uses the area beyond a radius of $9^\circ$ ($120\kpc$) to sample the contamination, and so provides an interpolation of the number of contaminants $\Sigma_{(g-i,i)}(\xi,\eta)$, as a function of color-magnitude position $(g-i,i)$ and sky position $(\xi,\eta)$.

\begin{figure}
\begin{center}
\includegraphics[width=\hsize]{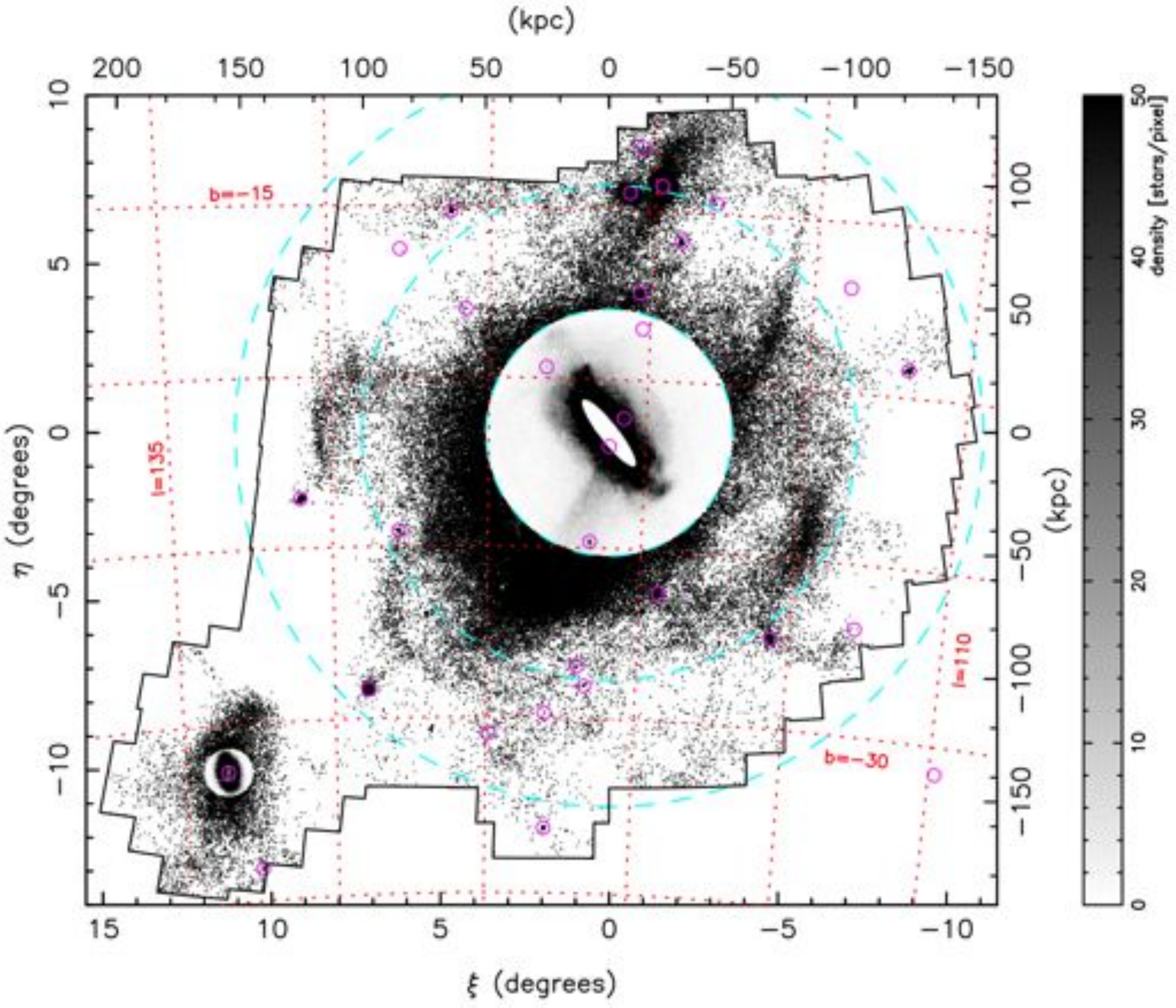}
\end{center}
\caption{Map of stars with $(g-i)_0<1.8$ and ${\rm -2.5 < [Fe/H] < 0}$ and $i_0<23.5$. The contamination from the foreground Milky Way as well as that from unresolved background galaxies has been removed in a statistical manner. The dense regions around M31 (radius $50\kpc$) and M33 (radius $10\kpc$) are shown as greyscale density images (with bin size $0\degg02\times0\degg02$) while the outer data are shown with points. The pink circles indicate the positions of the known satellites dwarf galaxies of M31.}
\label{fig:density}
\end{figure}

\begin{figure}
\begin{center}
\includegraphics[width=\hsize]{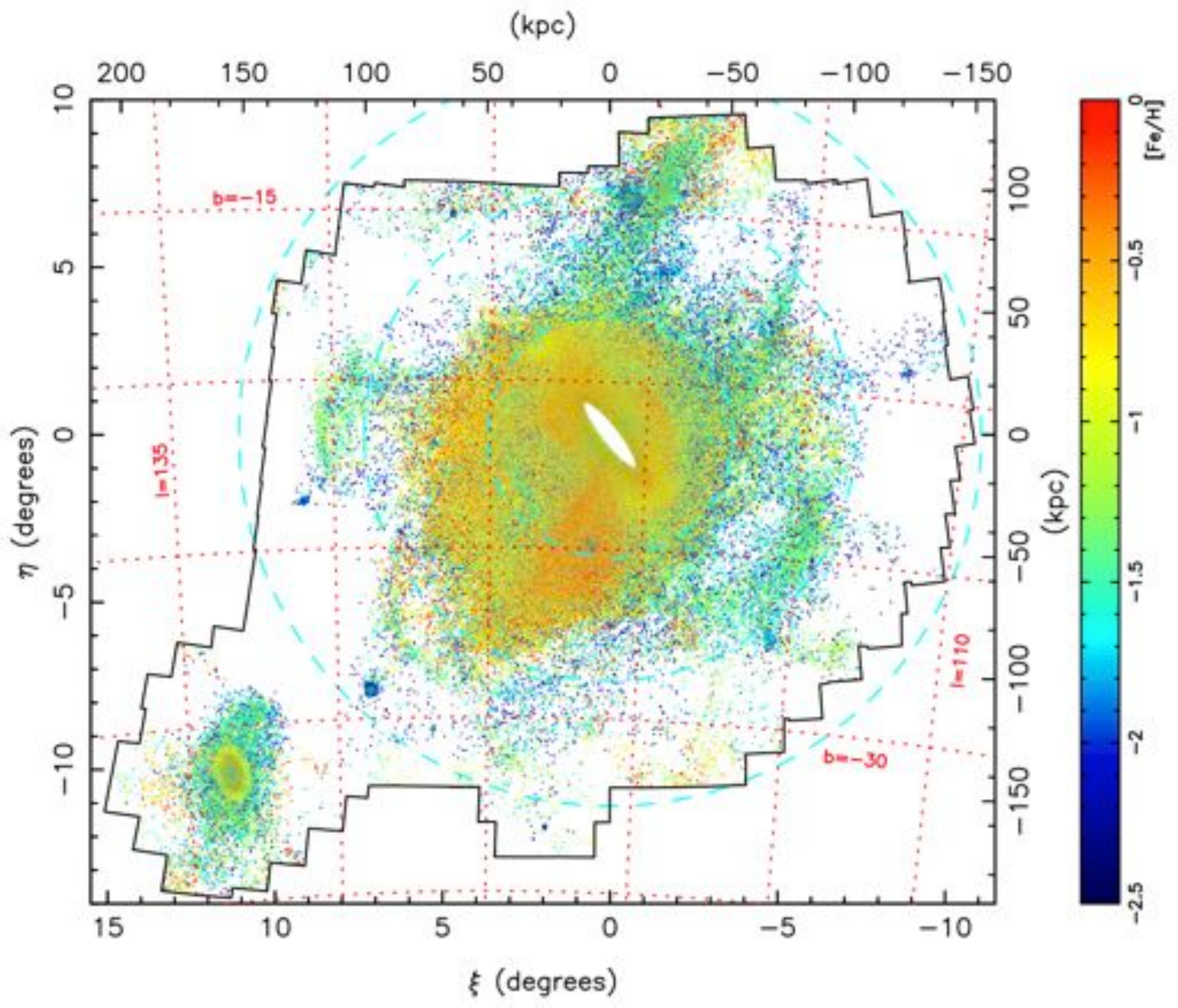}
\end{center}
\caption{Metallicity map of stars for the same parameter selection as Fig.~\ref{fig:density}. The rich tangle of substructure is seen to possess a wide range in metallicity.}
\label{fig:map_full_metallicity}
\end{figure}

\begin{figure*}
\begin{center}
{\hbox{
\includegraphics[angle=0, viewport= 150 60 715 580, clip,  width=9.0cm]{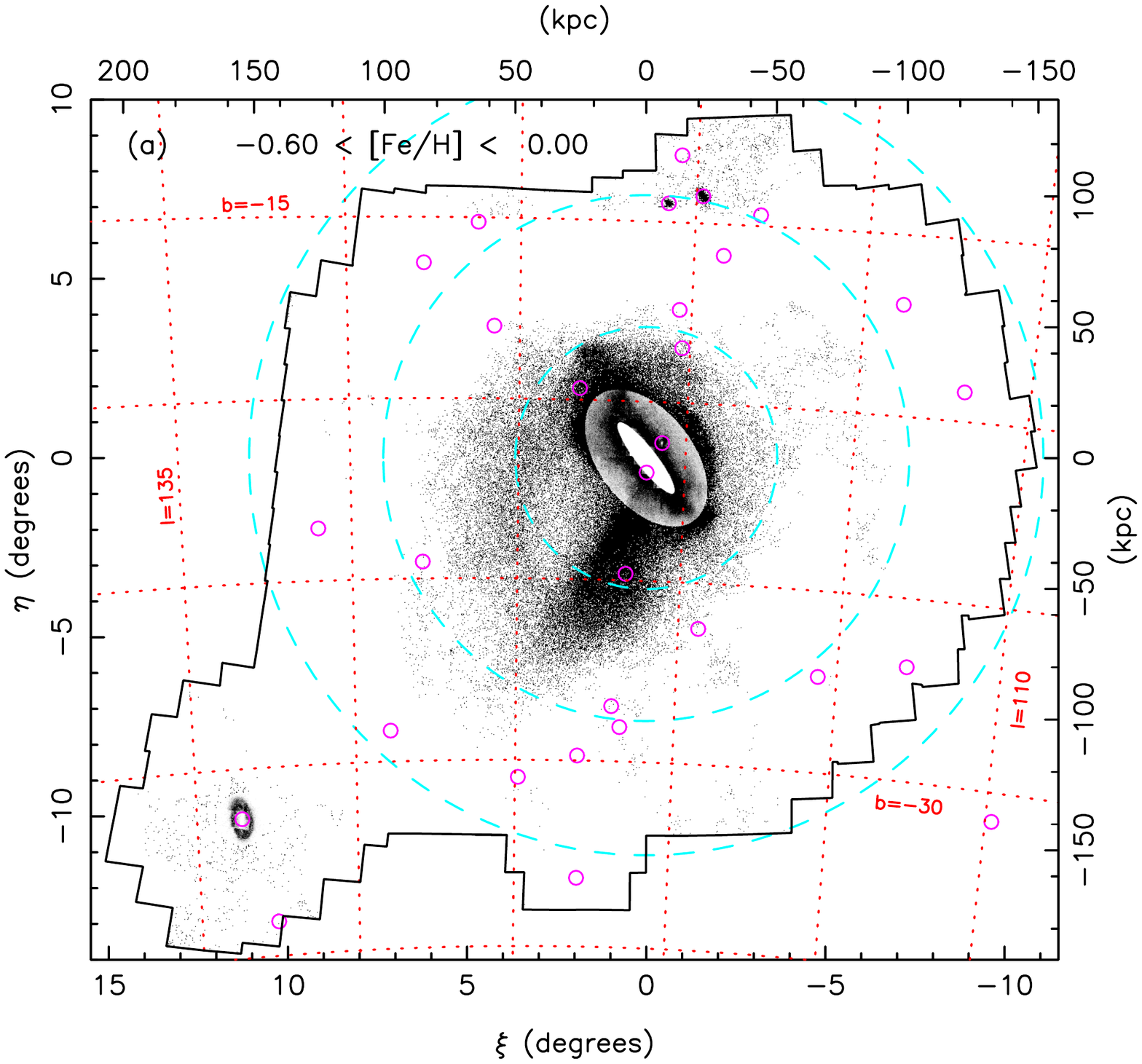}
\includegraphics[angle=0, viewport= 150 60 715 580, clip,  width=9.0cm]{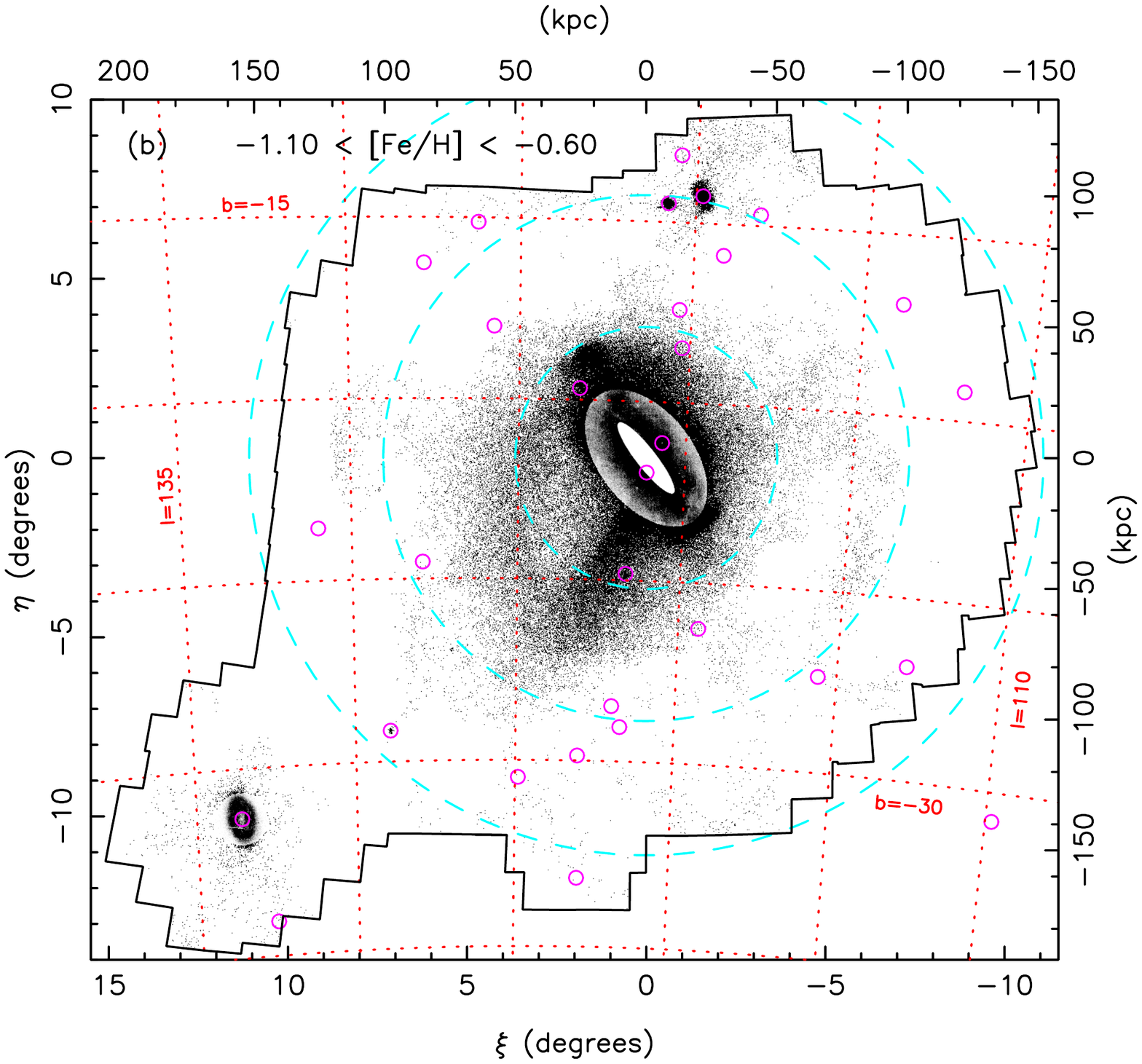}
}
\hbox{
\includegraphics[angle=0, viewport= 150 60 715 580, clip,  width=9.0cm]{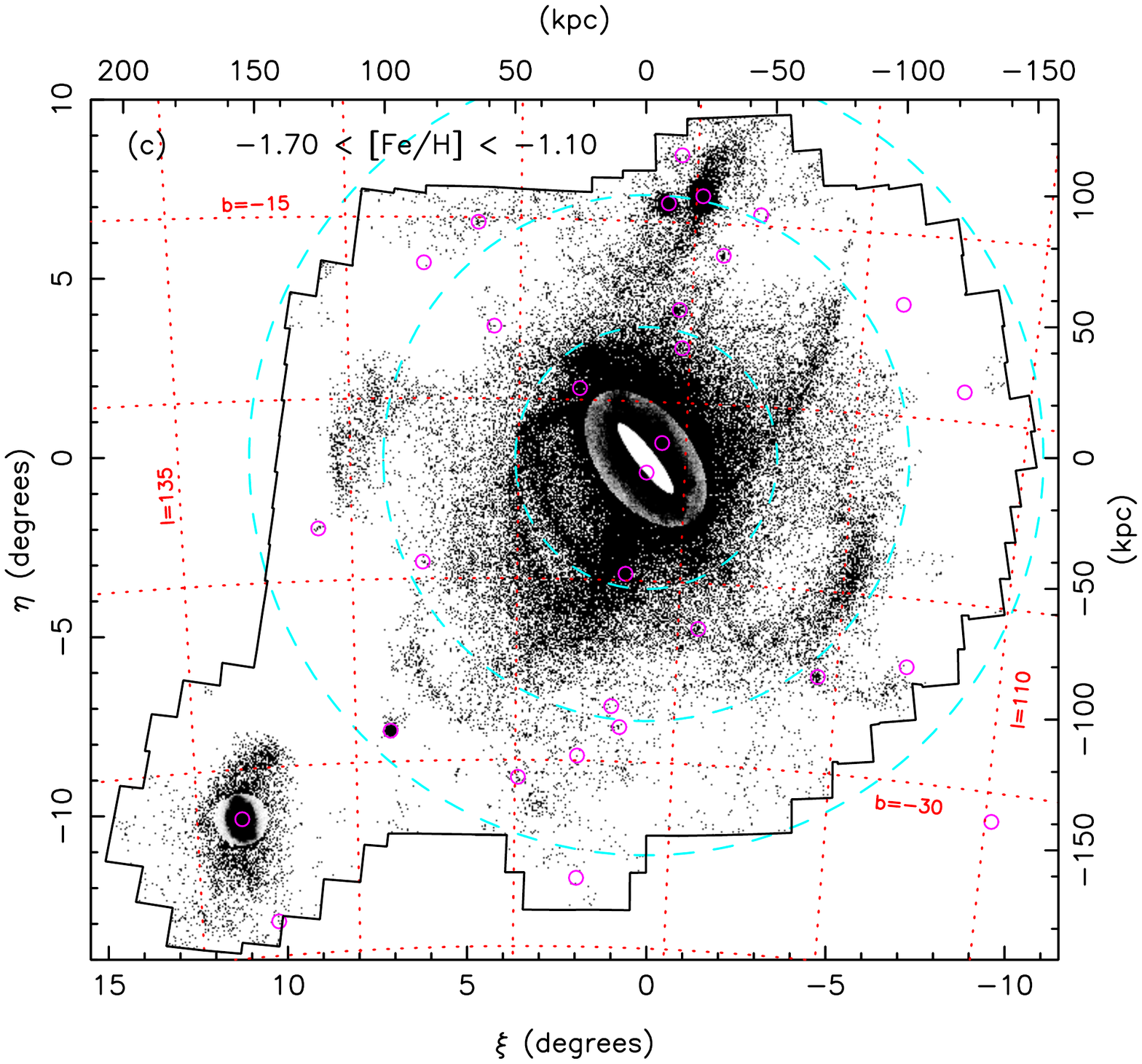}
\includegraphics[angle=0, viewport= 150 60 715 580, clip,  width=9.0cm]{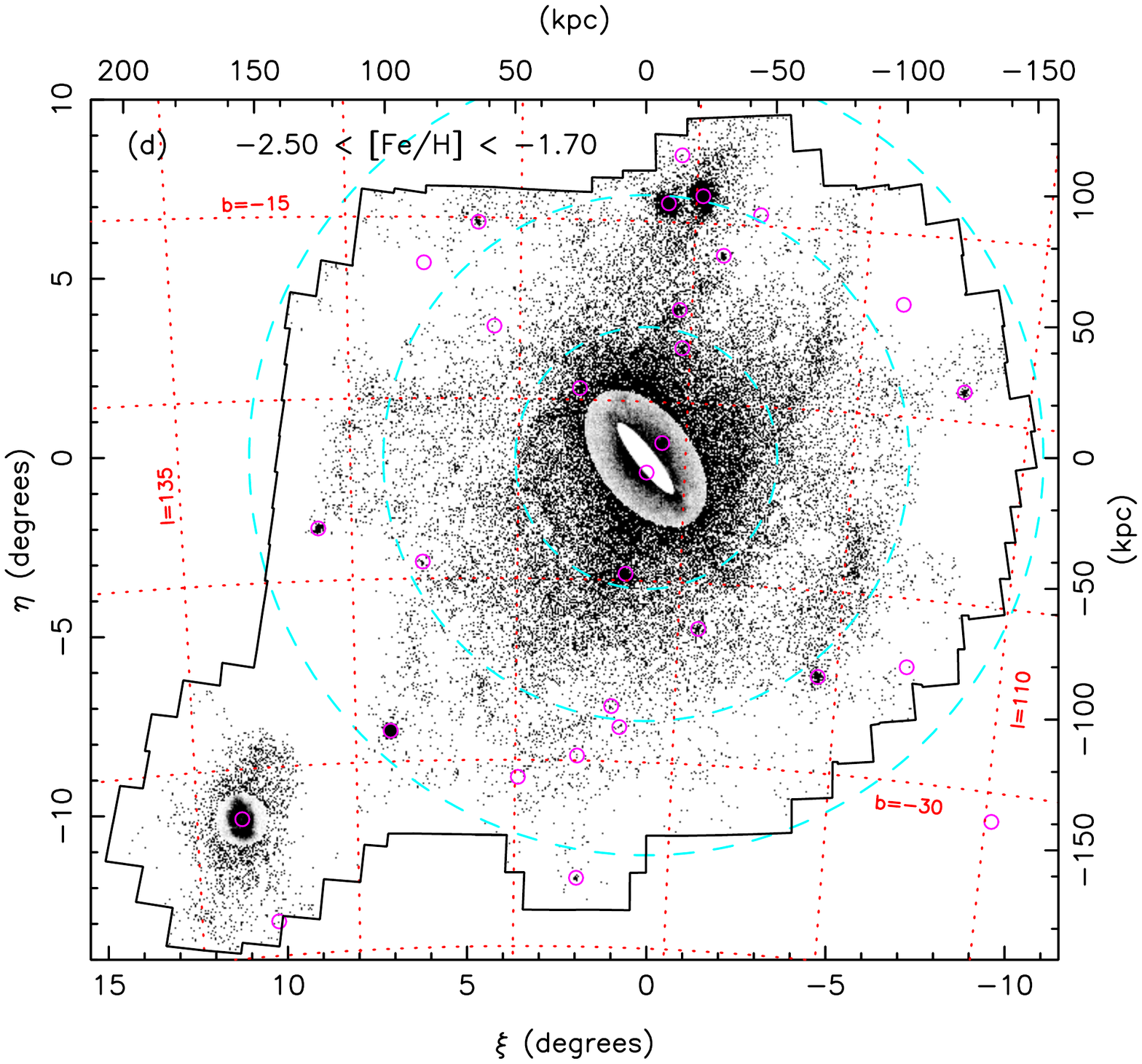}
}}
\end{center}
\caption{As Fig.~\ref{fig:density}, but showing four different metallicity cuts. The two high metallicity cuts (upper panels) are dominated by the southern ``Giant Stellar Stream'', although note that this structure changes shape slightly between the two panels. The low metallicity cuts in the bottom two panels display a more interesting spatial distribution. The ${\rm -1.7 < [Fe/H] < -1.1}$ selection is dominated by numerous streams, while the most metal poor selection (${\rm [Fe/H] < -1.7}$) appears, to first approximation, primarily smooth. In all panels we replace the stars within an ellipse of radius $30\kpc$ and axis ratio $0.6$ with a grayscale map (and similarly for M33).}
\label{fig:map_slices_metallicity}
\end{figure*}

In the remainder of this contribution we will be interested primarily in investigating the global properties of the M31 halo, and the analysis will be made more robust by considering a clean sample. After extensive tests we decided to limit the survey to $i_0<23.5$, which shows a smooth distribution of foreground contaminants while being typically more than a magnitude above the $5\sigma$ detection limit (see Fig.~\ref{fig:depth}). This will minimize field-to-field variations in counts due to photometric incompleteness.

A photometric metallicity was estimated for each star in the survey by comparing their colors and magnitudes to the Dartmouth isochrones \citep{2008ApJS..178...89D}, shifted into the MegaCam filter system. A common age of $13\Gyr$ was assumed for all stars, and $[\alpha/{\rm Fe}]=0$, which we consider a reasonable assumption for halo members. In order to estimate a photometric metallicity we also need to know the distance to the star under examination, yet this information is unavailable to us\footnote{The distance to a few substructures may be estimated from the tip of the RGB \citep{2012ApJ...758...11C}, but this is only probabilistic and not available for the bulk of the halo}. To overcome this problem, we further assumed that all the halo stars have the same Heliocentric distance as M31 (taken to be $785\kpc$, \citealt{McConnachie:2005hn}). Note however, that we observe major substructures at projected distances of up to $\sim 120\kpc$; a similar extension along the line of sight would result in changes in brightness of $\sim 0.35$~mag. Even though the isochrones on Fig.~\ref{fig:Hess_overall} are steep over the color interval of interest (especially for the metal-poor populations), changes of this size could substantially affect the inferred ${\rm [Fe/H]}$ and thus result in, for example, apparent metallicity gradients along streams. To give a quantitative appreciation of the size of this effect, consider a star with $i_0=22.6$ (this corresponds to the average magnitude of M31 candidates in our sample subject to the magnitude limit $i<23.5$) and $(g-i)_0=1.08$; at the nominal distance of M31 we would infer such a star to have ${\rm [Fe/H]=-1.5}$. However, if the star were in reality $120\kpc$ closer (farther) the actual metallicity should have been calculated to be ${\rm [Fe/H]=-1.22 \, (-1.81)}$. Thus shifts of up to $\Delta {\rm [Fe/H]} \sim 0.3$ can be expected in the derived photometric metallicity. But since the halo is highly centrally-concentrated (see \S4 below), most stars even in fields at large projected distances should be located close to the tangent point, so $\Delta {\rm [Fe/H]}$ should in general be substantially smaller than $0.3$.

We binned the M13 contamination model into a high resolution grid in color and magnitude (0.001~mag$\times$0.001~mag), and in an identical manner to the real stars, we calculated the metallicity that the contaminants would appear to have if they were misclassified as M31 stars. In this way we are able to account for the contamination given any sample selection in color, magnitude and photometric metallicity. 

Fig.~\ref{fig:Hess_overall} shows that the contamination becomes more manageable if we restrict the analysis to $(g-i)_0<1.8$ (see also the discussion in Paper~I). This has the effect of biassing our sample against the brightest metal-rich stars, but since such stars should be rare in the outer halo, we judge it is a sensible restriction to ensure that we obtain relatively clean structural maps. 

\begin{figure}
\begin{center}
{\hbox{
\includegraphics[angle=0, viewport= 150 60 715 580, clip,  width=4.5cm]{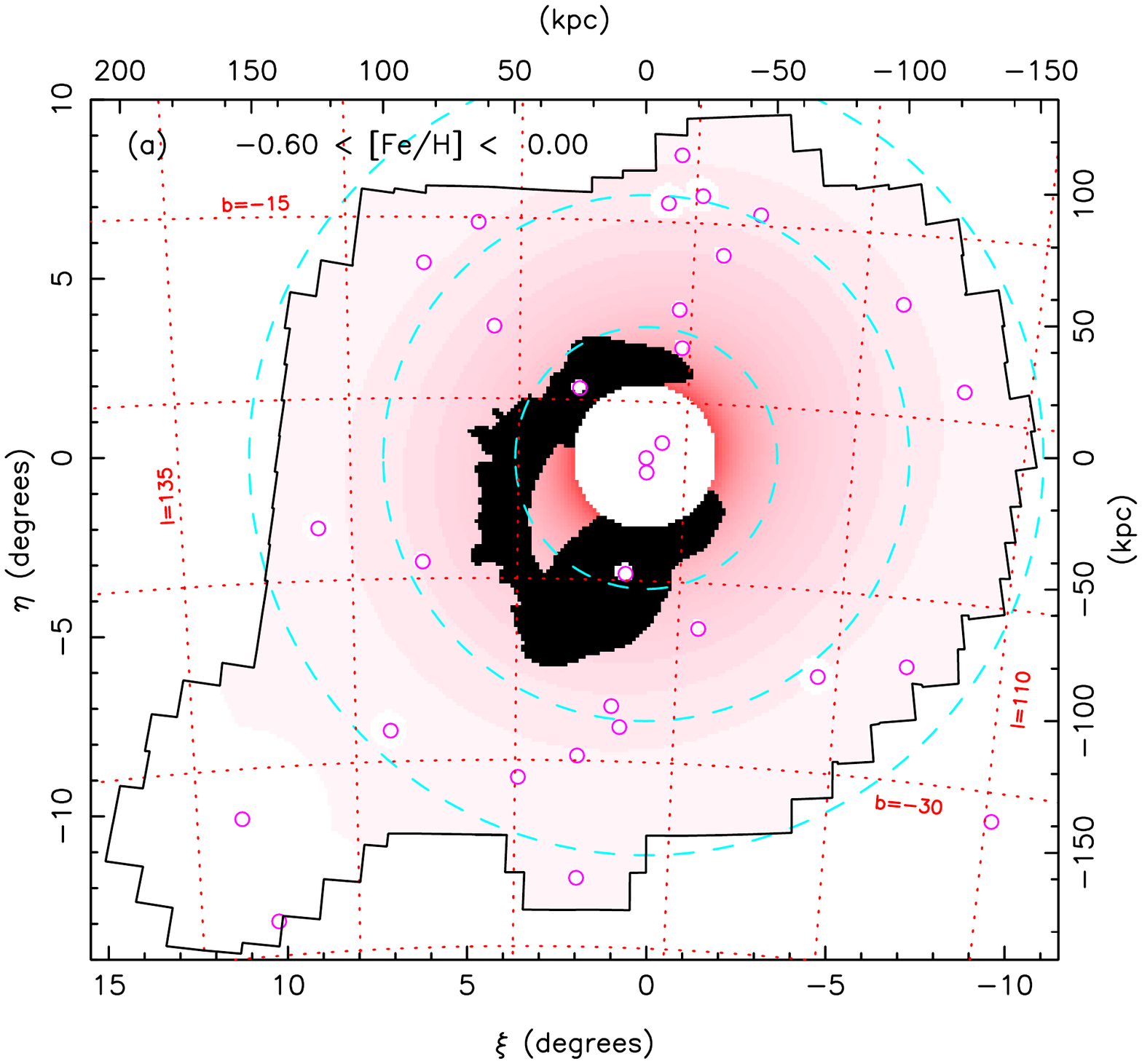}
\includegraphics[angle=0, viewport= 150 60 715 580, clip,  width=4.5cm]{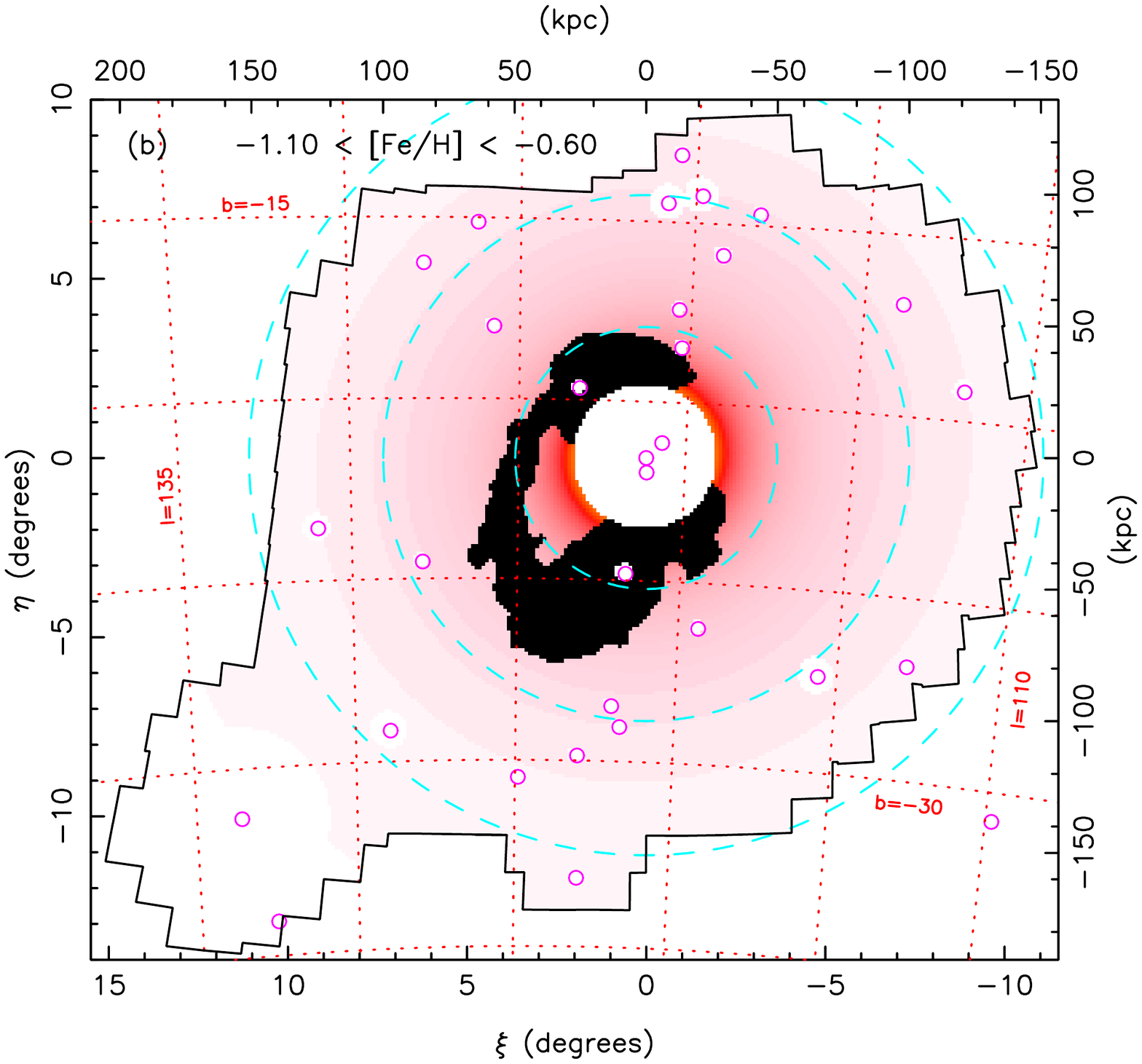}
}
\hbox{
\includegraphics[angle=0, viewport= 150 60 715 580, clip,  width=4.5cm]{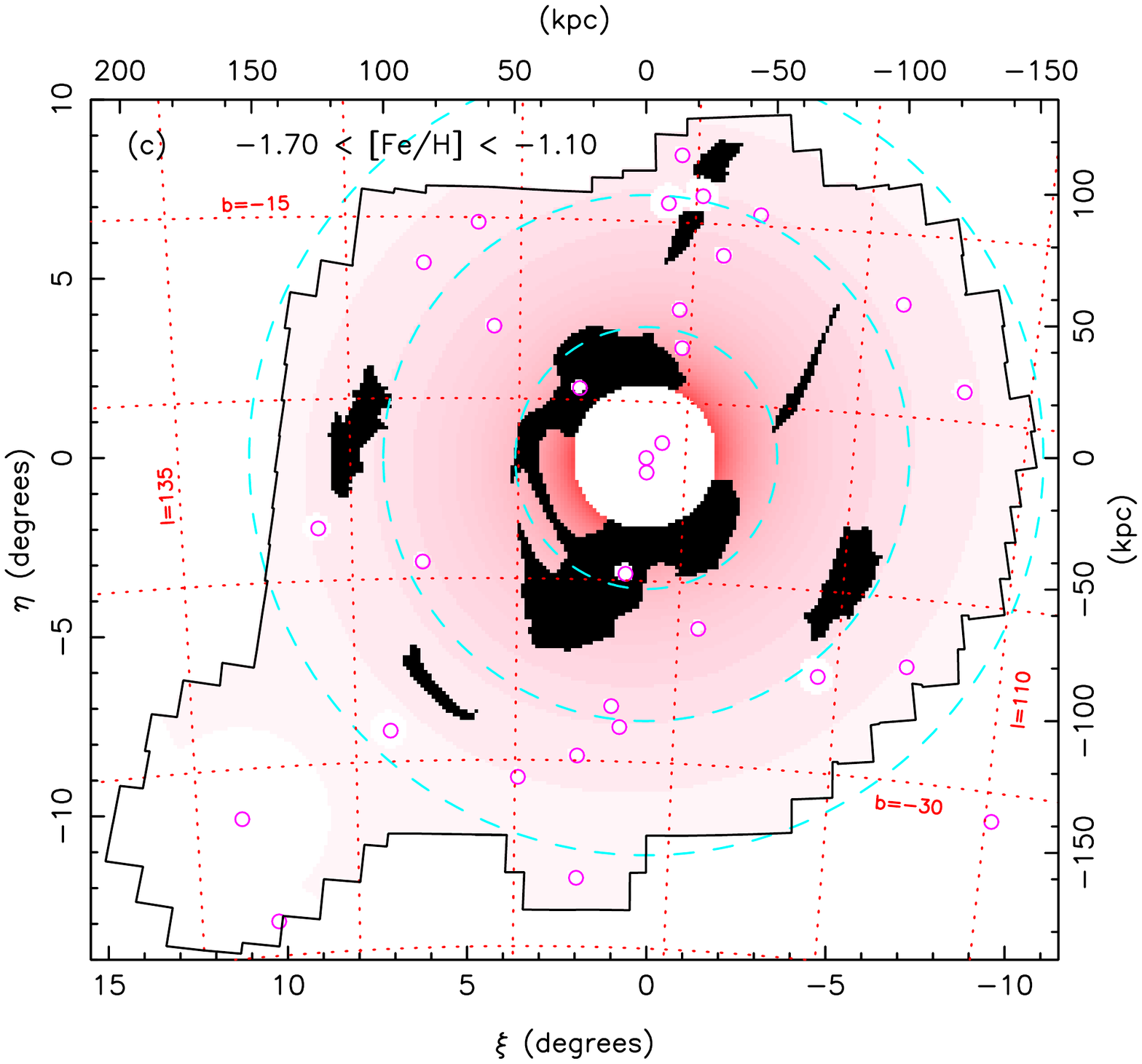}
\includegraphics[angle=0, viewport= 150 60 715 580, clip,  width=4.5cm]{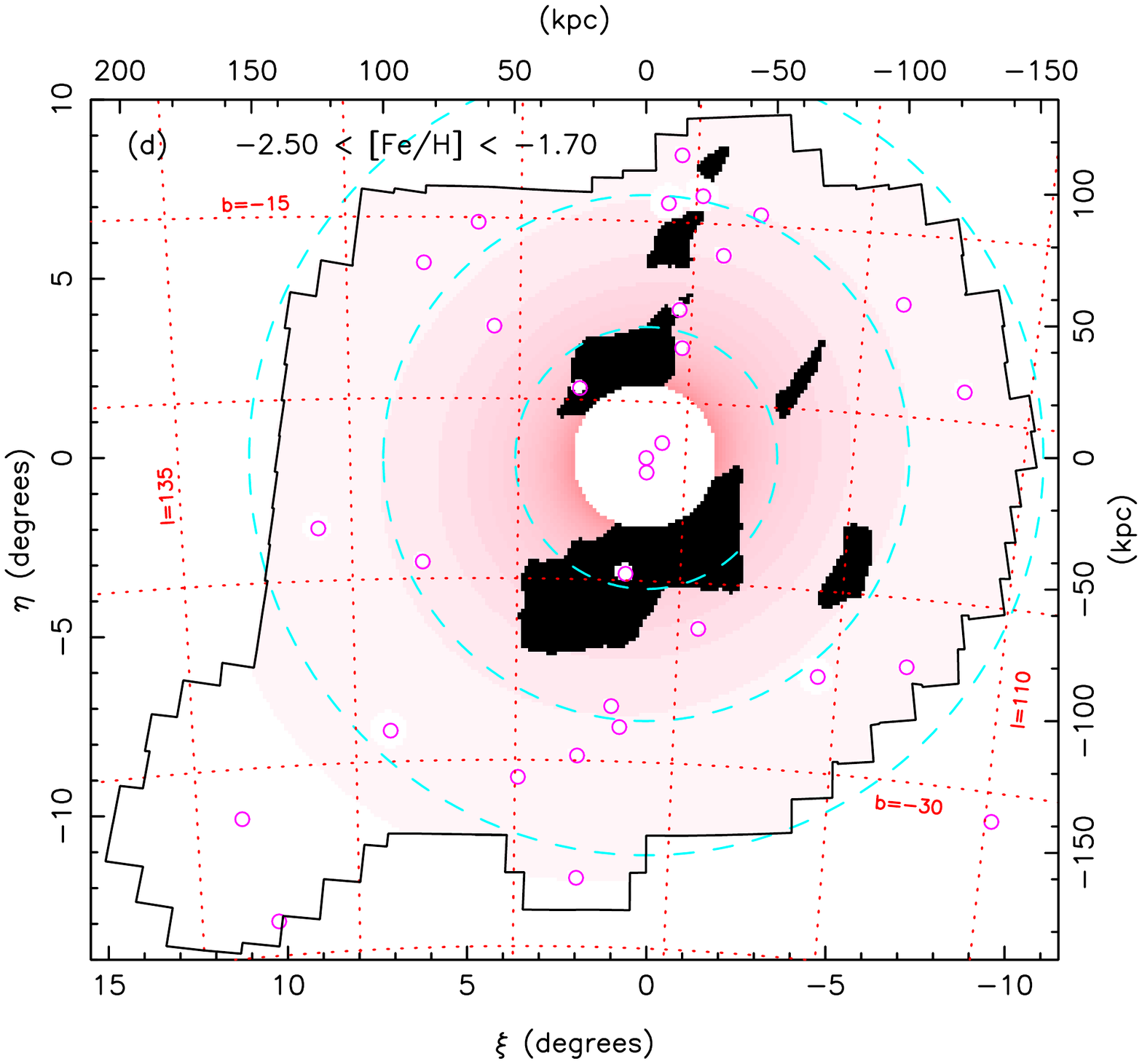}
}
\includegraphics[angle=0, viewport= 150 60 765 580, clip,  width=4.5cm]{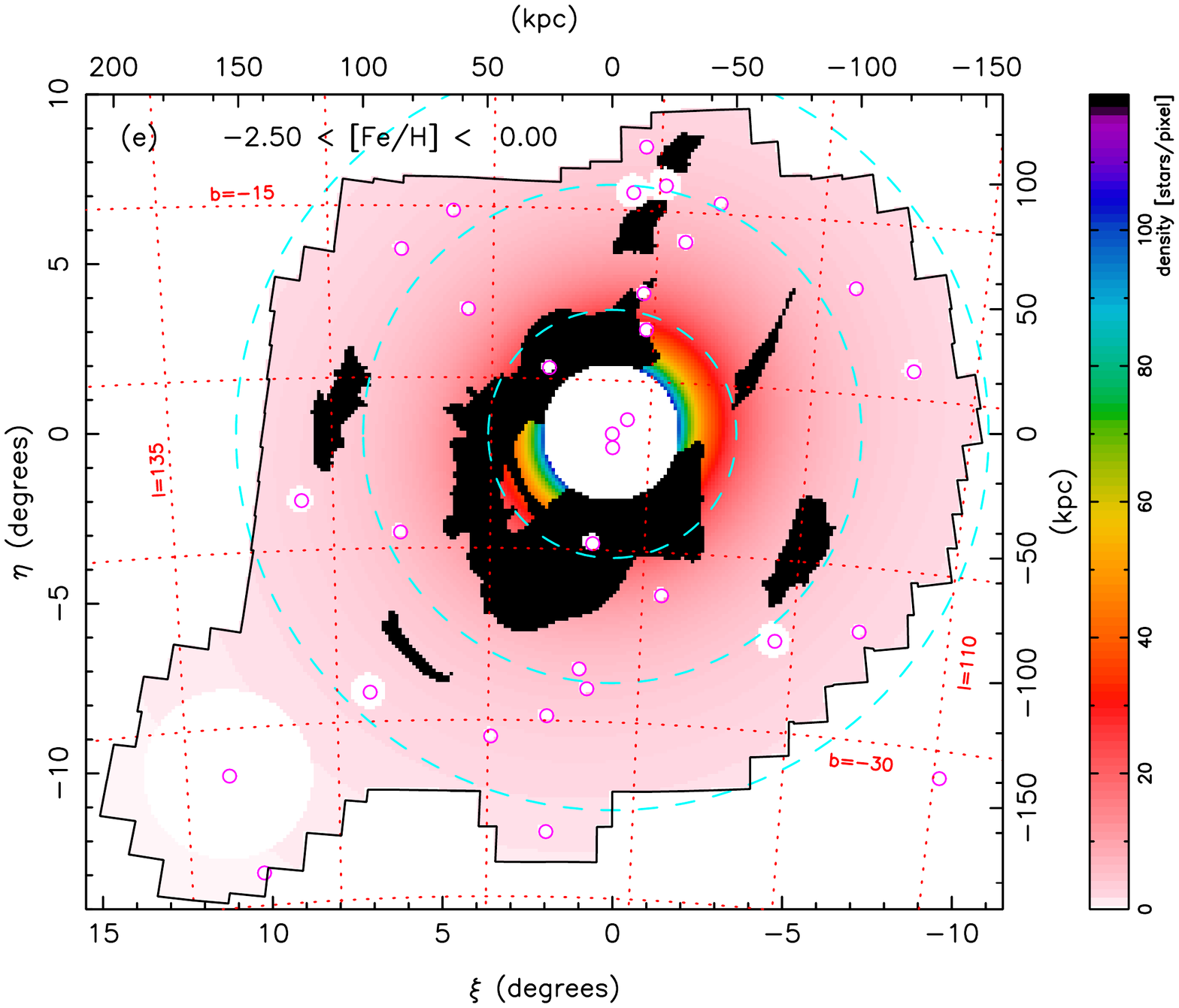}
}
\end{center}
\caption{The substructure masks generated by the halo fitting algorithm are displayed in black. Note that data within the central $2\deg$ of M31 was ignored, as we deemed the spatial distribution to be too complex in that region. Similarly, a $2\degg5$ region around M33 was excised. The bottom panel shows the union of the four upper masks. In each panel the underlying colored distribution shows the best-fit halo model for the corresponding metallicity selection. The halo fits (the parameters of which are listed in Table~\ref{tab:fits}) are clearly almost spherical.}
\label{fig:map_masks}
\end{figure}

The halo star distribution is shown in Fig.~\ref{fig:density}, limiting the sample to $i_0<23.5$, $(g-i)_0<1.8$ and ${\rm -2.5 < [Fe/H] < 0}$.  Here, we have subtracted off the contaminating populations; this is done in a statistical manner, treating the contamination as a distribution function in position, color, magnitude and metallicity. The algorithm generates a star at random from this distribution function, and subtracts the real star that lies closest in the parameter space to the artificial contaminant\footnote{This multi-dimensional space clearly has different units. To calculate a ``distance'' we adopt a scale of $0\degg25$ for both $\xi$ and $\eta$ positions, a scale of $0.25$~mag for the magnitude dimension, and $0.25$~dex for the metallicity dimension. Possible matches are considered up to a ``distance'' of 5 scale lengths.}. While the halo map changes slightly from one random realization to the next (depending on which stars were subtracted) the advantage of this technique for visualization is that it keeps the full resolution of the data. 

The corresponding metallicity distribution of these halo (and disk) stars is displayed in Fig.~\ref{fig:map_full_metallicity}. This colorful map reveals vividly the multitudinous accretions that have taken place during the formation of the Andromeda galaxy. The variations in metallicity, size, and orbit highlight differences in mass, age, and accretion time \citep[see, e.g.,][]{Johnston:2008jp}. By splitting the sample into the four wide bins shown in Fig.~\ref{fig:map_slices_metallicity}, one can further appreciate the variation in the behavior of the halo populations as a function of metallicity. It is interesting to note that by far the strongest signal of metal-rich stars (beyond the bound galaxies) is due to the so-called ``Giant Stellar Stream'' that protrudes down to the south of M31 \citep{Ibata:2001vs,Ibata:2004di,2006MNRAS.366.1012F,2007MNRAS.380...15F,2012MNRAS.423.3134F,2009ApJ...705.1275G}. This structure is likely the latest and most significant accretion event that has taken place in the last several ($\sim 2-3$) Gyr. While the progenitor (or its remnant) has not been identified, it is likely that its dismemberment also produced many of the metal rich features that surround M31 at distances of $\sim 30$--$50\kpc$ to the East and West of the host galaxy \citep{2013MNRAS.434.2779F}.

The maps at metallicity lower than ${\rm [Fe/H] = -1.1}$ are less contaminated by that single accretion event, and show multiple streams and shell structures of vast spatial extent. Some of the substructures that are strikingly evident in the maps are clearly directly related to the on-going disruption of Andromeda's companions: in particular the substructure around M33 and the tidal tails belonging to NGC 147. The most metal poor of the four selections, for ${\rm -2.5 < [Fe/H] < -1.7}$, is the only one of these maps that begins to approximate the classical picture (commonly accepted in the 1990s) of a smooth, ellipsoidal, stellar halo.

\section{Deconstructing the M31 halo}
\label{sec:deconstructing}

The maps presented in Fig.~\ref{fig:map_slices_metallicity} show clear evidence for a multitude of accretion structures overlaid on what appears to be a relatively smooth centrally-concentrated component. To some extent, this is exactly what is expected. Hierarchical halo formation scenarios are well-known to predict copious stream-like substructures \citep{Bullock:2005is}. But a smooth halo component is also expected. In a recent contribution, \citet{2013MNRAS.433.2576P} demonstrated that in hierarchical time-dependent potentials, stellar clumps tend to be effaced from the integral-of-motion space. The same process that leads to the formation of substructures, namely the accretion of satellites, induces an evolution of the host potential, which in turn speeds up the diffusion of pre-existing substructures. This is a dynamical process from which a ``smooth'' stellar halo must necessarily arise (in this picture the smooth stellar halo is nothing but diffused accreted substructures). 

Nevertheless, a reliable identification of this smooth halo poses significant practical problems: how does one distinguish the smooth and clumpy halo components? Indeed, how can these components be suitably defined in an observational survey, given signal to noise limitations at the spatial resolutions required to discriminate between residual faint substructure and the diffuse halo?

One could propose that the smooth halo is whatever is not in a ``clump''. However, there is no clear definition for ``clumps'', since streams can show an infinite number of phase-space configurations and metallicity distributions. Furthermore, these structures are transient, and will eventually diffuse away, so any distinction is strongly dependent on the system observed. Also even if we had access to perfect data, the analysis of \citet{2013MNRAS.433.2576P} indicates that there should not be a clear-cut distinction between ``smooth'' and ``clumpy'' components in integral-of-motion space, let alone in projected density.

These considerations would suggest that the best compromise is simply to ignore the distinction between smooth and clumpy halo, and to present the global properties of the population. This will be the approach adopted in \S\ref{sec:fit2D} below.

However, we will first develop in \S\ref{sec:fit3D} the alternative interpretation, despite the above concerns, and allow ourselves to be guided by the data, since we feel that the apparent smoothness of the metal-poor populations seen by PAndAS {\em demands} close examination. The caveat for this pragmatic alternative analysis is that what we will call a ``smooth'' halo, is only smooth to the limits of the PAndAS survey. While this observational definition may appear somewhat arbitrary, the well-defined selection function for PAndAS means that modelers may process their simulations in order to produce PAndAS-like mock catalogs, and so compare directly to our findings.

\subsection{Three-dimensional fits with masks}
\label{sec:fit3D}

For the purposes of this study we developed a new algorithm to help disentangle the stream and shell-like overdensities from a possible smooth large-scale component. Starting from a trial density profile $\rho(s)$ for the smooth stellar halo (which we implement to be either be a power-law or a spline interpolation function to $\log(\rho)$ vs. $\log{s}$), the algorithm integrates the density along the line of sight cone. In the following, we denote the line of sight coordinate $l$, the M31 tangent plane coordinates $(\xi,\eta)$ (aligned with the North and South directions, as displayed in all our maps), and we define an M31-centric elliptical coordinate $s=\sqrt{x^2+y^2+z^2/q^2}$. Here $x,y$ lie in the plane of the M31 disk, with $z$ perpendicular to the disk. The coordinate $s$ also depends on the flattening parameter $q$, which allows for oblateness or prolateness. The integration along a cone is necessary to account for the fact that the M31 halo extends over a huge distance range, such that the observed volume behind M31 is substantially larger than the observed volume in front of the galaxy. While the increased volume behind M31 will tend to favor the detection of distant stars, there is also an opposing tendency that these more distant stars will be fainter, and hence correspondingly less numerous in a magnitude limited survey like PAndAS. To account for the dimming of the RGB, we assume that the RGB luminosity function of all populations has the form \citep{2002AJ....124..213M}:
\begin{equation}
L(M \ge M_{\rm TRGB}) \propto 10^{0.3(M-M_{\rm TRGB})} ,
\label{eqn:TRGB}
\end{equation}
where $M_{\rm TRGB}$ is the absolute magnitude of the RGB tip. The density profile is defined as a spline profile in $\log({\rm density})$ versus $\log({\rm radius})$, constrained at an arbitrary number of radial anchor points. To allow for a flattening of the smooth halo, we permit $\rho$ to be a function of $s$, so that the ellipsoid has the same axis of symmetry as the disk of M31. The flattening $q$ is introduced as a global variable or, optionally, as a non-parametric splined function of ellipsoidal radius. The halo density model $\rho(s)$ is thus forced to have the same axis of symmetry as the disk of M31. 
We refrained from testing other axes of symmetry for the halo, since any flattening directed along (or close to) our line of sight would be almost entirely unconstrained, due to the lack of good distance information along the line of sight. In any case, we feel that it is a natural assumption for an axisymmetric halo to have the same axis of symmetry as the disk, as a non-spherical halo in another configuration would give rise to non-circular motions of gas in the disk.

We proceed to calculate the model over the ($\xi,\eta$) plane, subject to the spatial footprint $S$ of the survey, taking bins of $0\degg1\times0\degg1$ on the sky. With the above definitions, the normalized projected density of the halo model in a chosen range in ${\rm [Fe/H]}$ at position ($\xi,\eta$) is:
\begin{equation}
D_{\rm halo}(\xi,\eta) = {{\int_{0}^\infty \rho(s) L_c(l) l^2 dl}\over{\int\int_{S}\int_{0}^\infty \rho(s) L_c(l) l^2 dl \, d\xi \, d\eta}} \, ,
\end{equation}
where $L_c(l)$ is a correction function to account for the dimming of the population with distance. This correction function can be evaluated simply by integrating Eqn.~\ref{eqn:TRGB} between a bright apparent magnitude cutoff $i_b(=20.5)$ and the apparent magnitude limit of the survey $i_{l}(=23.5)$:
\begin{equation}
L_c(l)={{
\exp[ a ( i_{l}-i_{TRGB}^\prime(l)) ] -  \exp[ a ( i_{b}-i_{TRGB}^\prime(l)) ]  }
\over{
\exp[ a ( i_{l}-i_{TRGB}) ] - \exp[ a ( i_{b}-i_{TRGB}) ] }} \, ,
\end{equation}
if the population is in front of M31, or:
\begin{equation}
L_c(l)={{
\exp[ a ( i_{l}-i_{TRGB}^\prime(l)) ] -  1   }
\over{
\exp[ a ( i_{l}-i_{TRGB}) ] - \exp[ a ( i_{b}-i_{TRGB}) ] }} \, ,
\end{equation}
if the population is behind that galaxy. The quantity $i_{TRGB}$ is the i-band magnitude of the RGB tip at the nominal distance of M31 ($785\kpc$), while $i_{TRGB}^\prime(l)$ is the apparent magnitude of the RGB tip at the line of sight distance $l$. The numerical constant $a=0.3 \ln(10)$.

\begin{figure*}
\begin{center}
{\hbox{
\includegraphics[angle=0, viewport= 150 60 765 580, clip,  width=9.0cm]{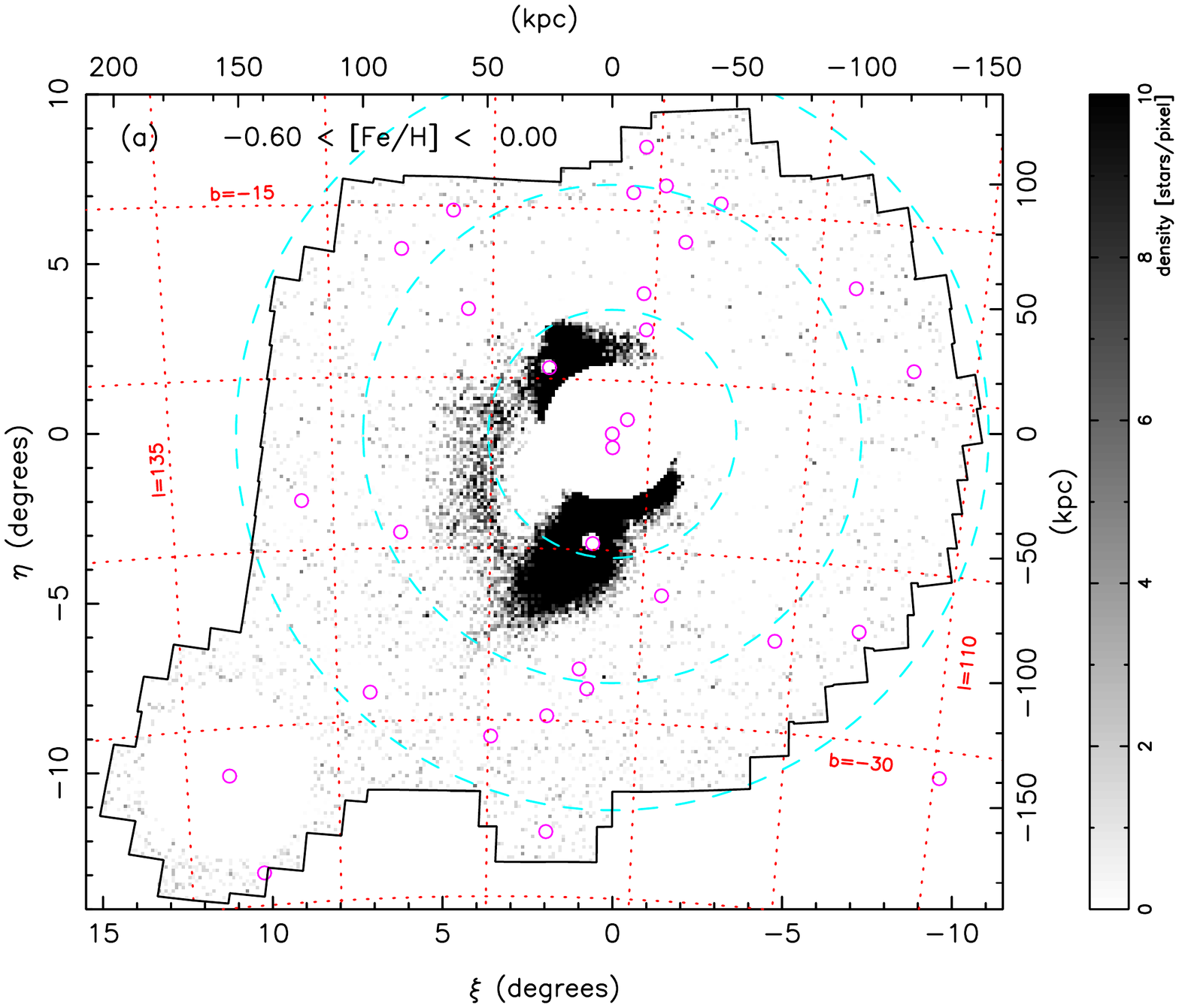}
\includegraphics[angle=0, viewport= 150 60 765 580, clip,  width=9.0cm]{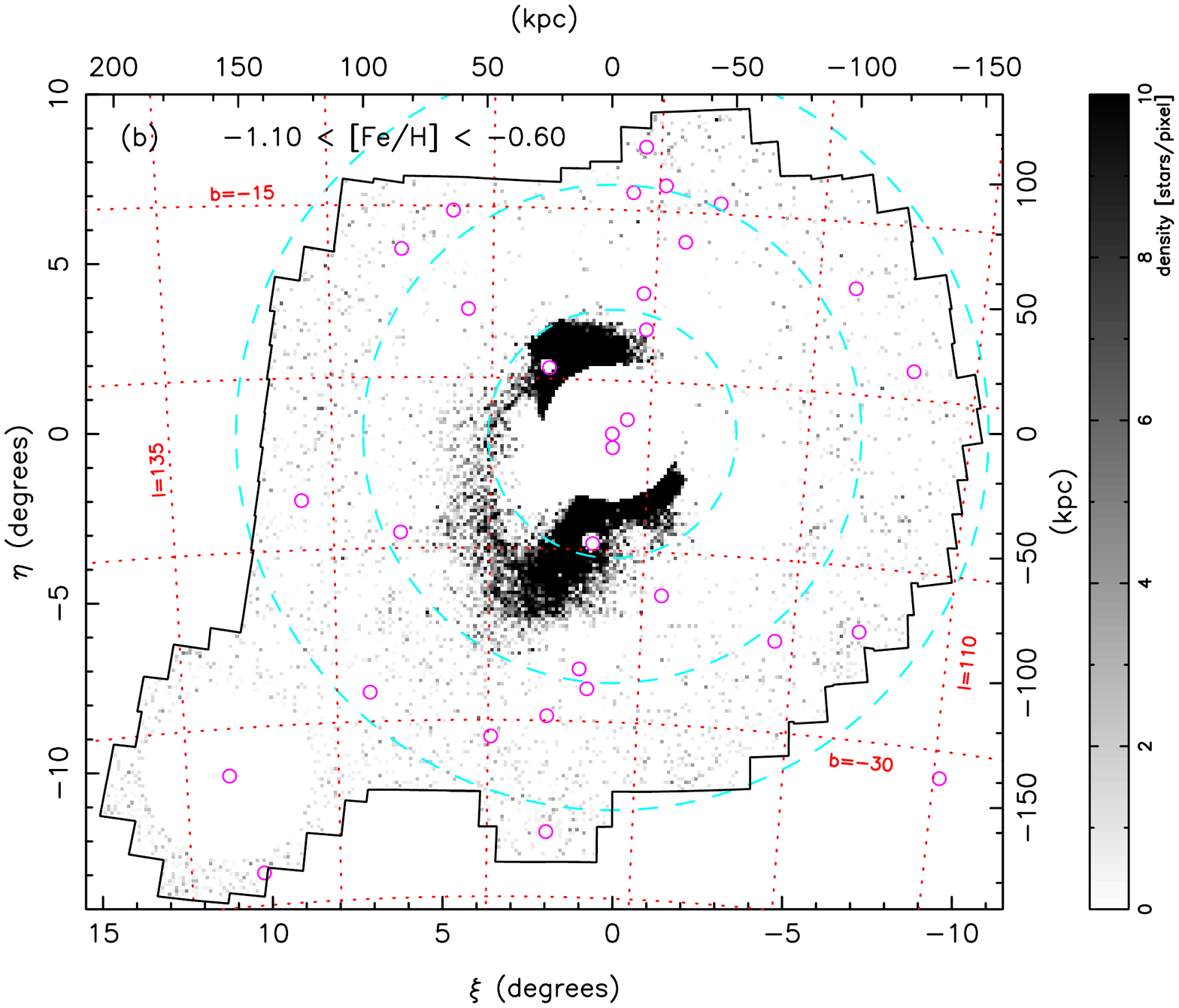}
}
\hbox{
\includegraphics[angle=0, viewport= 150 60 765 580, clip,  width=9.0cm]{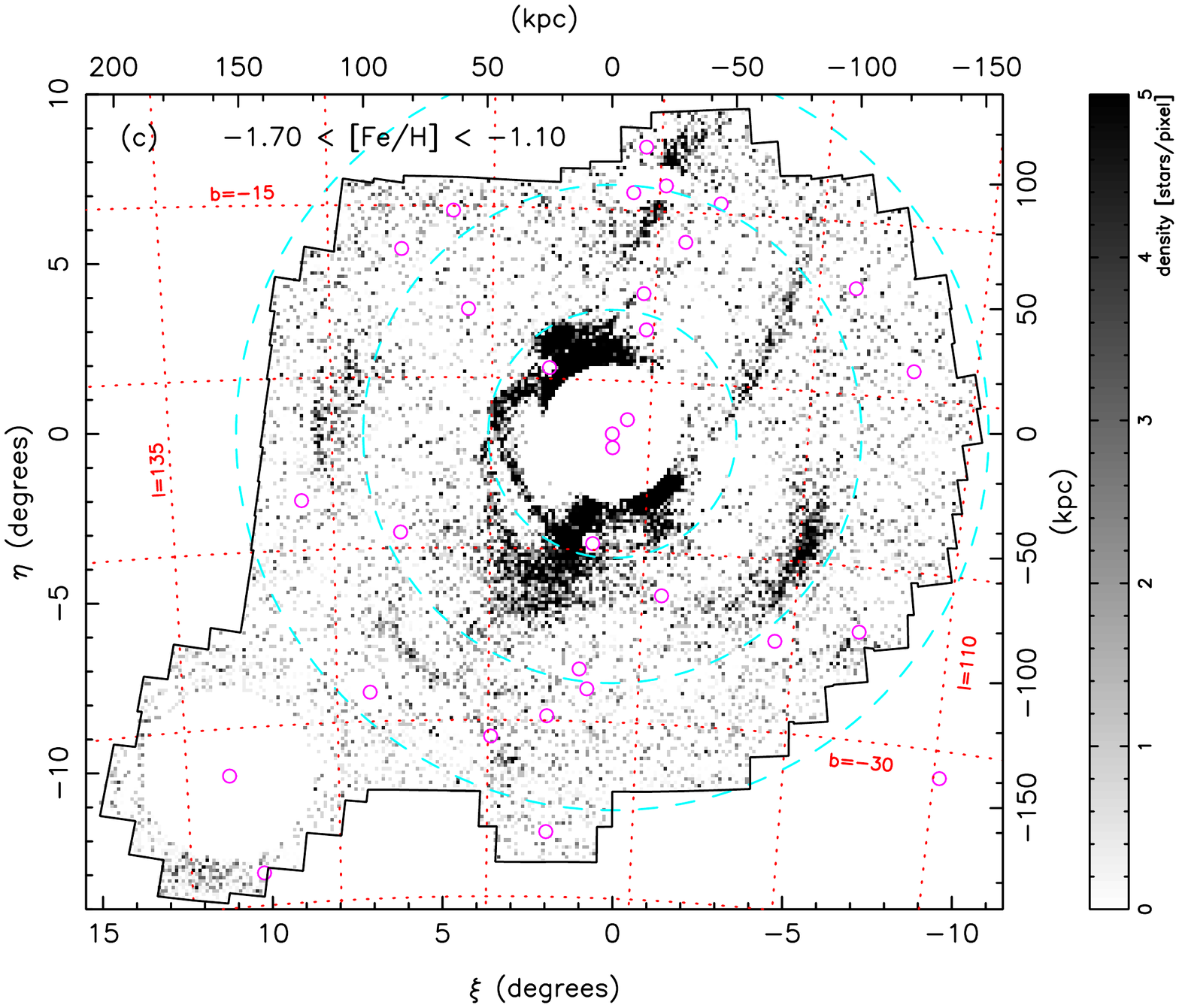}
\includegraphics[angle=0, viewport= 150 60 765 580, clip,  width=9.0cm]{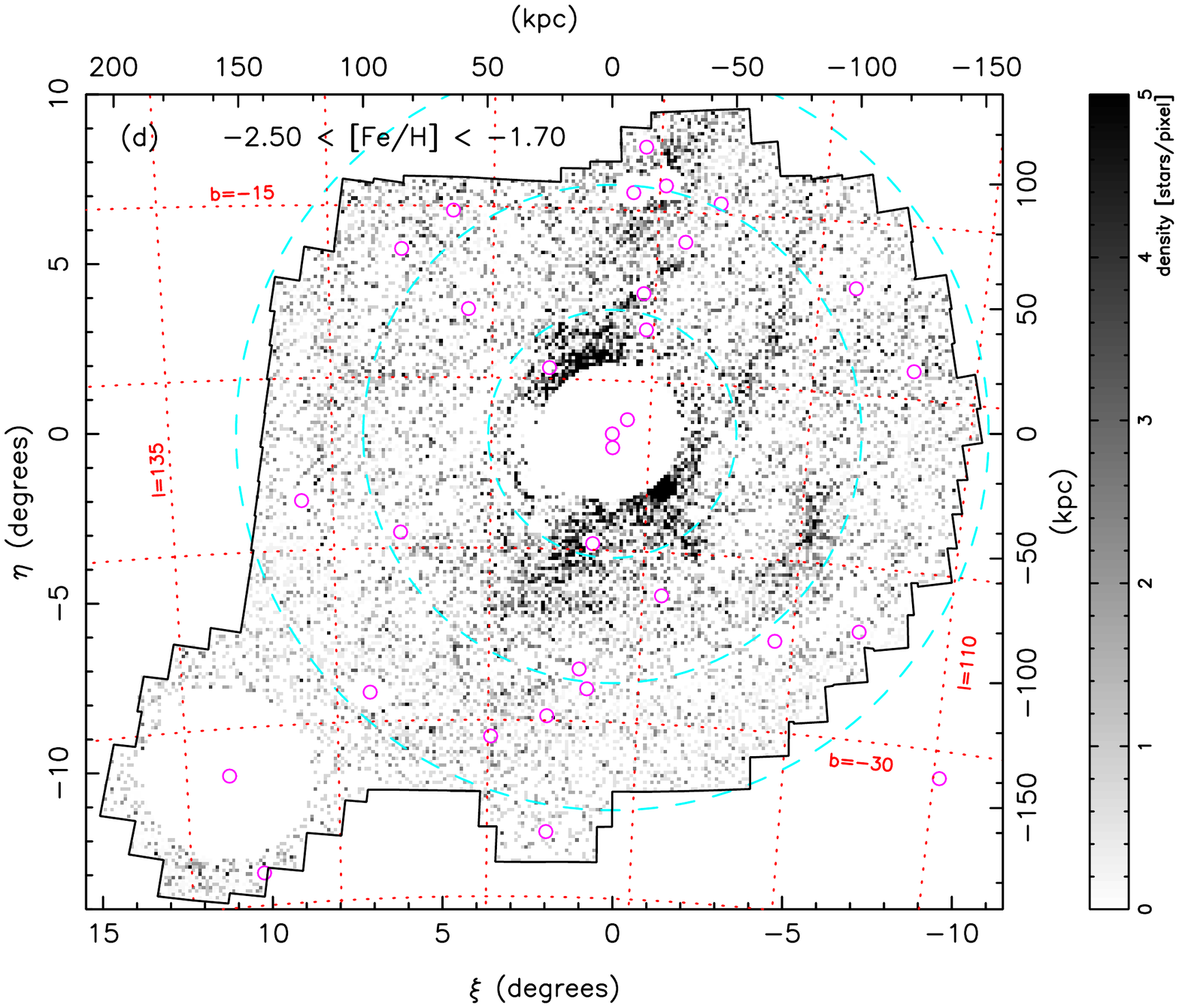}
}}
\end{center}
\caption{The residual maps of the smooth power-law halo fits (${\rm data-model}$). Each pixel is $0\degg1\times0\degg1$.}
\label{fig:map_residuals}
\end{figure*}

The M31 halo model thus created must be added to the contamination model before being compared to the selected stellar samples. However, the M13 contamination model $\Sigma_{(g-i,i)}(\xi,\eta)$ that we discussed above in \S\ref{sec:CMD_spatial} is necessarily an over-estimate in the color-magnitude bins occupied by the M31 halo, given that it used the region beyond $120\kpc$ to sample the background\footnote{This ``background'' is composed primarily of foreground Galactic contaminants.}. Unfortunately, there are no other suitable background fields in the CFHT archive that could be used to constrain better this contamination. We overcame this modeling problem by including an additional parameter $\mathcal{F}$ into our fit, in the form of a simple global factor to multiply the contamination model by. For convenience, we define a new quantity $C_{\rm bg}$ to be the sum over all color-magnitude bins of the M13 model that have metallicity in the chosen ${\rm [Fe/H]}$ range:
\begin{equation}
C_{\rm bg}(\xi,\eta)=\sum_{\rm CMD} \Sigma_{(g-i,i)}(\xi,\eta) \, .
\end{equation}
Thus the final model probability of finding a star in a small bin centered on $(\xi_i,\eta_i)$ becomes:
\begin{equation}
P_{\rm model}(\xi_i,\eta_i)={{(n-n_{\rm bg}) D_{\rm halo}(\xi_i,\eta_i) + \mathcal{F} C_{\rm bg}(\xi_i,\eta_i)}\over{n}} \, ,
\label{eqn:prob}
\end{equation}
where $n$ is the number of stars in the sample, and $n_{\rm bg} = \mathcal{F} \sum_S C_{\rm bg}$ is the sum of the background contamination model over the survey area $S$. Finally, we calculate the likelihood of the model given the data via:
\begin{equation}
\ln L = \sum_{i=1}^{n} \ln[P_{\rm model}(\xi_i,\eta_i)] \, .
\label{eqn:likelihood}
\end{equation}

We use the Markov Chain Monte Carlo (MCMC) engine described in \citet{2013MNRAS.428.3648I} to refine the initial guessed halo density model and the parameter $\mathcal{F}$. This MCMC solver uses multiple ``temperature'' parallel chains to probe the likelihood surface at different levels of smoothing, thereby greatly diminishing the chances that the algorithm will get caught in local maxima. The routine also uses a population of affine-invariant ``walkers'', which propagate across the likelihood surface; the scatter in the population adapts automatically to the steepness of the likelihood function in each dimension, and thereby probes the parameter space very efficiently. 

All runs consisted of $10^{5}$ iterations of the lowest temperature chain, and used a population of 100 walkers. The algorithm was tested extensively using artificial data, in which we included streams from N-body simulations as additional ``contamination'', mimicking the structures observed in our maps. For artificial samples of $10^6$ stars, our tests showed that the input parameters of the artificial smooth halo could be accurately recovered, including the flattening as a function of radius. 

Returning to the real PAndAS data, we used as before stars limited to $i_0<23.5$, $0.8<(g-i)_0<1.8$, and with color-magnitude consistent with stars that have a photometric metallicity in the range $-2.5 < [{\rm Fe/H}] < 0$, according to the above-mentioned Dartmouth isochrone ($13\Gyr$). Stars within circular regions around M31 and the known satellites were excluded ab initio (these regions can be discerned in Fig.~\ref{fig:map_masks}). 

A first run was undertaken to identify the spatial locations likely to contain streams and substructure. To this end, we imposed a prior to penalize negative residuals between the observed spatial distribution and the model (halo plus contamination). Despite having implemented the possibility of variable flattening in the fitting software, we decided for the present analysis to maintain the halo flattening as a global variable. Fig.~\ref{fig:map_masks} shows the areas (using a smoothing kernel of $0\degg15$) where the resulting residuals were found to be in excess of $2.5\sigma$; a visual comparison to Fig.~\ref{fig:map_slices_metallicity} shows that this procedure has nicely identified the substructures, as intended. The masked fractions of the available area (${\rm A}_{\rm masked}$, from which we have excluded the circular regions around the satellites) are listed in Table~\ref{tab:fractions} for the various metallicity selections; note that these remove only $\sim 10$\% of the available halo area. For completeness we also give the contamination fraction ($f_{\rm contam}$) in the metallicity interval according to the M13 model. 

The faintest of the masked structures is ``stream A'' previously reported in Paper~I, which has a surface brightness of $\Sigma_V \sim 32 {\rm \, mag \, arcsec^{-2}}$, and which is visible in Fig.~\ref{fig:map_slices_metallicity}c at ($\xi \sim 6\degg5$, $\eta\sim-6^\circ$). We propose this as an estimate of the limiting surface brightness beyond which we can no longer resolve the detections into separate structures (this is not a hard limit since it also depends on the size of the feature and the background level).

\begin{table}
\begin{center}
\caption{The foreground/background contamination fraction and the fraction of the total area suppressed by the masks shown in Fig.~\ref{fig:map_masks}.}
\label{tab:fractions}
\begin{tabular}{ccccr}
\tableline\tableline
Selection & panel & $f_{\rm contam}$ & ${\rm A}_{\rm masked}$ & $n$ \\
               &  & (fraction)\footnotemark[1] & (fraction) & \footnotemark[1]\\
\tableline
${\rm -0.6 < [Fe/H] <  0.0}$ & a & $0.56$ & $0.08$ &223107\\
${\rm -1.1 < [Fe/H] < -0.6}$ & b & $0.56$ & $0.08$ &288425\\
${\rm -1.7 < [Fe/H] < -1.1}$ & c & $0.66$ & $0.09$ &196259\\
${\rm -2.5 < [Fe/H] < -1.7}$ & d & $0.68$ & $0.09$ &94839\\
${\rm -2.5 < [Fe/H] <  0.0}$ & e & $0.56$ & $0.13$ &802630\\
\tableline\tableline
\end{tabular}
\end{center}
{\bf Notes.}
\footnotetext[1]{Fraction of the total number of stars (before masking, but after rejecting the circular regions around M31 and its satellites) considered to be contamination according to the M13 contamination model. The multiplicative factor $\mathcal{F}$ is not applied.}
\footnotetext[1]{Total number of stars in metallicity selection (unmasked and with no contaminants removed).}
\end{table}

Having identified the over-dense areas, we use them to define masks that will facilitate the study of the underlying smooth component. It is worth reiterating at this juncture that we are of course aware that with a sufficiently fine resolution and sufficiently deep data, it is possible that we would find that what we are calling the smooth component is composed of a (probably quite large) number of correlated families of stars. However, given our observational limitations, we feel that it is a useful approximation to group the unresolved structures into a single entity, and explore how this may be modeled as an ellipsoidal structure.

The algorithm was subsequently re-run, this time fixing the substructure masks defined from the initial iteration, and with the previous prior on negative residuals disabled. The halo fits to the masked samples using a simple power-law profile and a single global flattening are shown in color in Fig.~\ref{fig:map_masks}, the corresponding residuals in Fig.~\ref{fig:map_residuals}, and the details of the fits are listed in Table~\ref{tab:fits}. The table shows that the power-law exponent $\gamma$ of the smooth component becomes progressively shallower with decreasing metallicity. (We define $\gamma$ to be the exponent of the three-dimensional density distribution, while $\Gamma$ below will refer to the two-dimensional projected distribution. Likewise, $r$ below refers to 3-D radius, while $R$ is projected radius). In all cases, the algorithm finds a very close to spherical stellar halo. Note that the putative smooth component represents a relatively small fraction (14\%) of the total number of stars more metal rich than ${\rm [Fe/H]=-0.6}$. It is only for the most metal-poor selection that the smooth component becomes dominant (58\%). We note also that the contamination model correction factor $\mathcal{F}$ is, as expected, slightly lower than unity.

\begin{figure}
\begin{center}
\includegraphics[angle=0, viewport= 30 30 530 540, clip, width=\hsize]{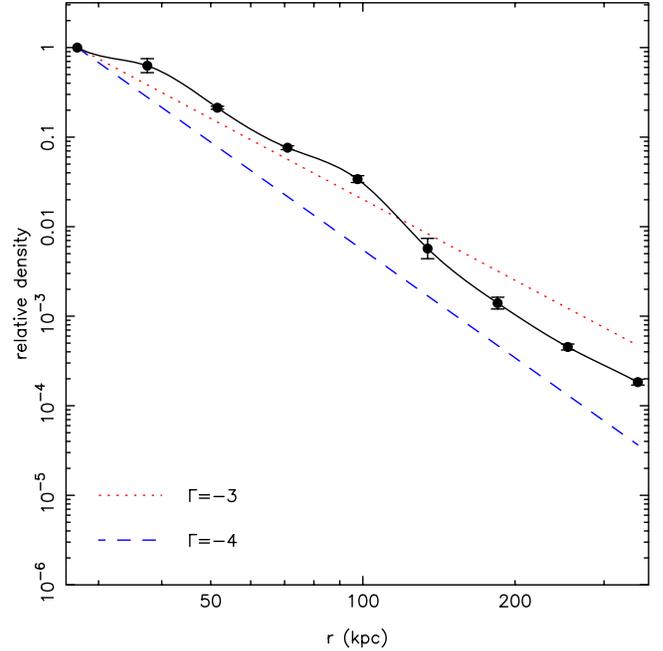}
\end{center}
\caption{The non-parametric density profile fit to the smooth halo for the metallicity range ${\rm -2.5 < [Fe/H] < -1.7}$, as a function of (three-dimensional) radius $r$. The filled circles show the anchors of the most likely halo model, with the corresponding error bars marking $1\sigma$ uncertainties derived from the Markov chain. The continuous line demonstrates the smooth spline function that the algorithm generates to interpolate the density. To aid visual interpretation, we have overlaid a $\gamma=-3$ profile (dotted line) and a $\gamma=-4$ profile (dashed line). The flattening of this model is $q=1.11\pm0.04$.}
\label{fig:nonparametric_profile}
\end{figure}

\begin{table}
\begin{center}
\caption{Parameters of the three dimensional fits to the smooth halo component (power law model), using the masked samples.}
\label{tab:fits}
\begin{tabular}{ccccc}
\tableline\tableline
Selection & $\gamma$ & $q$ & $f_{\rm smooth}$ & $\mathcal{F}$\\
               & \footnotemark[1] & \footnotemark[2] & (fraction)\footnotemark[3] & \footnotemark[4]\\
\tableline
${\rm -0.6 < [Fe/H] <  0.0}$ & $-3.34$      & $1.01$       & 0.14 & $0.93$       \\
                                            & $\pm0.04$ & $\pm0.07$ &         & $\pm0.04$  \\
${\rm -1.1 < [Fe/H] < -0.6}$ & $-3.62$      & $1.05$       & 0.22 & $0.91$       \\
                                            & $\pm0.08$ & $\pm0.04$ &         & $\pm0.04$  \\
${\rm -1.7 < [Fe/H] < -1.1}$ & $-3.16$      & $1.07$       & 0.42 & $0.93$      \\
                                            & $\pm0.06$ & $\pm0.04$ &         & $\pm0.04$ \\
${\rm -2.5 < [Fe/H] < -1.7}$ & $-3.08$      & $1.09$       & 0.58 & $0.93$       \\
                                            & $\pm0.07$ & $\pm0.03$ &         & $\pm0.03$ \\
${\rm -2.5 < [Fe/H] <  0.0}$ & $-3.59$      & $1.11$       & 0.21 & $0.93$      \\
                                            & $\pm0.08$ & $\pm0.07$ &         & $\pm0.03$ \\
\tableline\tableline
\end{tabular}
\end{center}
{\bf Notes.}
\footnotetext[1]{Power law exponent.}
\footnotetext[2]{Density flattening parameter.}
\footnotetext[3]{Fraction of the halo between $2\deg$($=27.2\kpc$) and $150\kpc$ present in the smooth component.}
\tablenotetext{4}{Multiplicative correction factor applied to the M13 ``background'' contamination model.}
\end{table}

\begin{table}
\begin{center}
\caption{Power-law fits to the projected distribution.}
\label{tab:fits_projected}
\begin{tabular}{ccc}
\tableline\tableline
Selection & $\Gamma_{\rm unmasked}$ & $\Gamma_{\rm masked}$ \\
               & \footnotemark[1] & \footnotemark[2] \\
\tableline
${\rm -1.1 < [Fe/H] <  0.0}$ & $-3.72$                     & $-2.66$ \\
                                            & $\pm0.01(\pm0.32)$ & $\pm0.02(\pm0.19)$ \\
${\rm -1.7 < [Fe/H] < -1.1}$ & $-2.71$                     & $-2.13$ \\
                                            & $\pm0.01(\pm0.22)$ & $\pm0.02(\pm0.12)$ \\
${\rm -2.5 < [Fe/H] < -1.7}$ & $-2.30$                     & $-2.08$ \\
                                            & $\pm0.02(\pm0.12)$ & $\pm0.02(\pm0.12)$ \\
\tableline\tableline
\end{tabular}
\end{center}
{\bf Notes.}
\footnotetext[1]{Power law exponent fit to the unmasked sample. The uncertainty in brackets reflects the uncertainty due to the azimuthal scatter in the profiles.}
\footnotetext[2]{Power law exponent fit to the masked sample.}
\end{table}

Given the paucity of candidate stars that could belong to the smooth halo in the higher metallicity bins, we decided to only attempt the spline profile fitting of the halo density profile with the ${\rm -2.5 < [Fe/H] < -1.7}$ selection. The fitted density profile is shown in Fig.~\ref{fig:nonparametric_profile}, for which the corresponding flattening is $q=1.11\pm0.04$. It is interesting to note that the algorithm finds a profile that falls off as $\gamma \sim -3$ out to approximately $100\kpc$, beyond which the profile steepens towards $\gamma=-4$.

\begin{figure*}
\begin{center}
\includegraphics[angle=0, viewport= 40 35 780 425, clip, width=\hsize]{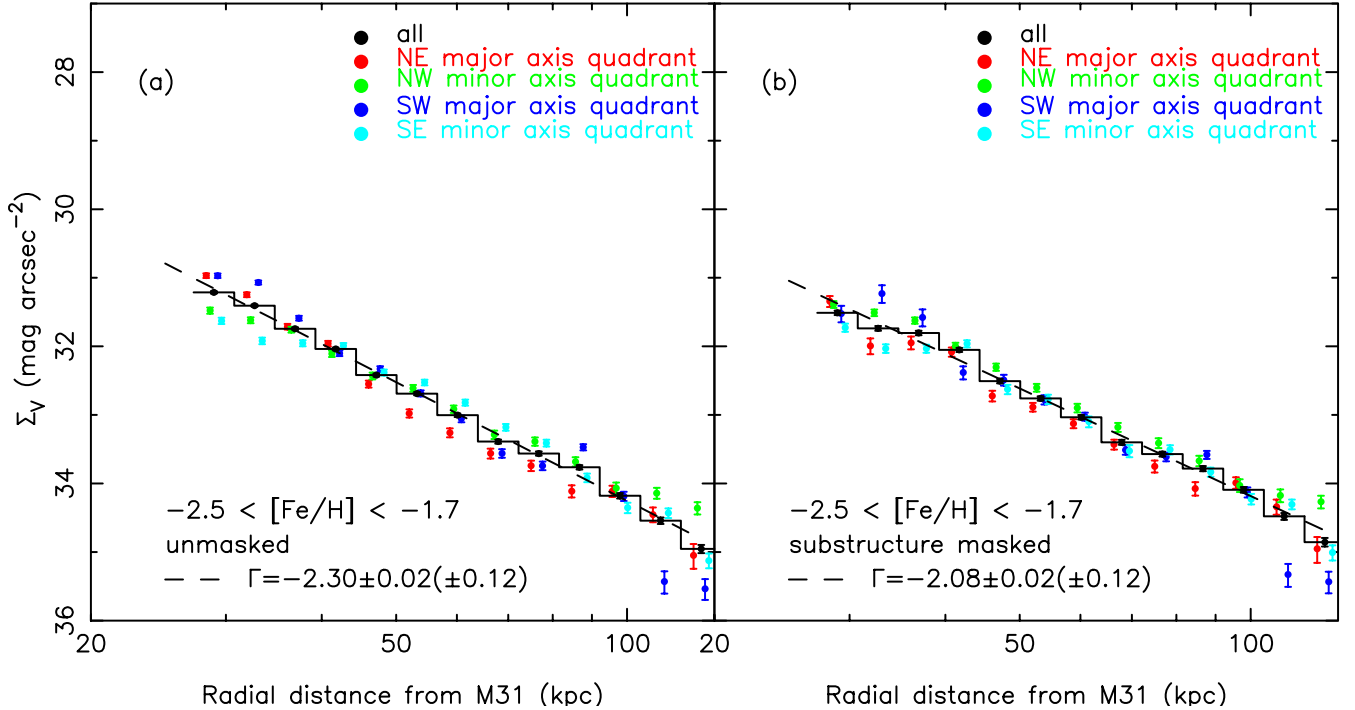}
\end{center}
\caption{The star-counts profile of the smooth metal poor (${\rm -2.5 < [Fe/H] < -1.7}$) population as a function of projected radial distance. Panel (a) corresponds to the full unmasked survey (but regions around known satellites are excluded), while panel (b) is masked with the mask shown in Fig.~\ref{fig:map_masks}d. In both panels the data for all azimuthal angles is shown in black. We also show the profiles of the four quadrants individually, color-coded according to quadrant as indicated in the diagram. A small radial offset has been applied to the colored data points in order to make them easier to see. The overall normalization (i.e the vertical offset) has been set by comparison to Dartmouth stellar population models. For the models we assumed an age of $13\Gyr$, $[\alpha/Fe]=0.0$, and a log-normal initial mass function; we also examined $9\Gyr$ models, but the differences (which can be appreciated from Table~\ref{tab:properties}) were found to be relatively small. The dashed line is a linear fit to the full profile, implying a (projected) power-law slope of $\Gamma=-2.30\pm0.02$ and $\Gamma=-2.08\pm0.02$, respectively for the unmasked and masked samples. The larger uncertainties marked in brackets on the diagrams are derived from taking the root mean square scatter in azimuthal bins as an estimate of the uncertainty in the profile.}
\label{fig:projected_profile_sel4}
\end{figure*}
\begin{figure*}
\begin{center}
\includegraphics[angle=0, viewport= 40 35 780 425, clip, width=\hsize]{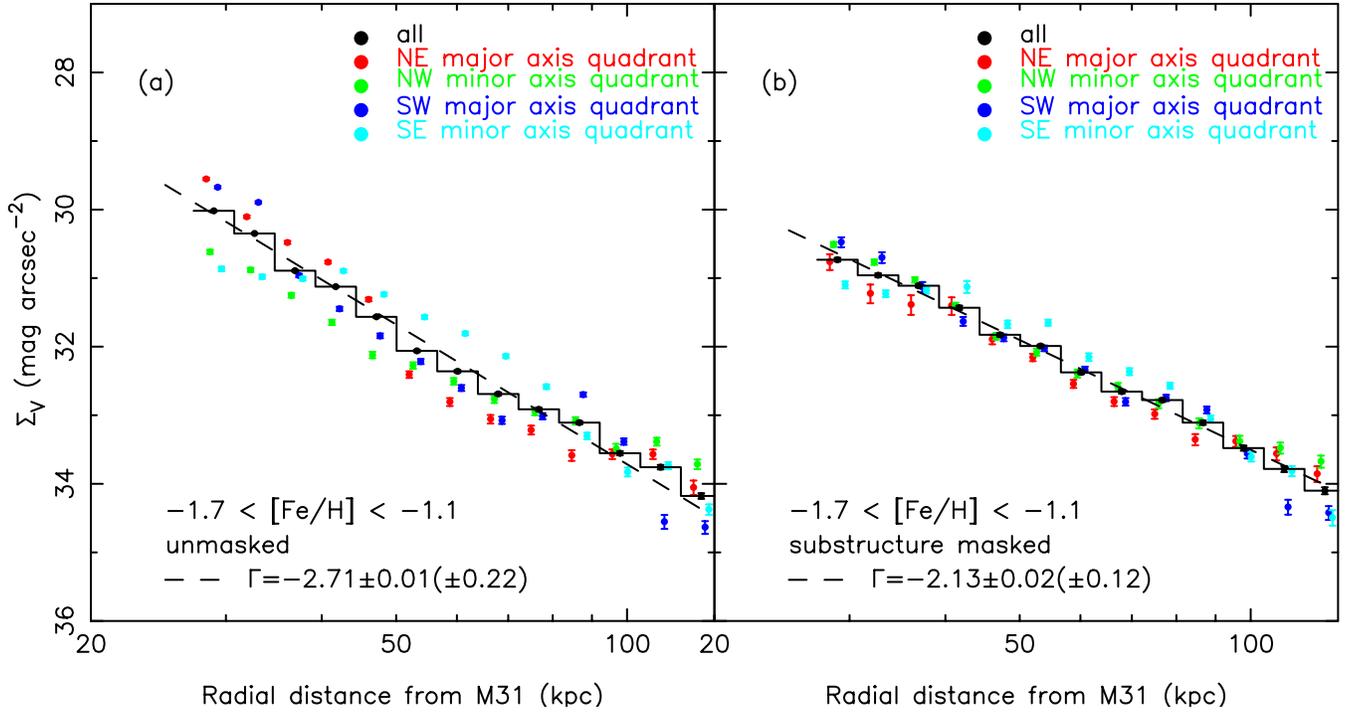}
\end{center}
\caption{As Fig.~\ref{fig:projected_profile_sel4}, but for the ${\rm -1.7 < [Fe/H] < -1.1}$ population. Over this vast radial range the projected profile of the masked population closely approximates a power-law.}
\label{fig:projected_profile_sel3}
\end{figure*}

\begin{figure*}
\begin{center}
\includegraphics[angle=0, viewport= 40 35 780 425, clip, width=\hsize]{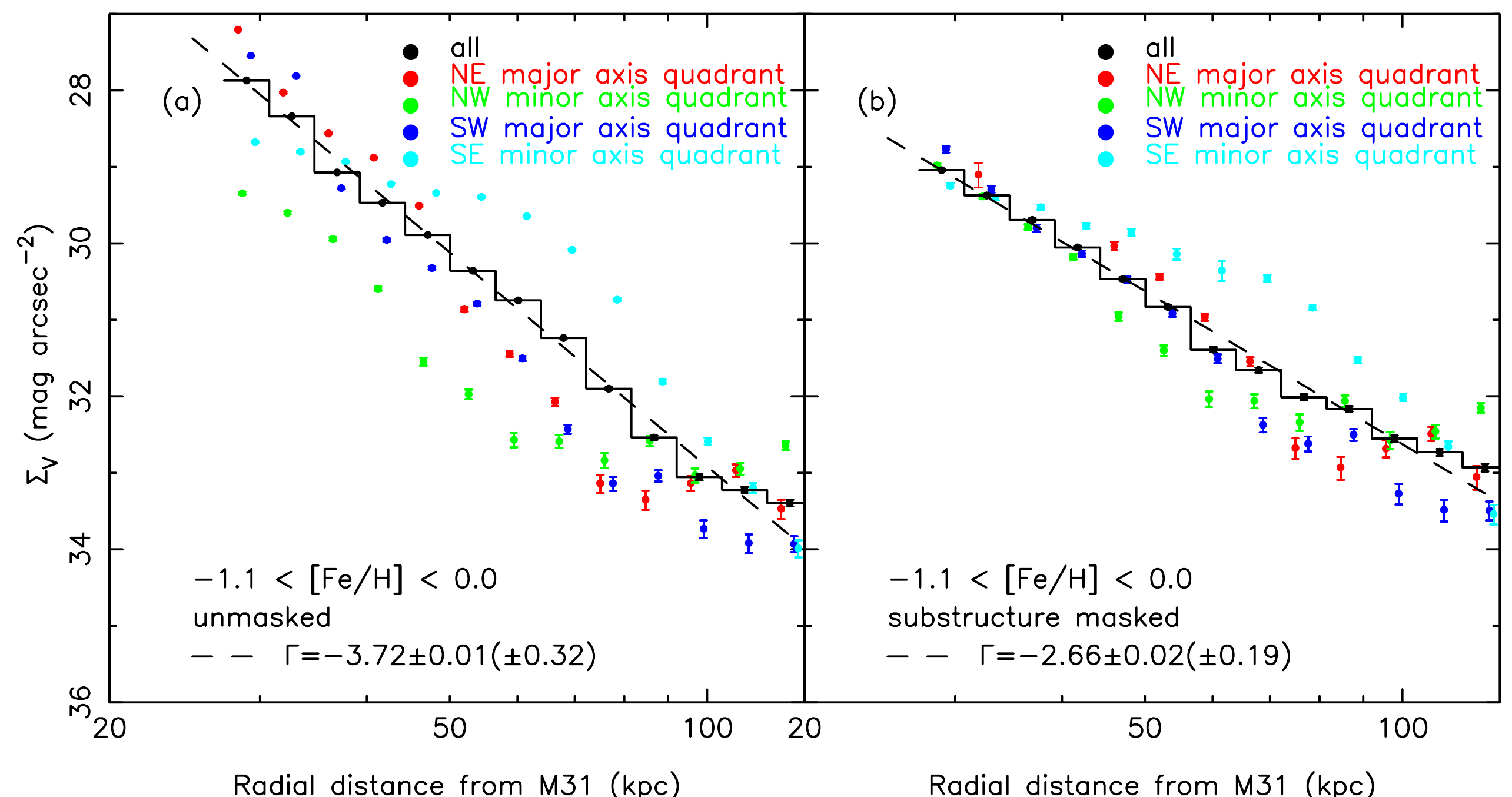}
\end{center}
\caption{As Fig.~\ref{fig:projected_profile_sel4}, but for stars more metal rich than ${\rm [Fe/H] =-1.1}$. The very large variations between quadrants in the profile derived from the masked sample (panel b) are almost certainly due to residual substructures that have not been eliminated by the masks (Figs.~\ref{fig:map_masks}a and \ref{fig:map_masks}b), the Giant Southern Stream in particular. The marked power law has slope $\Gamma=-2.66$, but it is clearly not applicable for the profile in all quadrants beyond the inner halo.}
\label{fig:projected_profile_sel1and2}
\end{figure*}

\subsection{Two-dimensional fits}
\label{sec:fit2D}

As an alternative to the three-dimensional halo model fits discussed in the previous section, we present in Fig.~\ref{fig:projected_profile_sel4} the direct (i.e. projected) star-counts profile for the ${\rm -2.5 < [Fe/H] < -1.7}$ selection. For the full data (panel a) over all azimuthal angles (black), the power-law fit (dashed line) has a slope of $\Gamma=-2.30\pm0.02$. When the substructure is masked out (panel b, using the mask in Fig.~\ref{fig:map_masks}), the slope is $\Gamma=-2.08\pm0.02$; given that this has been fit to a {\em projected} profile, it is fairly similar to the corresponding MCMC model fit listed in Table~\ref{tab:fits}.  In Fig.~\ref{fig:projected_profile_sel4} we also show the data split into four quadrants (colored points); clearly there are significant differences between the profiles, but the radial profile appears similar between quadrants. Fig.~\ref{fig:projected_profile_sel3} shows the same information for the ${\rm -1.7 < [Fe/H] < -1.1}$ selection; the somewhat larger scatter around the power-law fit in panel `b' compared to Fig.~\ref{fig:projected_profile_sel4}b is not surprising given the substantial substructure evident in this metallicity interval, which our simple spatial mask (Fig.~\ref{fig:map_masks}c) has evidently not entirely removed. The power law fit has an exponent $\Gamma=-2.13\pm0.02$, identical within the uncertainties with that of the most metal poor selection. It is interesting to note that both of these profiles show no steepening of the profile at large radius.

Due to the small number of stars in the putative smooth halo component at metallicities greater than ${\rm [Fe/H] =-1.1}$, we combined the two metal-rich bins to construct the profile shown in Fig.~\ref{fig:projected_profile_sel1and2}b. Given the large variations between quadrants, it is obvious that our masks have not removed all of the substructure, and the residual presence of the Giant Stellar Stream in the SE minor axis quadrant is particularly striking. In the unmasked profiles displayed in Fig.~\ref{fig:projected_profile_sel1and2}a, it is surprising to find that the azimuthally-averaged profile follows quite closely a (very steep) power law whereas the profiles from individual quadrants show huge scatter.

\subsection{Metallicity structure}
\label{sec:metallicity_structure}

We now turn our attention to the variation of metallicity with radius. The stellar sample with $i_0<23.5$ and $0.8<(g-i)_0<1.8$ was divided into 25 equally spaced bins in metallicity between ${\rm [Fe/H]=-2.5}$ and ${\rm [Fe/H]=0.0}$, using the $13\Gyr$ Dartmouth isochrone model, as discussed above. The upper color limit was imposed on the sample as we deemed that it was the only practical way of avoiding overwhelming contamination by foreground dwarf stars, but it can be appreciated from an inspection of Fig.~\ref{fig:Hess_overall} that the limit excludes most of the evolved metal-rich RGB stars from our sample. Fortunately, it is possible to use the same isochrone models to estimate the number of stars that were missed by the cut and correct for the absent members. The model estimates that no stars are missing below ${\rm [Fe/H]=-1.1}$,  but for more metal rich populations the correction ramps up to a factor of $3.8$ at ${\rm [Fe/H]=0.0}$. For younger populations the correction is somewhat smaller: for instance a $9\Gyr$ old model predicts a correction of a factor of $3.3$ at ${\rm [Fe/H]=0.0}$.

\begin{figure}
\begin{center}
\includegraphics[angle=0, viewport= 35 35 570 780, clip, width=\hsize]{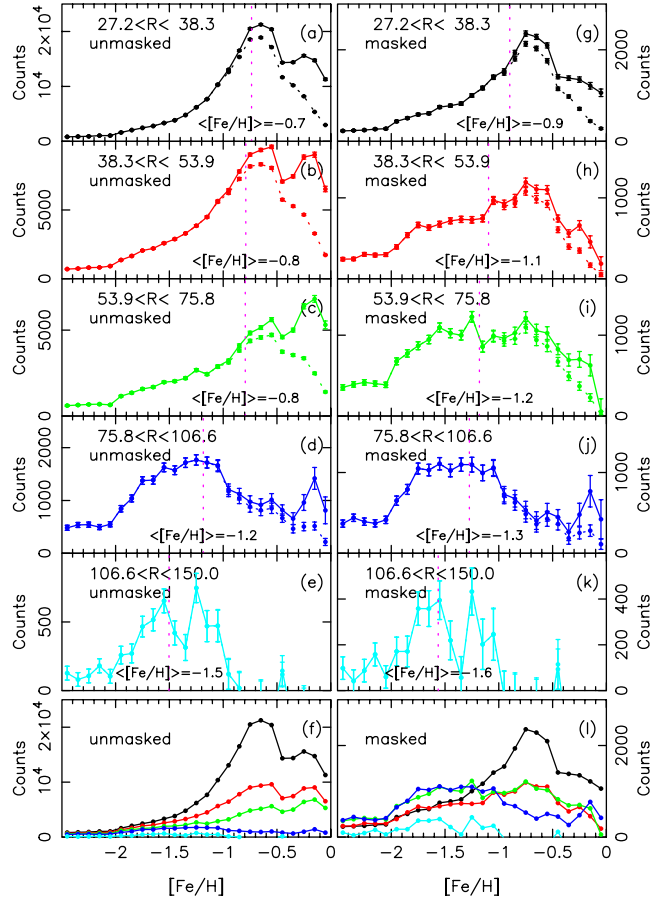}
\end{center}
\caption{The metallicity distribution of the M31 halo in five logarithmically-spaced intervals covering the range $27.2 < R < 150\kpc$. All stars within small circular regions of all the known satellites have been removed. The left column presents the metallicity distributions of the full sample, while the column on the right shows the sample subject to the mask of Fig.~\ref{fig:map_masks}e (which is appropriate for the full metallicity range $-2.5 < {\rm [Fe/H]} < 0.0$). No correction is applied to the masked sample to account for the absent areas of sky. The dashed line connects points that represent the raw counts in each metallicity bin. Due to the imposed color cut of $(g-i)_0<1.8$, there is significant incompleteness at the metal-rich end; we correct for this using a $13\Gyr$ Dartmouth population model, as described in the text. The resulting corrected counts are connected with the full line curve. The error-bars shown correspond to Poisson counting uncertainties. The dotted vertical line in magenta shows the mean metallicity of each distribution, whose value is indicated. The bottom two panels reproduce the previous rows on the same vertical scale to aid their comparison. A prominent trend towards lower metallicity as a function of radius is revealed by these data, both for the full and masked samples, with the masked distributions being more metal-poor than the unmasked distributions at the same radial position.}
\label{fig:metallicity_distribution}
\end{figure}

In the discussion above we have seen that what we have dubbed the ``smooth component'' follows a power law with index $\Gamma \sim -2$ in projection, with the metal-poor stars following this behavior quite closely. This motivates the use of logarithmic distance bins, since $\Gamma=-2$ would imply equal numbers of stars present in each interval. Figure~\ref{fig:metallicity_distribution} shows the sample distributed into five such intervals between $R=27.2\kpc$ ($2\deg$) and $R=150\kpc$. The raw star counts are displayed as points with uncertainties connected by a dashed line. We also show the counts corrected for the incompleteness of the red stars  beyond $(g-i)_0=1.8$; these are connected with the continuous line. 

The left column of Fig.~\ref{fig:metallicity_distribution} shows the full sample (minus satellites), while the distributions in the right column have had the substructure masked out (using the mask displayed in Fig.~\ref{fig:map_masks}e). These data show a clear metallicity gradient in both the full and masked samples. The bottom row of  Fig.~\ref{fig:metallicity_distribution} displays the different radial bins on the same scale to allow the reader an appreciation of the relative density present in each interval. Note that for the masked populations (panel l), the metal-poor populations (${\rm Fe/H]} \sim -1.5$) have roughly the same count level, consistent with $\Gamma\sim -2$ given the logarithmic radial bins.

\subsection{Mass and luminosity of the halo}

\begin{figure}
\begin{center}
\includegraphics[angle=0, viewport= 30 30 530 540, clip, width=\hsize]{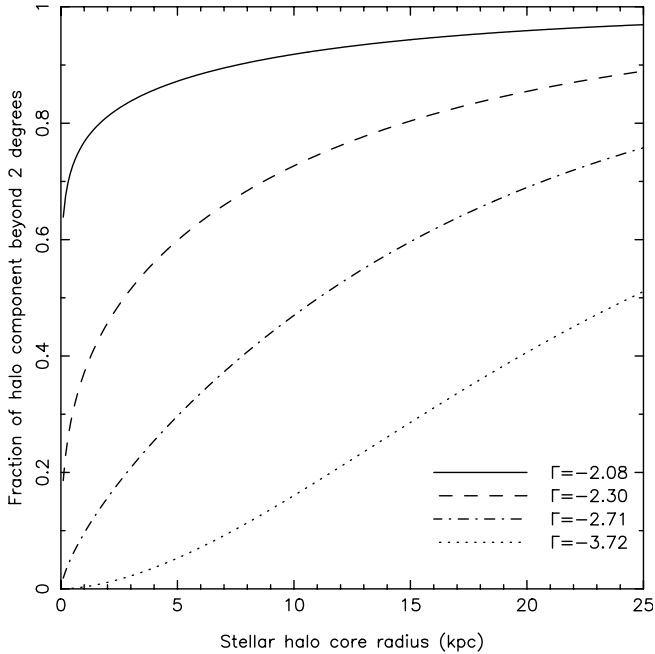}
\end{center}
\caption{The fraction of the halo present in the PAndAS sample beyond $2\deg$. Assuming that the halo has a cored power-law form $(1+(R/R_c)^2)^{\Gamma/2}$, we calculate the fraction of the halo we have measured beyond $2\deg$, as a function of the core radius $R_c$. The lines show the relations for various $\Gamma$ values measured in Figs.~\ref{fig:projected_profile_sel4}-\ref{fig:projected_profile_sel1and2}. The reader may hereby easily use their preferred core radius value to correct our halo mass measurements for the missing central $2\deg$.}
\label{fig:central_correction}
\end{figure}

The total mass and luminosity of stellar halos are among their most fundamental properties, and it is clearly important to provide this information on the M31 halo derived from a panoramic study such as PAndAS. But while the present data set is superb for quantifying the populations at large radius, the superposition of the central  components towards the inner galaxy precludes a true global analysis.

Obviously it makes little sense to extrapolate the power-laws we have fitted in \S\ref{sec:fit2D} to $R=0$, since they will diverge. Furthermore, very little real information on the stellar halo exists in the inner few kpc of M31 that is not highly dependent on the modeling choices (and in particular, on the dominant bulge and disk components that must be subtracted to reveal the halo). So we again take a pragmatic approach and measure the stellar halo's properties where we have detected this component with confidence in PAndAS, namely from $R=27.2\kpc$ to $R=150\kpc$.

To calculate the mass and luminosity of the stellar populations, we need to correct again for the stars that are not present in the color-magnitude selection box (Fig.~\ref{fig:Hess_overall}), notably all of the stars below $i_0=23.5$ in addition to the metal-rich stars with colors redder than $(g-i)_0=1.8$. This is accomplished with the aid of the previously-used $13\Gyr$ Dartmouth model, as well as a younger $9\Gyr$ model for comparison. The results are listed in Table~\ref{tab:properties}, where we give the absolute V-band magnitude, the i-band magnitude, the mass and the luminosity of this radial selection of the halo. It can seen that for these plausible choices of population age, the derived quantities do not change substantially. The alternative measurements for the halo sample with substructure masked out are given in Table~\ref{tab:properties_masked}, where we have accounted for the missing area under the masks.

It can be seen that for the unmasked population, the halo beyond $2\deg$ amounts to $\sim 10^{10}\msun$, or approximately 10\% of typical estimates of the total baryonic mass of M31 \citep{Klypin:2002bm}. The smooth halo in the same distance range amounts to approximately one third of this value.

How much mass or light are we likely missing by rejecting the central $2\deg$? Figs.~\ref{fig:projected_profile_sel4}-\ref{fig:projected_profile_sel1and2} show that the azimuthally-smoothed stellar halo follows a power law quite closely, so it may be instructive to consider a cored power law $(1+(R/R_c)^2)^{\Gamma/2}$ behavior in the central regions, as is commonly assumed (see, e.g., \citealt{Courteau:2011fr}). Clearly the core radius $R_c$ must be significantly smaller than $27\kpc$, since we do not see a turnover in the star counts profiles. In Fig.~\ref{fig:central_correction} we integrate this cored power law to calculate the fraction of the total light or mass we have measured in our samples beyond $2\deg$ as a function of the value of $R_c$. One can see, for instance, that for the smooth metal-poor population with $\Gamma=-2.08$, most of the mass is already accounted for in Table~\ref{tab:properties_masked}, even if $R_c$ is as small as $1\kpc$. On the other hand, for the unmasked metal-rich selection with $\Gamma=-3.72$, the values listed in Table~\ref{tab:properties} would only be a small fraction of the total  population, if the true profile really were a cored power-law all the way to the galaxy centre.

To take a concrete example, if we adopt a core radius value of $R_c=5.2\kpc$, which is the best-fit value of the preferred M31 model by \citet{Courteau:2011fr}, and extrapolate inwards using the fitted power laws, we estimate the total stellar mass in the smooth component over all radii to be $\sim 8\times10^{9}\msun$. This is substantially larger than the total mass of the Galactic halo, which is estimated at $\sim 2\times10^9\msun$ (see discussion in \citealt{Bullock:2005is}). We refrain from attempting to extrapolate inwards the unmasked clumpy halo, since the answer we would obtain would be non-sensical.

\begin{table*}
\begin{center}
\caption{Properties of the full M31 halo sample in the projected radial range from $27.2\kpc$ ($2\deg$) to $150\kpc$.}
\label{tab:properties}
\begin{tabular}{ccccccccc}
\tableline\tableline
Selection                        & 
$M_{V,Vega}^{13\Gyr}$ &
$M_{V,Vega}^{9\Gyr}$ &
$M_{i,AB}^{13\Gyr}$ &
$M_{i,AB}^{9\Gyr}$ &
$({{\rm mass}\over{10^9 \msun}})^{13\Gyr}$ &
$({{\rm mass}\over{10^9 \msun}})^{9\Gyr}$ &
$({{L}\over{10^9 \lsun}})^{13\Gyr}$ &
$({{L}\over{10^9 \lsun}})^{9\Gyr}$ \\
\tableline
$-2.5 < {\rm [Fe/H]} < -1.7$ & -14.5 & -14.7 & -15.3 & -15.4 &  0.44 &  0.44 & 0.09 & 0.10 \\
$-1.7 < {\rm [Fe/H]} < -1.1$ & -15.2 & -15.3 & -16.0 & -16.1 &  1.1  &  1.2  & 0.17 & 0.19 \\
$-1.1 < {\rm [Fe/H]} < -0.6$ & -16.0 & -16.1 & -17.0 & -17.0 &  3.0  &  3.2  & 0.49  & 0.47  \\
$-0.6 < {\rm [Fe/H]} <  0.0$ & -16.1 & -16.0 & -17.4 & -17.0 &  6.0  &  4.0  & 1.2  & 0.27  \\
$-2.5 < {\rm [Fe/H]} <  0.0$ & -17.1 & -17.2 & -18.2 & -18.0 & 10.5  &  8.8  & 1.9  & 1.0  \\
\tableline\tableline
\end{tabular}
\end{center}
\end{table*}

\begin{table*}
\begin{center}
\caption{Properties of the masked M31 halo sample in the projected radial range from $27.2\kpc$ ($2\deg$) to $150\kpc$, corrected for the masked-out area.}
\label{tab:properties_masked}
\begin{tabular}{ccccccccc}
\tableline\tableline
Selection                        & 
$M_{V,Vega}^{13\Gyr}$ &
$M_{V,Vega}^{9\Gyr}$ &
$M_{i,AB}^{13\Gyr}$ &
$M_{i,AB}^{9\Gyr}$ &
$({{\rm mass}\over{10^9 \msun}})^{13\Gyr}$ &
$({{\rm mass}\over{10^9 \msun}})^{9\Gyr}$ &
$({{L}\over{10^9 \lsun}})^{13\Gyr}$ &
$({{L}\over{10^9 \lsun}})^{9\Gyr}$ \\
\tableline
$-2.5 < {\rm [Fe/H]} < -1.7$\footnotemark[1] & -14.3 & -14.5 & -15.1 & -15.2 &  0.35 &  0.35 & 0.08 & 0.08 \\
$-1.7 < {\rm [Fe/H]} < -1.1$ & -14.7 & -14.8 & -15.5 & -15.6 &  0.66  &  0.71  & 0.11 & 0.12 \\
$-1.1 < {\rm [Fe/H]} < -0.6$ & -14.9 & -15.0 & -16.1 & -16.1 &  1.1  &  1.2  & 0.22  & 0.21  \\
$-0.6 < {\rm [Fe/H]} <  0.0$ & -14.5 & -14.1 & -15.8 & -15.0 &  1.1  &  0.64  & 0.14  & 0.06  \\
$-2.5 < {\rm [Fe/H]} <  0.0$ & -16.0 & -16.0 & -17.1 & -16.8 &  3.0  & 2.2  & 0.52  & 0.28  \\
\tableline\tableline
\end{tabular}
\end{center}
{\bf Notes.}
\footnotetext[1]{The masks of Fig.~\ref{fig:map_masks} corresponding to each of these metallicity intervals are used, hence the final row is not simply the sum of the previous rows.}
\end{table*}

\section{Discussion}

\subsection{What is the nature of the smooth halo component?}

The Keck/DEIMOS spectroscopic survey of G12 (a collaboration known as ``Spectroscopic and Photometric Landscape of Andromeda's Stellar Halo'' or SPLASH) presented kinematics from a total of 38 DEIMOS fields (covering approximately $1/500$ of the area of the present CFHT survey) distributed in an irregular fashion around M31. Their fields probe the halo primarily along the minor axis, but they also targeted dwarf galaxies and other substructures (see Figure~1 of G12). In terms of the region surveyed, all fields except one (targeting the satellite And~VII) are contained within a projected circle of $150\kpc$ of M31, which makes their study a very useful complement to PAndAS.

Both PAndAS and SPLASH are dominated by contaminants, but the two studies identify and correct for these in different ways. G12 de-contaminated their sample by using kinematics and several diagnostic parameters to minimize the number of Milky Way dwarf stars. However, since not every candidate star in their color-magnitude selection box within the DEIMOS fields could be observed, they further used the Besan{\c c}on Galactic model \citep{Robin:2003jk} to weight the relative number of M31 RGB stars in each field. In the present analysis, we use an empirical fit to the contamination, as we found that the Besan{\c c}on model does not give a sufficiently good representation of the foreground populations over the entire PAndAS survey (with errors exceeding $\sim 50$\% for some populations).

G12 measure the velocity distribution in each field, and remove features in velocity space that are related to the known substructure. They also performed a multi-Gaussian fit to the velocity distribution in all fields to identify any additional kinematically cold components. Slightly under a third of their fields (12/38) contained such kinematically identified substructure. In addition, a high velocity dispersion ``hot'' spheroid was fit in all fields, though they note that beyond a projected distance of $90\kpc$ the statistics were not sufficient to disentangle kinematic substructure if it were present.

While we do not have the information to relate directly our smooth halo component to the hot kinematic component detected in the G12 fields, it is highly plausible that the two populations are closely related. They appear to have the same power-law exponent (further discussion below), and in both analyses they are what is left over after the most obvious substructure has been removed. However, it is unclear whether cosmological halo formation simulations (e.g., \citealt{Bullock:2005is,2010MNRAS.406..744C}) are compatible with the significant fraction of stars in this ``smooth'' component, given that all stars at these distances in the simulations will have been accreted from disrupted satellite galaxies, and such debris has not had enough time to phase-mix. It is of course completely plausible that future deeper photometric and spectroscopic surveys will manage to resolve this ``smooth'' halo component into further substructure (beyond $\Sigma_V \simgt 32 {\rm \, mag \, arcsec^{-2}}$). Given the apparent smoothness in the spatial distribution and the high dispersion of the kinematic distributions, it is likely that this would have to involve a large number of separate accretions.

However, it seems difficult to accommodate these many accretions into this smooth component, since the total luminosity integrated over projected radii between $30\kpc$ and $150\kpc$ is modest ($\sim 1.9 \times 10^8\lsun$ for ${\rm [Fe/H]} < -1.1$). If what we have identified as a smooth halo was built in this way, its building blocks must have been structures of very low stellar luminosity accreted at very early times in the formation of M31. For comparison, in the simulations of \citet{Johnston:2008jp} it is only the accreted material of age $\sim 12\Gyr$ that is relatively free of stream-like structure.

\subsection{Radial profile of the halo}

Despite the above-mentioned differences in the elimination of contamination between PAndAS and SPLASH, the power-law fits to the projected counts of metal-poor stars (Figs.~\ref{fig:projected_profile_sel4}b and \ref{fig:projected_profile_sel3}b, with index of $-2.08\pm0.02$ and $-2.13\pm0.02$, respectively) are in excellent agreement with the \citet{2012ApJ...760...76G} fit of $-2.2\pm0.2$ to their ``substructure-removed'' sample. This provides a powerful cross-check of our analyses. The agreement suggests that the G12 halo analysis was not substantially biassed by the large fraction of their fields that targeted substructure. This was a concern, since we now know that globular clusters preferentially lie in substructure-dominated regions \citep{2010ApJ...717L..11M}, while dwarf galaxies are themselves obvious sources of stream material. Evidently, the choice made by G12 to consider objects beyond the tidal radius of their targeted satellite galaxies as M31 halo stars did not significantly affect their halo profile, probably due to the care they took to remove kinematic spikes.

PAndAS is the first wide-field survey of M31 where it has been possible to properly remove the spatially coherent substructure (at least down to $\Sigma_V \sim 32 {\rm \, mag \, arcsec^{-2}}$) to identify the properties of the underlying ``smooth'' halo population. The analysis in Paper~I was hampered by the presence of copious sub-structures in the Southern quadrant of M31, but excising the obvious over-densities we found $\Sigma_V(R) \propto R^{-1.9\pm0.1}$, consistent with the present work. A number of other studies that attempted to measure the halo profile are reviewed in G12, but these either probe small fields (and hence are highly prone to unidentified substructure) or probe only the inner halo region, and so are not directly comparable to G12 and the present contribution.

It is fascinating that in addition to the spatial smoothness, and apparent high velocity dispersion, the ``smooth'' metal-poor halo component additionally possesses a smooth power-law profile (Figs.~\ref{fig:projected_profile_sel4} and~\ref{fig:projected_profile_sel3}). This reinforces the notion that it is probably an ancient, well-mixed structure (or that it is composed of many ancient, smoothed-out remnants).

In the simulations of \citet{Bullock:2005is}, who used a hybrid semi-analytic plus N-body approach to model galaxy formation, the halo density profiles tend to steepen with radius from approximately $-3$ to $-3.5$ beyond a radius of $50$--$100\kpc$. They argue that if the internal structure of the satellites that contributed to the buildup of the halo was more extended, this would give rise to a more extended halo (i.e. with a larger radius at which the fall-off to higher power-law index begins).  Figures~\ref{fig:projected_profile_sel4} and~\ref{fig:projected_profile_sel3} show no sign of such a fall-off, either in the masked, or unmasked samples, possibly implying that the disrupted satellites that formed the M31 halo were fluffier than in the simulations of \citet{Bullock:2005is}. It is interesting to note that if the satellites that formed the halo were indeed more extended relative to their dark matter sub-halo and so less tightly-bound, it is almost certainly easier --- in the available time --- to form a smoother halo distribution from their stellar material. 

Presumably another means to generate the difference in profile between the \citet{Bullock:2005is} simulations and our observations is if the ancient (now defunct) satellites of the real M31 system had more radial (plunging) orbits with smaller peri-centers, which would have been disrupted more readily. A similar behavior of steepening of the density profile is reported in the simulations of \citet{2010MNRAS.406..744C}, which were also based on semi-analytic schemes, where the sharp drop-off occurs in all their models interior to $100\kpc$. Our non-parametric modeling (Fig.~\ref{fig:nonparametric_profile}) aimed to measure the three-dimensional density profile as well as possible with current data. Interestingly, the data prefer a slight steepening of the density profile beyond $100\kpc$ to $\sim -3.5$, although the outer power-law appears more in-line with the predictions of \citet{Bullock:2005is} than the very abrupt drop-off apparent in the \citet{2010MNRAS.406..744C} simulations.

\citet{2013ApJ...763..113D} interpret breaks in the halo profile as an indication of the range of apocenters of the halo progenitors. A continuous profile, such as what we observe in M31, would then suggest that no single accretion event formed the bulk of the halo.

The power-law exponent measured here is similar to those found in studies of the Milky Way. In a spectroscopic survey of many small fields that was rather similar in design to G12, \citet{Morrison:2000ew} reported $\gamma=-3$, while \citet{Juric:2008ie} using SDSS star-counts (and so similar to the present study) found $\gamma=-2.78\pm0.03$. While these comparisons show surprising similarity, we caution that they are probably not referring to the same structures in the two galaxies. Large-scale studies of the Milky Way halo so far have been heavily biassed to what we would call the inner halo in M31 --- a region that is characterized by a rich mixture of complex substructures in in Andromeda. In the Milky Way several studies have shown evidence for a break in the halo profile: this is seen in RRLyrae stars at a Galactocentric distance of $\sim 25\kpc$ \citep{2009MNRAS.398.1757W}, in main sequence turn-off stars at $\sim 28\kpc$ \citep{2011ApJ...731....4S}, and in BHB stars at $27\kpc$ \citep{2011MNRAS.416.2903D}. Yet at these distances in M31, the halo is clearly highly contaminated, especially by the Giant Stellar Stream. We suspect that a useful comparative study of the outer halos of the two Local Group giants will have to await the advent of the LSST and, more imminently, HyperSuprimeCam.

\subsection{Metallicity distribution}
\label{sec:metallicity_distribution}

The metallicity measurements presented in \S\ref{sec:metallicity_structure} are in reasonable agreement with the results of earlier spectroscopic analyses that sampled the outer halo of M31 at a small number of locations \citep{Chapman:2006ia,Kalirai:2006ix,2008ApJ...689..958K}, and which reported mean metallicities in the range $\langle{\rm [Fe/H]}\rangle \sim -1.26$ to $-2.0$. The differences between authors are accountable to differences in the techniques of the measurement and calibration of the \ion{Ca}{2} triplet equivalent widths, and probably also due in part to field-to-field variance. While no similar panoramic survey has been conducted in another more distant giant spiral galaxy, we find that our measurements are qualitatively similar to results in other systems: e.g., extra-planar stars resolved with HST in NGC~891 show $\langle{\rm [Fe/H]}\rangle = -1.3$ at $R=20\kpc$ \citep{2009MNRAS.395..126I}, while integrated light at a distance of $R=8.5\kpc$ above the plane of the disk of NGC~3957 has colors consistent with a metallicity of $\langle{\rm [Fe/H]}\rangle = -1.3$  \citep{Jablonka:2010gp}.

At a qualitative level, the complex metallicity structure seen in Fig.~\ref{fig:map_full_metallicity} agrees well with the simulations of \citet[][see their Figure~1]{Font:2006hx}, who studied the chemical abundance variations due to hierarchical accretion in the formation of galaxy halos. The agreement can be seen in the visual prominence and number of streams, as well as in the variations of metallicity from one stream to another.

\citet{Font:2006hx} suggested that a single massive metal-rich accretion could be responsible for high metallicity stars in the halo of a galaxy. This is indeed the case in M31, as the metal-rich GSS is responsible for the majority of the metal rich stars in the halo region.

They also find that the Milky Way is an unusual galaxy given its paucity of old metal-rich stars, which are commonly accreted in their simulations at ancient times from massive satellites. The detection of a modest fraction of metal-rich stars at large radius in M31 distributed in a smooth component suggests that their models are consistent, however, with the abundance in the Andromeda halo. We find that the profile of the metal-rich stars (${\rm [Fe/H] > -1.1}$; Fig.~\ref{fig:projected_profile_sel1and2}) is substantially steeper than that of more metal poor stars (${\rm [Fe/H] < -1.1}$; Figs.~\ref{fig:projected_profile_sel4} and \ref{fig:projected_profile_sel3}), which of course implies that there is a large metallicity gradient through the halo. This broad property was again predicted by the \citet{Font:2006hx} simulations.

However, it is the combination of the smooth power-law profiles (Figs.~\ref{fig:projected_profile_sel4}b and \ref{fig:projected_profile_sel3}b) and the metallicity gradient (Figs.~\ref{fig:metallicity_distribution}g-k) in the ``smooth'' component that is particularly interesting. It indicates the presence of different families of stars. As we proceed inwards from the outer boundary of the survey, we detect more and more metal-rich families with smaller apocenters. A large number of families is required since there is no break in the profiles. This is remarkable fossil evidence of the formation of the halo, and is consistent with a picture in which more massive satellites, which due to their deeper potential wells are able to hold on to more stellar ejecta and hence enrich further in ${\rm [Fe/H]}$, are also those satellites that suffer more orbital decay due to dynamical friction.

\subsection{Extent of halo}

The PAndAS survey shows that the M31 halo is a vast structure, extending out in every direction to at least a projected distance of $150\kpc$, i.e. approximately half of the virial radius (estimated to be $290\kpc$, \citealt{Klypin:2002bm}). Indeed, our three-dimensional fit is consistent with stars being distributed along the line of sight all the way out to the virial radius.

Of course, given that orbital time-scales are close to a Hubble time at such distances, it is inconceivable that the outermost reaches of the stellar halo should be smooth and azimuthally uniform. This is indeed what we see in the projected star-counts profiles of the smooth metal-poor populations (Figs.~\ref{fig:projected_profile_sel4}b and~\ref{fig:projected_profile_sel3}b), where the scatter between quadrants appears to increase beyond $100\kpc$ in projection. (It has not escaped our attention that the scatter is also larger at small radius). The variance between the different quadrants in the outermost two radial bins is clearly very large. However, we will return to the statistics of substructure, and the quantification of the variance in the halo in a future contribution (McConnachie et al., in preparation, Paper II in this series).

\subsection{Shape of the stellar halo}

One of the most interesting results presented above is the finding that the smooth halo component underlying the identifiable substructure is approximately spherical. Formally, the solutions are slightly prolate, with mass flattening $q$ progressing from $q=1.01\pm0.07$ for the most metal-rich selection to $q=1.09\pm0.03$ for the most metal-poor. While our analysis method differed from that of G12 (we integrated M31 halo models along the line of sight, accounting for the dimming luminosity function, and took the disk plane as the plane of symmetry of the halo), our results are again very similar. G12 also found that their spectroscopic sample favored a prolate halo\footnote{There is a typo on the lower uncertainty limit on page 17 of G12.} with $q=1.06^{+0.11}_{-0.09}$. The fact that a kinematically selected sample from a small number of pencil-beam fields yields the same result as a panoramic photometrically-selected sample lends strong support to both studies. A robust result from these studies is that at large radii the M31 stellar halo is closely spherical. 

In contrast, measurements of the flattening of the Milky Way stellar halo find a much more oblate structure, with $q \sim 0.5$--$0.8$ \citep{Juric:2008ie}, and hints of an even more flattened halo sub-component with $q \sim 0.2$ have been reported \citep{Morrison:2009bb}. However, we suspect that these differences with respect to Andromeda are probably due to the fact that most Milky Way studies (and in particular the SDSS) probe much smaller radii where a low number of large accretions have deposited their stars. It is clear that the M31 halo would also appear very flattened to an observer located in Andromeda studying their halo with a survey like the SDSS, given that the inner halo is dominated by a flattened structure produced by the disruption of the Giant Stellar Stream.

\subsection{Shape of the dark matter distribution}

Our decomposition of the M31 halo into ``lumpy'' and ``smooth'' components carries all the caveats we laid out in \S\ref{sec:deconstructing}. Nevertheless, the apparent large-scale homogeneity of the smooth component, as well as the high velocity dispersion seen in the individual G12 spectroscopic fields, strongly suggests that this component is a reasonably well phase-mixed population and hence a reasonable tracer of the underlying gravitational potential.

In collisionless cold dark matter simulations, cosmological halos possess an approximate triaxial shape, with axis ratios $c/a \sim 0.6$, $c/b \sim 0.7$ \citep[][where $c \le b \le a$]{2002ApJ...574..538J}. However, these triaxial structures become approximately oblate when baryonic physics is included in the simulations \citep[see, e.g.][]{2010MNRAS.407..435A}, with $c/a\sim 0.8$, $c/b\sim 0.9$ for the dark matter component, at a radial distance corresponding to $\sim 30\kpc$ in M31 \citep[][see their Figure~11]{2012ApJ...748...54Z}. Beyond that radius the dark matter halo is expected to become (on average) more triaxial, attaining $c/a\sim 0.7$, $c/b\sim 0.8$ at $\sim 100\kpc$. It has been extremely difficult to test these predictions observationally, and unfortunately our best measurements of the shape of dark matter halos on galaxy scales, which came from weak lensing studies, have recently been shown to be fraught with difficulties \citep[see][and references therein]{2010MNRAS.407..891H,2012MNRAS.420.3303B}. Obtaining constraints on the galactic dark matter distribution from nearby galaxies is therefore of great interest.

The approximate sphericity of the smooth stellar halo measured in this contribution is surprising in the context of the above predictions of dark matter morphology. In principle, there is a large amount of freedom to populate the dark matter halo with a stellar tracer population, and undoubtedly one could find equilibrium solutions with peculiar velocity dispersion tensors that would allow for the observed spherical stellar distribution. Nevertheless, we suspect that such solutions are contrived in the sense that there will be very few ways to form a spherical stellar system within a triaxial dark matter halo, whereas there is a vast number of ways to form a triaxial stellar system in the same potential. Moreover, the sphericity of the stellar halo is detected in several bins in metallicity, and as we have seen in \S\ref{sec:metallicity_distribution}, this together with the power-law profiles, implies that several sub-populations fill the halo, each possessing different apocenter and metallicity. Unless the sub-populations were formed from the disruption of progenitors that had correlated orbital properties, there would have to be a similarly contrived spherical solution for each sub-population. This of course compounds the improbability of the scenario.

Hence we conclude that our observations are much more naturally explained if the M31 dark matter halo is also close to spherical. It will be very interesting to confront this possibility by analyzing the kinematics of a spatially well-selected sample of stars belonging to the smooth halo component.

\subsection{The M31 satellite alignment in light of the smooth halo}

In a recent study that was also based on PAndAS data, \citet{2013Natur.493...62I} showed that 50\% of the satellite galaxies of Andromeda (13 of the then known 27) are contained within a very thin plane \citep{2013ApJ...766..120C} and possess a common sense of rotation about M31. Although the exact formation mechanism of this structure is unknown, and indeed poses quite a puzzle \citep{2013MNRAS.tmp.1111H,2013arXiv1307.2102G}, it is clear that it implies that the formation of this sub-set of M31 satellites was highly correlated. The nature of the remaining 50\% of satellites is unclear, but one could imagine an extreme scenario in which the satellites that are not part of the plane might also originate in a similar way, perhaps coming from a small number of additional families.

However, the detection in the present contribution of a smooth, spherical halo with a well-behaved power-law profile that is the dominant component at low metallicity, requires, almost certainly, many progenitors with different angular momenta to fill out the vast volume in this regular manner. The alternative would be that the metal poor smooth halo formed in situ, but this is not possible in standard galaxy formation models. If there was a significant number of uncorrelated satellites in the past, it would therefore seem likely that the present-day satellites that are not in the correlated planar structure are mostly not related to each other.

\section{Conclusions}

We have analyzed the large-scale properties of the halo of the Andromeda galaxy, using photometric data from the Pan-Andromeda Archaeological Survey. This large endeavor has measured stellar sources down to $g=26.0$, $i=24.8$ covering M31 and M33 out to a projected radius of $150\kpc$ and $50\kpc$, respectively. This is the first survey of M31 with which it has been possible to properly remove the substructure to recover the wide-field panoramic properties of the underlying halo.

We have assumed that the stellar populations beyond the inner disk are predominantly old ($\simgt 10 \Gyr$); with this assumption the age-metallicity degeneracy is largely avoided, and a simple conversion from color and magnitude to metallicity is possible using stellar population models. In order to examine the population properties we divided the sample into four wide metallicity bins. Our main findings are:
\begin{enumerate}
\item For stars more metal-rich than ${\rm [Fe/H]} = -1.1$ the halo is almost completely dominated by the debris of a single accretion, known as the Giant Stellar Stream. This pervades the halo out to $100\kpc$, especially in the Southern quadrant, and is responsible for the apparently flattened inner ``halo'' at radii of up to $\sim 50\kpc$. 
\item The populations with metallicity in the range ${\rm -1.7 < [Fe/H] < -1.1}$ show a rich web of intercrossing streams (at least in projection), testifying that numerous intermediate mass accretion events took place during the formation of M31. In the most metal-poor selection, with ${\rm -2.5 < [Fe/H] < -1.7}$, the contrast of the streams is greatly diminished and a clear sign of what appears to be a diffuse smooth halo is perceived.
\item Despite the numerous substructures, and substantial azimuthal variations in density, the azimuthally-averaged projected star-counts profiles are remarkably featureless and possess power-law behavior. The power-law fits become steeper with increasing metallicity, being $\Gamma=-2.3\pm0.02$ for stars in the range ${\rm -2.5 < [Fe/H] < -1.7}$ and $\Gamma=-3.72\pm0.01$ for ${\rm -1.1 < [Fe/H] < 0}$.
\item In order to disentangle the substructure from what appears to be a smooth component (but which we argue may just be a consequence of our observational limitations) we developed a new algorithm that fits a galaxy halo model and a contamination model to the data. Given the fact that the depth of the M31 halo ($\pm \sim 300\kpc$) is a large fraction of the distance to that galaxy, it was necessary to integrate the model in a cone along the line of sight, taking into account the dimming of the stellar populations with distance. The algorithm automatically identifies regions with overdense substructure. Applying this technique to the data, we find that a ``smooth'' halo component is present in all of our metallicity subsamples, albeit in a minor fraction ($\sim14$\%) for the most metal-rich stars, but rising to $\sim58$\% for the most metal-poor stars. 
\item This ``smooth'' halo follows closely a power-law profile in projection, with an exponent of $\Gamma=-2.08\pm0.02$ for the most metal-poor stars. Fits using our three-dimensional non-parametric model have a space density that follows a power law with exponent $\gamma \sim -3$ out to $100\kpc$ beyond which the profile steepens slightly to $\gamma \sim -3.5$, in good agreement with the simulations of \citet{Bullock:2005is}.
\item The fits also show that the shape of the smooth halo population is close to spherical in the four metallicity sub-samples. The global structure we are calling a smooth halo can be identified with the hot kinematic component detected with Keck/DEIMOS spectroscopy by \citet{2012ApJ...760...76G} in 38 pencil-beam fields. This strongly suggests a formation from a large number of satellites arriving in an uncorrelated way with a range of angular momentum, energy and metallicity. We argue that the sphericity of the stellar distribution in each metallicity sub-sample suggests that the dark matter distribution is not strongly triaxial, but also close to spherical.
\item By summing the stars in our survey and accounting for the stars outside of the color-magnitude selection box, we estimate that the total stellar mass in the halo beyond $2\deg$ is $\sim 1.1\times 10^{10}\msun$, while that of the ``smooth'' component is $\sim 3\times 10^{9}\msun$. If the smooth halo follows a cored-power law profile into the center of M31, we estimate (very roughly) a total mass of $\sim 8\times 10^{9}\msun$ for this component. These are considerable fractions ($\sim 10$\%) of the baryonic mass of M31.
\item A substantial metallicity gradient is observed, both in the full halo sample and in the smooth halo sample. The mean metallicity of the full halo drops monotonically from $\langle{\rm [Fe/H]}\rangle = -0.7$ at $R=27.2\kpc$ down to $\langle{\rm [Fe/H]}\rangle = -1.5$ at $R=150\kpc$. The smooth halo follows the same trend, but approximately 0.2~dex more metal poor at a given radius.
\end{enumerate}

So we find that Andromeda's halo is indeed in rough agreement with the expectations of hierarchical galaxy formation, but to go beyond these relatively broad-brush tests, we now need to confront a new generation of models to our data. This will be the subject of a subsequent contribution.

\acknowledgments

We thank the staff of the Canada-France-Hawaii Telescope for taking the PAndAS data, and for their continued support throughout the project.
R.A.I. gratefully acknowledges support from the Agence Nationale de la Recherche though the grant POMMME (ANR 09-BLAN-0228).
G.F.L thanks the Australian research council for support through his Future Fellowship (FT100100268) and Discovery Project (DP110100678).


\begin{thebibliography}{69}
\expandafter\ifx\csname natexlab\endcsname\relax\def\natexlab#1{#1}\fi


\bibitem[{Abadi {et~al.}(2010)Abadi, Navarro, Fardal, Babul, \&
  Steinmetz}]{2010MNRAS.407..435A}
Abadi, M.~G., Navarro, J.~F., Fardal, M., Babul, A., \& Steinmetz, M. 2010,
  MNRAS, 407, 435

\bibitem[{Abadi {et~al.}(2006)Abadi, Navarro, \&
  Steinmetz}]{2006MNRAS.365..747A}
Abadi, M.~G., Navarro, J.~F., \& Steinmetz, M. 2006, MNRAS, 365, 747

\bibitem[{Bailin {et~al.}(2011)Bailin, Bell, Chappell, Radburn-Smith, \&
  de~Jong}]{2011ApJ...736...24B}
Bailin, J., Bell, E.~F., Chappell, S.~N., Radburn-Smith, D.~J., \& de~Jong,
  R.~S. 2011, ApJ, 736, 24

\bibitem[{Barker {et~al.}(2009)Barker, Ferguson, Irwin, Arimoto, \&
  Jablonka}]{2009AJ....138.1469B}
Barker, M.~K., Ferguson, A. M.~N., Irwin, M., Arimoto, N., \& Jablonka, P.
  2009, AJ, 138, 1469

\bibitem[{Bell {et~al.}(2008)Bell, Zucker, Belokurov, Sharma, Johnston,
  Bullock, Hogg, Jahnke, \& {et al.}}]{Bell:2008bc}
Bell, E.~F., {et~al.} 2008, ApJ, 680, 295

\bibitem[{Bernard {et~al.}(2012)Bernard, Ferguson, Barker, Hidalgo, Ibata,
  Irwin, Lewis, McConnachie, \& {et al.}}]{Bernard:2012kz}
Bernard, E.~J., {et~al.} 2012, MNRAS, 420, 2625

\bibitem[{Bett(2012)}]{2012MNRAS.420.3303B}
Bett, P. 2012, MNRAS, 420, 3303

\bibitem[{Brown {et~al.}(2007)Brown, Smith, Ferguson, Guhathakurta, Kalirai,
  Rich, Renzini, Sweigart, \& {et al.}}]{Brown:2007gq}
Brown, T.~M., {et~al.} 2007, ApJ, 658, L95

\bibitem[{Brown {et~al.}(2006)Brown, Smith, Ferguson, Rich, Guhathakurta,
  Renzini, Sweigart, \& Kimble}]{Brown:2006iq}
Brown, T.~M., Smith, E., Ferguson, H.~C., Rich, R.~M., Guhathakurta, P.,
  Renzini, A., Sweigart, A.~V., \& Kimble, R.~A. 2006, ApJ, 652, 323

\bibitem[{Bullock \& Johnston(2005)}]{Bullock:2005is}
Bullock, J.~S., \& Johnston, K.~V. 2005, ApJ, 635, 931

\bibitem[{Carollo {et~al.}(2007)Carollo, Beers, Lee, Chiba, Norris, Wilhelm,
  Sivarani, Marsteller, \& {et al.}}]{Carollo:2007fw}
Carollo, D., {et~al.} 2007, Nature, 450, 1020

\bibitem[{Chapman {et~al.}(2006)Chapman, Ibata, Lewis, Ferguson, Irwin,
  McConnachie, \& Tanvir}]{Chapman:2006ia}
Chapman, S.~C., Ibata, R., Lewis, G.~F., Ferguson, A. M.~N., Irwin, M.,
  McConnachie, A., \& Tanvir, N. 2006, ApJ, 653, 255

\bibitem[{Cockcroft {et~al.}(2013)Cockcroft, McConnachie, Harris, Ibata, Irwin,
  Ferguson, Fardal, Babul, \& {et al.}}]{2013MNRAS.428.1248C}
Cockcroft, R., {et~al.} 2013, MNRAS, 428, 1248

\bibitem[{Conn {et~al.}(2012)Conn, Ibata, Lewis, Parker, Zucker, Martin,
  McConnachie, Irwin, \& {et al.}}]{2012ApJ...758...11C}
Conn, A.~R., {et~al.} 2012, ApJ, 758, 11

\bibitem[{Conn {et~al.}(2013)Conn, Lewis, Ibata, Parker, Zucker, McConnachie,
  Martin, Valls-Gabaud, \& {et al.}}]{2013ApJ...766..120C}
---. 2013, ApJ, 766, 120

\bibitem[{Cooper {et~al.}(2010)Cooper, Cole, Frenk, White, Helly, Benson,
  De~Lucia, Helmi, \& {et al.}}]{2010MNRAS.406..744C}
Cooper, A.~P., {et~al.} 2010, MNRAS, 406, 744

\bibitem[{Courteau {et~al.}(2011)Courteau, Widrow, McDonald, Guhathakurta,
  Gilbert, Zhu, Beaton, \& Majewski}]{Courteau:2011fr}
Courteau, S., Widrow, L.~M., McDonald, M., Guhathakurta, P., Gilbert, K.~M.,
  Zhu, Y., Beaton, R.~L., \& Majewski, S.~R. 2011, ApJ, 739, 20

\bibitem[{Crnojevic {et~al.}(2013)Crnojevic, Ferguson, Irwin, Bernard, Arimoto,
  Jablonka, \& Kobayashi}]{Crnojevic2013}
Crnojevic, D., Ferguson, A. M.~N., Irwin, M.~J., Bernard, E.~J., Arimoto, N.,
  Jablonka, P., \& Kobayashi, C. 2013, arXiv

\bibitem[{Dalcanton {et~al.}(2009)Dalcanton, Williams, Seth, Dolphin, Holtzman,
  Rosema, Skillman, Cole, \& {et al.}}]{2009ApJS..183...67D}
Dalcanton, J.~J., {et~al.} 2009, ApJS, 183, 67

\bibitem[{Deason {et~al.}(2011)Deason, Belokurov, \&
  Evans}]{2011MNRAS.416.2903D}
Deason, A.~J., Belokurov, V., \& Evans, N.~W. 2011, MNRAS, 416, 2903

\bibitem[{Deason {et~al.}(2013)Deason, Belokurov, Evans, \&
  Johnston}]{2013ApJ...763..113D}
Deason, A.~J., Belokurov, V., Evans, N.~W., \& Johnston, K.~V. 2013, ApJ, 763,
  113

\bibitem[{Dotter {et~al.}(2008)Dotter, Chaboyer, Jevremovi{\'c}, Kostov, Baron,
  \& Ferguson}]{2008ApJS..178...89D}
Dotter, A., Chaboyer, B., Jevremovi{\'c}, D., Kostov, V., Baron, E., \&
  Ferguson, J.~W. 2008, ApJS, 178, 89

\bibitem[{Fardal {et~al.}(2006)Fardal, Babul, Geehan, \&
  Guhathakurta}]{2006MNRAS.366.1012F}
Fardal, M.~A., Babul, A., Geehan, J.~J., \& Guhathakurta, P. 2006, MNRAS, 366,
  1012

\bibitem[{Fardal {et~al.}(2007)Fardal, Guhathakurta, Babul, \&
  McConnachie}]{2007MNRAS.380...15F}
Fardal, M.~A., Guhathakurta, P., Babul, A., \& McConnachie, A.~W. 2007, MNRAS,
  380, 15

\bibitem[{Fardal {et~al.}(2012)Fardal, Guhathakurta, Gilbert, Tollerud,
  Kalirai, Tanaka, Beaton, Chiba, \& {et al.}}]{2012MNRAS.423.3134F}
Fardal, M.~A., {et~al.} 2012, MNRAS, 423, 3134

\bibitem[{Fardal {et~al.}(2013)Fardal, Weinberg, Babul, Irwin, Guhathakurta,
  Gilbert, Ferguson, Ibata, \& {et al.}}]{2013MNRAS.434.2779F}
---. 2013, MNRAS, 434, 2779

\bibitem[{Font {et~al.}(2006)Font, Johnston, Bullock, \&
  Robertson}]{Font:2006hx}
Font, A.~S., Johnston, K.~V., Bullock, J.~S., \& Robertson, B.~E. 2006, ApJ,
  646, 886

\bibitem[{Font {et~al.}(2008)Font, Johnston, Ferguson, Bullock, Robertson,
  Tumlinson, \& Guhathakurta}]{Font:2008jh}
Font, A.~S., Johnston, K.~V., Ferguson, A. M.~N., Bullock, J.~S., Robertson,
  B.~E., Tumlinson, J., \& Guhathakurta, P. 2008, ApJ, 673, 215

\bibitem[{Gilbert {et~al.}(2012)Gilbert, Guhathakurta, Beaton, Bullock, Geha,
  Kalirai, Kirby, Majewski, \& {et al.}}]{2012ApJ...760...76G}
Gilbert, K.~M., {et~al.} 2012, ApJ, 760, 76

\bibitem[{Gilbert {et~al.}(2009)Gilbert, Guhathakurta, Kollipara, Beaton, Geha,
  Kalirai, Kirby, Majewski, \& {et al.}}]{2009ApJ...705.1275G}
---. 2009, ApJ, 705, 1275

\bibitem[{Goerdt \& Burkert(2013)}]{2013arXiv1307.2102G}
Goerdt, T., \& Burkert, A. 2013, arXiv, 2102

\bibitem[{Hammer {et~al.}(2013)Hammer, Yang, Fouquet, Pawlowski, Kroupa, Puech,
  Flores, \& Wang}]{2013MNRAS.tmp.1111H}
Hammer, F., Yang, Y., Fouquet, S., Pawlowski, M.~S., Kroupa, P., Puech, M.,
  Flores, H., \& Wang, J. 2013, MNRAS, 1111

\bibitem[{Helmi {et~al.}(2011)Helmi, Cooper, White, Cole, Frenk, \& Navarro}]{2011ApJ...733L...7H}
Helmi, A., Cooper, A.~P., White, S.~D.~M., Cole, S., Frenk, C.~S., \& Navarro, J. 2011, ApJ, 733, L7

\bibitem[{Howell \& Brainerd(2010)}]{2010MNRAS.407..891H}
Howell, P.~J., \& Brainerd, T.~G. 2010, MNRAS, 407, 891

\bibitem[{Ibata {et~al.}(2004)Ibata, Chapman, Ferguson, Irwin, Lewis, \&
  McConnachie}]{Ibata:2004di}
Ibata, R., Chapman, S., Ferguson, A. M.~N., Irwin, M., Lewis, G., \&
  McConnachie, A. 2004, MNRAS, 351, 117

\bibitem[{Ibata {et~al.}(2001)Ibata, Irwin, Lewis, Ferguson, \&
  Tanvir}]{Ibata:2001vs}
Ibata, R., Irwin, M., Lewis, G., Ferguson, A. M.~N., \& Tanvir, N. 2001,
  Nature, 412, 49

\bibitem[{Ibata {et~al.}(2007)Ibata, Martin, Irwin, Chapman, Ferguson, Lewis,
  \& McConnachie}]{Ibata:2007jr}
Ibata, R., Martin, N.~F., Irwin, M., Chapman, S., Ferguson, A. M.~N., Lewis,
  G.~F., \& McConnachie, A.~W. 2007, ApJ, 671, 1591

\bibitem[{Ibata {et~al.}(2009)Ibata, Mouhcine, \&
  Rejkuba}]{2009MNRAS.395..126I}
Ibata, R., Mouhcine, M., \& Rejkuba, M. 2009, MNRAS, 395, 126

\bibitem[{Ibata {et~al.}(2013{\natexlab{a}})Ibata, Nipoti, Sollima, Bellazzini,
  Chapman, \& Dalessandro}]{2013MNRAS.428.3648I}
Ibata, R., Nipoti, C., Sollima, A., Bellazzini, M., Chapman, S.~C., \&
  Dalessandro, E. 2013{\natexlab{a}}, MNRAS, 428, 3648

\bibitem[{Ibata {et~al.}(2013{\natexlab{b}})Ibata, Lewis, Conn, Irwin,
  McConnachie, Chapman, Collins, Fardal, \& {et al.}}]{2013Natur.493...62I}
Ibata, R.~A., {et~al.} 2013{\natexlab{b}}, Nature, 493, 62

\bibitem[{Irwin \& Lewis(2001)}]{Irwin:2001eq}
Irwin, M., \& Lewis, J. 2001, New Astronomy Reviews, 45, 105

\bibitem[{Jablonka {et~al.}(2010)Jablonka, Tafelmeyer, Courbin, \&
  Ferguson}]{Jablonka:2010gp}
Jablonka, P., Tafelmeyer, M., Courbin, F., \& Ferguson, A. M.~N. 2010, A\&A,
  513, A78

\bibitem[{Jing \& Suto(2002)}]{2002ApJ...574..538J}
Jing, Y.~P., \& Suto, Y. 2002, ApJ, 574, 538

\bibitem[{Johnston {et~al.}(2008)Johnston, Bullock, Sharma, Font, Robertson, \&
  Leitner}]{Johnston:2008jp}
Johnston, K.~V., Bullock, J.~S., Sharma, S., Font, A., Robertson, B.~E., \&
  Leitner, S.~N. 2008, ApJ, 689, 936

\bibitem[{Juri{\'c} {et~al.}(2008)Juri{\'c}, Ivezi{\'c}, Brooks, Lupton,
  Schlegel, Finkbeiner, Padmanabhan, Bond, \& {et al.}}]{Juric:2008ie}
Juri{\'c}, M., {et~al.} 2008, ApJ, 673, 864

\bibitem[{Kalirai {et~al.}(2006)Kalirai, Gilbert, Guhathakurta, Majewski,
  Ostheimer, Rich, Cooper, Reitzel, \& {et al.}}]{Kalirai:2006ix}
Kalirai, J.~S., {et~al.} 2006, ApJ, 648, 389

\bibitem[{Klypin {et~al.}(2002)Klypin, Zhao, \& Somerville}]{Klypin:2002bm}
Klypin, A., Zhao, H., \& Somerville, R.~S. 2002, ApJ, 573, 597

\bibitem[{Koch {et~al.}(2008)Koch, Rich, Reitzel, Martin, Ibata, Chapman,
  Majewski, Mori, \& {et al.}}]{2008ApJ...689..958K}
Koch, A., {et~al.} 2008, ApJ, 689, 958

\bibitem[{Mackey {et~al.}(2010)Mackey, Huxor, Ferguson, Irwin, Tanvir,
  McConnachie, Ibata, Chapman, \& {et al.}}]{2010ApJ...717L..11M}
Mackey, A.~D., {et~al.} 2010, ApJ, 717, L11

\bibitem[{Magnier \& Cuillandre(2004)}]{2004PASP..116..449M}
Magnier, E.~A., \& Cuillandre, J.-C. 2004, PASP, 116, 449

\bibitem[{Martin {et~al.}(2013)Martin, Ibata, McConnachie, Dougal~Mackey,
  Ferguson, Irwin, Lewis, \& Fardal}]{2013arXiv1307.7626M}
Martin, N.~F., Ibata, R.~A., McConnachie, A.~W., Dougal~Mackey, A., Ferguson,
  A. M.~N., Irwin, M.~J., Lewis, G.~F., \& Fardal, M.~A. 2013, arXiv, 7626

\bibitem[{Mart{\'\i}nez-Delgado {et~al.}(2010)Mart{\'\i}nez-Delgado, Gabany,
  Crawford, Zibetti, Majewski, Rix, Fliri, Carballo-Bello, \& {et
  al.}}]{2010AJ....140..962M}
Mart{\'\i}nez-Delgado, D., {et~al.} 2010, AJ, 140, 962

\bibitem[{Mart{\'\i}nez-Delgado {et~al.}(2008)Mart{\'\i}nez-Delgado,
  Pe{\~n}arrubia, Gabany, Trujillo, Majewski, \& Pohlen}]{2008ApJ...689..184M}
Mart{\'\i}nez-Delgado, D., Pe{\~n}arrubia, J., Gabany, R.~J., Trujillo, I.,
  Majewski, S.~R., \& Pohlen, M. 2008, ApJ, 689, 184

\bibitem[{McConnachie {et~al.}(2005)McConnachie, Irwin, Ferguson, Ibata, Lewis,
  \& Tanvir}]{McConnachie:2005hn}
McConnachie, A.~W., Irwin, M.~J., Ferguson, A. M.~N., Ibata, R.~A., Lewis,
  G.~F., \& Tanvir, N. 2005, MNRAS, 356, 979

\bibitem[{McConnachie {et~al.}(2009)McConnachie, Irwin, Ibata, Dubinski,
  Widrow, Martin, C{\^o}t{\'e}, Dotter, \& {et al.}}]{2009Natur.461...66M}
McConnachie, A.~W., {et~al.} 2009, Nature, 461, 66

\bibitem[{M{\'e}ndez {et~al.}(2002)M{\'e}ndez, Davis, Moustakas, Newman,
  Madore, \& Freedman}]{2002AJ....124..213M}
M{\'e}ndez, B., Davis, M., Moustakas, J., Newman, J., Madore, B.~F., \&
  Freedman, W.~L. 2002, AJ, 124, 213

\bibitem[{Morrison {et~al.}(2009)Morrison, Helmi, Sun, Liu, Gu, Norris,
  Harding, Kinman, \& {et al.}}]{Morrison:2009bb}
Morrison, H.~L., {et~al.} 2009, ApJ, 694, 130

\bibitem[{Morrison {et~al.}(2000)Morrison, Mateo, Olszewski, Harding,
  Dohm-Palmer, Freeman, Norris, \& Morita}]{Morrison:2000ew}
Morrison, H.~L., Mateo, M., Olszewski, E.~W., Harding, P., Dohm-Palmer, R.~C.,
  Freeman, K.~C., Norris, J.~E., \& Morita, M. 2000, AJ, 119, 2254

\bibitem[{Mouhcine {et~al.}(2010)Mouhcine, Ibata, \& Rejkuba}]{Mouhcine:2010cz}
Mouhcine, M., Ibata, R., \& Rejkuba, M. 2010, ApJ, 714, L12

\bibitem[{Paudel {et~al.}(2013)Paudel, Duc, C{\^o}t{\'e}, Cuillandre,
  Ferrarese, Ferriere, Gwyn, Mihos, \& {et al.}}]{Paudel:2013ww}
Paudel, S., {et~al.} 2013, arXiv

\bibitem[{Pe{\~n}arrubia(2013)}]{2013MNRAS.433.2576P}
Pe{\~n}arrubia, J. 2013, MNRAS, 433, 2576

\bibitem[{Regnault {et~al.}(2009)Regnault, Conley, Guy, Sullivan, Cuillandre,
  Astier, Balland, Basa, \& {et al.}}]{Regnault:2009bk}
Regnault, N., {et~al.} 2009, A\&A, 506, 999

\bibitem[{Richardson {et~al.}(2009)Richardson, Ferguson, Mackey, Irwin,
  Chapman, Huxor, Ibata, Lewis, \& {et al.}}]{2009MNRAS.396.1842R}
Richardson, J.~C., {et~al.} 2009, MNRAS, 396, 1842

\bibitem[{Robin {et~al.}(2003)Robin, Reyl{\'e}, Derri{\`e}re, \&
  Picaud}]{Robin:2003jk}
Robin, A.~C., Reyl{\'e}, C., Derri{\`e}re, S., \& Picaud, S. 2003, A\&A, 409,
  523

\bibitem[{Schlafly {et~al.}(2012)Schlafly, Finkbeiner, Juri{\'c}, Magnier,
  Burgett, Chambers, Grav, Hodapp, \& {et al.}}]{2012ApJ...756..158S}
Schlafly, E.~F., {et~al.} 2012, ApJ, 756, 158

\bibitem[{Schlegel {et~al.}(1998)Schlegel, Finkbeiner, \&
  Davis}]{Schlegel:1998fw}
Schlegel, D.~J., Finkbeiner, D.~P., \& Davis, M. 1998, ApJ, 500, 525

\bibitem[{Sch{\"o}nrich {et~al.}(2011)Sch{\"o}nrich, Asplund, \&
  Casagrande}]{2011MNRAS.415.3807S}
Sch{\"o}nrich, R., Asplund, M., \& Casagrande, L. 2011, MNRAS, 415, 3807

\bibitem[{Sesar {et~al.}(2011)Sesar, Juri{\'c}, \&
  Ivezi{\'c}}]{2011ApJ...731....4S}
Sesar, B., Juri{\'c}, M. \& Ivezi{\'c}, Z. 2011, ApJ, 731, 4

\bibitem[{Stetson(1987)}]{Stetson:1987fx}
Stetson, P.~B. 1987, PASP, 99, 191

\bibitem[{Tanaka {et~al.}(2011)Tanaka, Chiba, Komiyama, Guhathakurta, \&
  Kalirai}]{Tanaka:2011hm}
Tanaka, M., Chiba, M., Komiyama, Y., Guhathakurta, P., \& Kalirai, J.~S. 2011,
  ApJ, 738, 150

\bibitem[{Watkins {et~al.}(2009)Watkins, Evans, Belokurov, Smith, Hewett, Bramich, Gilmore, Irwin, \& {et al.}}]{2009MNRAS.398.1757W}
Watkins, L.~ L., {et~al.} 2009, MNRAS, 398, 1757

\bibitem[{Xue {et~al.}(2011)Xue, Rix, Yanny, Beers, Bell, Zhao, Bullock, Johnston, \& {et al.}}]{2011ApJ...738...79X}
Xue, X.-X., {et~al.} 2011, ApJ, 738, 79

\bibitem[{Zemp {et~al.}(2012)Zemp, Gnedin, Gnedin, \&
  Kravtsov}]{2012ApJ...748...54Z}
Zemp, M., Gnedin, O.~Y., Gnedin, N.~Y., \& Kravtsov, A.~V. 2012, ApJ, 748, 54

\bibitem[{Zolotov {et~al.}(2010)Zolotov, Willman, Brooks, Governato, Hogg,
  Shen, \& Wadsley}]{2010ApJ...721..738Z}
Zolotov, A., Willman, B., Brooks, A.~M., Governato, F., Hogg, D.~W., Shen, S.,
  \& Wadsley, J. 2010, ApJ, 721, 738

\end{thebibliography}

\end{document}